\documentclass[12pt]{article}
\usepackage{amsmath}
\usepackage{amssymb,amsfonts}
\usepackage{amsthm}
\usepackage[all]{xy}
\usepackage{graphicx}
\usepackage{setspace}
\usepackage{hyperref}
\usepackage{tikz-cd}
\usepackage{verbatim}
\usepackage{multirow}
\usepackage{mathtools}

\usepackage[parfill]{parskip}
\usepackage{tabu}
\usepackage{float}
\usepackage[utf8]{inputenc}

\usepackage{soul}
\usepackage{comment}
\usepackage{mathtools}
\usepackage{stmaryrd}

\usepackage{diagbox}
\definecolor{myblue}{rgb}{0,0,0.8}

\numberwithin{equation}{section}
\numberwithin{table}{section}

\def\Z{\mathbb Z}

\def\abSolLong{t_{k,L}}
\def\abSolShort{t_{k,S}}
\def\shortNorm{\kappa}
\def\interMatrix{\Omega}
\newcommand{\be}{\begin{equation}}
\newcommand{\ee}{\end{equation}}
\def\beq{\begin{eqnarray}}
\def\eeq{\end{eqnarray}}
\def\ba{\begin{eqnarray}}
\def\ea{\end{eqnarray}}
\def\a{\alpha}
\def\b{\beta}
\def\ep1{\epsilon_1}
\def\eps2{\epsilon_2}

\newcommand{\zero}{{\tiny 0}}

\newcommand{\IZ}{\mathbb{Z}}
\newcommand{\IC}{\mathbb{C}}
\newcommand{\IP}{\mathbb{P}}
\newcommand{\IN}{\mathbb{N}}
\newcommand{\IR}{\mathbb{R}}
\newcommand{\IQ}{\mathbb{Q}}

\newcommand{\IF}{\mathbb{F}}
\newcommand{\IE}{\mathbb{E}}
\usepackage{marginnote}
\newcommand{\Bn}{C} 

\newcommand{\tr}{\mathrm{Tr\, }}

\newcommand{\rk}{\mathrm{rk\, }}

\newcommand{\nn}{\nonumber}

\newcommand{\cN}{{\cal N}}

\newcommand{\cO}{{\cal O}}

\newcommand{\cD }{{\cal D}}

\newcommand{\cZ }{{\cal Z}}

\newcommand{\mg}{\mathfrak{g}}

\newcommand{\LambdaCR}{\Lambda_{cr}}
\newcommand{\LambdaR}{\Lambda_{r}}
\newcommand{\LambdaCW}{\Lambda_{cw}}
\newcommand{\LambdaW}{\Lambda_{w}}

\newcommand{\weyl}{W}
\newcommand{\dynkinsym}{\text{DynkinSym}}
\newcommand{\Eucl}{\IE}

\newcommand{\gs}{g_{top}}

\begin{document}
	\begin{titlepage}
		{}~ \hfill\vbox{ \hbox{} }\break
		
		\rightline{KIAS-P20070}
		
		\vskip 1 cm

		\begin{center}
			\Large \bf Weyl invariant Jacobi forms along Higgsing trees
		\end{center}
		
		\vskip 0.8 cm
		\centerline{Zhihao Duan${}^{a}$, David Jaramillo Duque${}^{b}$,} 
		
		\centerline{Amir-Kian Kashani-Poor${}^{b}$}

		\vskip 0.2in
		\begin{center}{\footnotesize
				\begin{tabular}{c}
					${}^{\, a}${\em School of Physics, Korea Institute for Advanced Study, Seoul 02455, Korea}\\[0ex]
				\end{tabular}
		}\end{center}
		\begin{center}{\footnotesize
				\begin{tabular}{c}
					${}^{\, b}${\em LPENS, CNRS, PSL Research University, Sorbonne Universit\'{e}s, UPMC, 75005 Paris, France}\\[0ex]
				\end{tabular}
		}\end{center}

		\setcounter{footnote}{0}
		\renewcommand{\thefootnote}{\arabic{footnote}}
		\vskip 60pt
		\begin{abstract} 
			{}{\bf Abstract:} Using topological string techniques, we compute BPS counting functions of 5d gauge theories which descend from 6d superconformal field theories upon circle compactification. Such theories are naturally organized in terms of nodes of Higgsing trees. We demonstrate that the specialization of the partition function as we move from the crown to the root of a tree is determined by homomorphisms between rings of Weyl invariant Jacobi forms. Our computations are made feasible by the fact that symmetry enhancements of the gauge theory which are manifest on the massless spectrum are inherited by the entire tower of BPS particles. In some cases, these symmetry enhancements have a nice relation to the 1-form symmetry of the associated gauge theory. 
		\end{abstract}
		
		{\let\thefootnote\relax
			\footnotetext{\tiny xduanz@gmail.com, david.jaramillo-duque@phys.ens.fr,  kashani@lpt.ens.fr}}
		
	\end{titlepage}

	\pagebreak
	
	\vfill \eject

	\tableofcontents
	\newpage
	
	\section{Introduction}
	
	Theories with extended supersymmetry have interesting massive spectra protected against decay by the BPS constraint (see e.g. \cite{Seiberg:1994rs}). Many techniques have been developed for computing exact counting functions on the BPS spectrum, notably topological string methods and localization, see \cite{Hori:2003ic} and \cite{Pestun:2016zxk} for reviews. Rather than starting with a set of fundamental fields and their interactions and deriving the resulting spectrum of the theory, the starting point of many of these approaches is a wholesale construction of the theory from the outset, in terms of its embedding in string theory. In such cases, a question with far-reaching consequences is to what extent the low lying spectrum constrains the full spectrum of the theory. Ultimately, we would like to understand, if and if yes then in how many ways a given low lying spectrum can be completed to a consistent spectrum of a UV complete theory. The swampland program, as initiated in \cite{Vafa:2005ui}, is an overarching term for work attempting to address this question, with the emphasis that the UV complete theory encompass gravity; see \cite{Brennan:2017rbf, Palti:2019pca} for recent reviews.
	
	In this work, we will be concerned with the BPS spectra of 5d gauge theories with 8 supercharges obtained via compactification on a circle of rank 1 6d theories with (1,0) supersymmetry. These theories are dubbed 5d KK theories in \cite{Jefferson:2017ahm}. Much work has been expended in studying their superconformal limits \cite{DelZotto:2017pti,Bhardwaj:2018yhy, Bhardwaj:2018vuu,Bhardwaj:2019fzv}. Here, we will be interested in the theories at a generic point on their Coulomb branch and the associated spectrum of BPS particles. 
	
	Rank 1 theories are readily constructed via F-theory compactifications \cite{Morrison:1996na,Morrison:1996pp} on elliptically fibered Calabi-Yau manifolds $X$ \cite{Candelas:1996su,Bershadsky:1996nh,Perevalov:1997vw}. The rank in the context of 6d theories indicates the dimension of the tensor branch, which geometrically corresponds to the number of compact homology 2-cycles in the base of the fibration. Requiring that an elliptic fibration $X$ over a given base be a Calabi-Yau manifold imposes a minimal singularity on the elliptic fiber, which maps to the gauge algebra $\mathfrak{g}$ of the 6d theory obtained by F-theory compactification on $X$. Resolving the singularity via blow-ups does not change the 6d physics, but moves the associated 5d KK theory away from the origin of its Coulomb branch. Upon specializing the complex structure of $X$, the singularity can be enhanced, leading to a larger gauge algebra $\mathfrak{g}'$ of the compactified theory. Physically, the transition $\mathfrak{g}' \rightarrow \mathfrak{g}$ corresponds to Higgsing. The graph with nodes consisting of elliptically fibered Calabi-Yau manifolds and links describing the process of specializing complex structure to a singularity, then resolving it by blowing up, is referred to as a Higgsing tree. Rank 1 Higgsing trees, constructed over the base surfaces $\cO(-n) \rightarrow \IP^1$, for $n = 0, \ldots , 8, 12$, provide the setting for our computations.
	
	A counting function for 5d BPS states is famously given by the topological string partition function $Z_{top}$ \cite{Gopakumar:1998ii,Gopakumar:1998jq,Iqbal:2012xm}. In \cite{Candelas:1993dm,Katz:1996ht}, examples of the type of transformation encoded in the links of Higgsing trees were dubbed geometric transitions and studied in detail for several 2 and 3 parameter Calabi-Yau manifolds. The genus 0 Gromov-Witten invariants of the 1 parameter Calabi-Yau models obtained after the transition were shown to be related to those of the parent theory by summing over the K\"ahler parameters associated to the blown-down curves. In this work, we will obtain a stronger form of this result for the nodes and links of rank 1 Higgsing trees. By invoking modularity results \cite{Huang:2015sta,Gu:2017ccq,DelZotto:2016pvm,DelZotto:2017mee}, the computation of $Z_{top}$ at a given order in an expansion in the base curve exponentiated K\"ahler parameter $Q_B$, schematically $Z_{top} \sim \sum_k Z_k Q_B^k$, reduces to the determination of a holomorphic Weyl invariant Jacobi form of determined index and weight, with the coarsest choice of Weyl group possible being that of the gauge group of the engineered theory, $\weyl(\mathfrak{g})$. We conjecture that the topological string partition functions of nodes of the Higgsing tree specialize upon moving towards the root of the tree according to maps $J(\mathfrak{g}') \rightarrow J(\mathfrak{g})$ relating the ring of Jacobi forms of the associated Weyl groups.\footnote{The connection between different nodes of rank 1 Higgsing trees was studied for some examples in a different presentation of $Z_{top}$ in \cite{Kim:2018gjo,Gu:2020fem}.} We provide ample evidence for this conjecture by computing $Z_k$ at $k=1$ for a host of examples and demonstrating the specialization explicitly.
	
	Note that the topological string is insensitive to complex structure deformations\footnote{Here and in the following, we speak from the perspective of the so-called $A$-model \cite{Witten:1991zz}.\label{A-model}}. From its vantage point, all nodes of a Higgsing tree can hence be seen as singular geometries lying on subslices of the K\"ahler moduli space. One parent theory (which on infinite length Higgsing trees would depend on infinitely many parameters) should hence capture all geometries subsumed in a Higgsing tree. This perspective could offer a path towards the proof of our conjecture; the point to be addressed is that the topological string partition function exhibits singularities on the subslices of moduli space corresponding to the singular geometries. For the models that we consider, we demonstrate by computation that these singularities are only apparent.
	
	The elements of $J(\mathfrak{g})$ at given weight and index span a finite dimensional vector space over $\IQ$.
	The computation of $Z_k$ in the class of models we are considering is reduced to obtaining the expansion coefficients in an appropriate basis of this space. The reduction of the computation to a finite dimensional problem is conceptually important. Practically, the number of coefficients grows rapidly with the rank of $\mathfrak{g}$ and the base degree $k$. Luckily, $Z_k$ for many of the geometries we consider turns out to exhibit a higher symmetry than merely the Weyl group $\weyl(\mathfrak{g})$. These enhanced symmetries have various origins. The most straight-forward cases are 1-form symmetries as well as constraints arising from the fact that a theory arises via Higgsing. Somewhat more surprisingly, the fact that some maps $J(\mathfrak{g}') \rightarrow J(\mathfrak{g})$ are injective, hence invertible on their image, can lead to enhanced symmetries motivated by moving opposite the Higgsing arrow. We call this phenomenon ``inverse inheritance." This latter class of enhanced symmetries can also be explained intrinsically, without reference to Higgsing, via a cancellation mechanism described in \cite{Kashani-Poor:2019jyo}. Notably, we argue for symmetry enhancement at the level of the massless 6d spectrum of the parent theory of the 5d theory, yet find computationally that the symmetry extends to the full BPS spectrum, begging the question (which we leave for future study) whether this had to be the case. The elliptic genus in 6d or 4d theories as expressed in terms of Jacobi forms or related structures is enlisted to study various conjectures relating to the swampland program in the works \cite{Lee:2018spm,Lee:2018urn,Lee:2019tst,Lee:2020blx,Lee:2020gvu,1837146}. 
	
	We conclude this introduction with a summary of the ensuing sections. We begin with a rapid review of the physical setting, and discuss the general structure of $Z_k$ in section \ref{sec:how_to}, completing the discussion in the literature to encompass all rank 1 theories. In section \ref{sec:moving}, we take a first look at the constraints arising from being a node of a Higgsing tree, and discuss in detail the specialization of $Z_k$ upon descending a Higgsing tree towards its root. We also derive how our specialization results for $Z_k^{\mg}$ manifest themselves at the level of Gromov-Witten invariants. Section \ref{sec:enhanced} is dedicated to the discussion of symmetry enhancements of $Z_k$ derived both from the perspective of neighboring nodes in the Higgsing tree and intrinsically. Several technical appendices conclude the paper: in appendix \ref{app:explicit}, we explicitly give $Z_1$ for one among the many models for which we have computed it, to convey the general flavor of our results. Appendix \ref{app:GWinv} contains tables of Gromov-Witten data which exemplify the specialization results derived in section \ref{sec:moving}. Appendix \ref{app:root_systems} summarizes data on simple Lie algebras which is relevant for the discussion in the main text. Appendix \ref{app:Jacobi_forms} gives explicit formulae for the generators of the ring of Jacobi forms for all simple groups other than $E_n$, $n=6,7,8$. Appendix \ref{app:specializationformulas} provides further details regarding the specialization maps between these. Appendix \ref{app:Weyl_invariant_polynomials} provides a brief introduction to Weyl invariant polynomials, and explains how these enter in our calculations. Finally, appendix \ref{app:Higgsing_trees} provides a very brief review of elliptically fibered Calabi-Yau manifolds, and reproduces some rank 1 Higgsing trees for the reader's convenience.

    \section{How to capture BPS degeneracies via Weyl invariant Jacobi forms} \label{sec:how_to}
    \subsection{The topological string and BPS states of the 5d theory}
    The topological string was born as a worldsheet theory \cite{Witten:1991zz,Bershadsky:1993cx}. To a Calabi-Yau manifold $X$ and to each worldsheet genus $g$, it assigns a power series $F_g$ in K\"ahler parameters $Q_i$ associated to the homology 2-cycles of $X$ (see footnote \ref{A-model}). The coefficients are the celebrated Gromov-Witten invariants of $X$. Upon introducing a formal parameter $\gs$, a putative theory tying together all $F_g$ is assigned the partition function
    \be
    Z_{top}(X) = \exp\left( \sum_g F_g\, \gs^{2g-2} \right) \,.
    \ee
    In \cite{Gopakumar:1998ii,Gopakumar:1998jq}, it was shown that $Z_{top}(X)$ is not merely a formal construct; it captures the non-perturbative BPS spectrum of M-theory compactified on $X$, arising from M2 branes wrapping holomorphic curves in $X$ (see \cite{Iqbal:2012xm} for a succinct summary of these matters).
    
    The 5d theories of interest in this paper arise upon compactifying M-theory on a certain class of elliptically fibered, interlinked Calabi-Yau manifolds $X$ constituting the nodes of so-called Higgsing trees, as we review in appendix \ref{app:Higgsing_trees}. We will call the theory obtained upon compactification on $X$ $\mathrm{M}[X]$ for the purposes of this section. The massless perturbative spectrum of $\mathrm{M}[X]$ is captured by the compactification of 11 dimensional supergravity on $X$. For non-compact $X$, this spectrum consists of $h^{1,1}(X)$ massless vector fields and $h^{2,1}(X)+1$ uncharged massless hypermultiplets. 
    
    The symmetries constraining $Z_{top}(X)$ become manifest at special points in the K\"ahler moduli space of $X$ in which this massless spectrum is enhanced. The description of these points of enhancement is most natural from a 6d perspective, arising from compactifying F-theory on $X$. In this description, only the complex structure of the elliptic fiber of $X$ is physical. Considering a representative of the geometry in which all exceptional curves in the fiber are blown down brings the gauge symmetry $\mathfrak{g}$ associated to the resulting singular fiber to light, together with hypermultiplets charged under $\mathfrak{g}$. This theory yields $\mathrm{M}[X]$ upon circle compactification, as follows from M-theory/F-theory duality \cite{Vafa:1996xn}. In contrast to the F-theory compactification, the elliptic fiber now is fully physical, with its size inversely proportional to the size of the compactification circle \cite{Vafa:1996xn}. Away from the singular limit of the fiber, we recover the perturbative massless spectrum described above.

    \subsection{The topological string and Jacobi forms}
    On an elliptically fibered Calabi-Yau manifold $X$, an astute rewriting of the holomorphic anomaly equations \cite{Bershadsky:1993ta,Bershadsky:1993cx} can be used to demonstrate that the topological string partition function inherits modular properties of the elliptic fiber \cite{Hosono:1999qc,Klemm:2012,Huang:2015sta,Cota:2019cjx}: upon extracting a universal fiber independent contribution $Z_0$ and expanding in the base class $Q_B$,
	\be \label{eq:Zk_from_Ztop}
	Z_{top} = Z_0 \left( 1 + \sum_{k > 0} c_k(\mathbf{Q}) Q_B^k Z_k \right) \,,
	\ee
    with $c_k(\mathbf{Q})$ a coefficient on which we shall comment momentarily, the expansion coefficients $Z_k$ can be shown to be meromorphic Jacobi forms (with a simple multiplier due to a contribution from the Dedekind $\eta$ function). An alternative route towards unearthing this modular structure proceeds by identifying $Z_k$ with the elliptic genera of $k$ non-critical strings in the spectrum of the 6d theory describing F-theory compactified on $X$ \cite{Benini:2013a,Benini:2013xpa,Haghighat:2014vxa,Haghighat:2015ega,DelZotto:2016pvm,DelZotto:2017mee,Kim:2018gjo}. From either route, the following ansatz for $Z_k$ can be motivated:
    \be
	Z_k = \frac{1}{\eta^{n(k)}(q)} \frac{\cN}{\cD}(q,\mathbf{Q},\gs) \,,
	\label{eq:ZkAnsatz}
	\ee
    where $\cN$ and $\cD$ are holomorphic Jacobi forms with modular parameter the K\"ahler parameter $q$ of the generic fiber,
    \be \label{eq:generic_fiber}
	q = Q_0 \prod_i (Q_i)^{a_i} \,.
	\ee
	Here, $Q_0$ and $Q_i$, $i=1, \ldots, \rk(\mathfrak{g})$, denote exponentiated K\"ahler parameters of fibral curves ($Q_0$ being associated to the only curve among these which intersects the base $B$ of the fibration).\footnote{Note that in addition to these parameters, dependence on flavor fugacities can be introduced in the elliptic genus \cite{Kim:2018gjo,Gu:2020fem}. At the level of the geometry and the topological string partition function, this requires including additional divisors in $X$. For further discussion of flavor symmetry in the SCFT limit of 5d theories, see \cite{Bhardwaj:2020ruf,Bhardwaj:2020avz}.} The $a_i$ coincide with the marks\footnote{Note that this equation appears with the $a_i$ identified as comarks in several previous works. The distinction is of course irrelevant for simply laced groups.} of $\mathfrak{g}$. The elliptic parameters of the Jacobi forms are given by $Q_i$ as well as $g_{top}$, with the Weyl group $\weyl(\mathfrak{g})$ acting on the former.
    
    The denominator $\cD$ has a universal contribution present for all rank 1 models which depends only on $\tau$ and $\gs$. Its form is largely fixed by comparison with the Gopakumar-Vafa expansion \cite{Gopakumar:1998ii,Gopakumar:1998jq} of the topological string:
	\be \label{eq:denom_uni}
	\cD_{univ} = \prod_{m=1}^k \phi_{-2,1}(m \gs) \,.
	\ee
	For all $X$ leading to gauge symmetry, the denominator also depends on the K\"ahler parameters of the resolved curves in the fiber. The expression for this contribution that we will use is derived in \cite{Kim:2018gak} by lifting the result for $Z_{top}$ in the gauge theory limit \cite{Nekrasov:2002qd} to 6d \cite{Hollowood:2003cv}. Recall \cite{Bernard:1977nr} that an instanton solution for the gauge group $SU(2)$ \cite{Belavin:1975fg} can be embedded into the gauge group $G$  via the embedding of the gauge algebra $\mathfrak{a}_1$ into $\mathfrak{g}$, with image a generator $T^{\alpha}$ of the Cartan subalgebra associated to a given root $\alpha$ and the corresponding lowering and raising operators. The bilinear form on $\mathfrak{g}$ takes the form
    \begin{equation}
        \text{tr}(T_{a}^{\alpha} T_{b}^{\alpha}) = c_\alpha \delta_{ab}
    \end{equation}
    on these three generators. Choosing the normalization of the bilinear form such that $c_\alpha$ is equal to 1 for all long roots, the constant takes the following values for short roots:
    \begin{equation} \label{eq:def calpha}
    \begin{aligned}
        c_\alpha &= 2, \quad G = B_n, C_n\ \text{and}\ F_4,\\
        c_\alpha &= 3, \quad G = G_2.
    \end{aligned}
    \end{equation}
    In the following, we will drop the index $\alpha$: $c$ will refer to the appropriate value given in \eqref{eq:def calpha}.
    An $SU(2)$ instanton with instanton number $k_{SU(2)}$ maps under this embedding to a $G$ instanton with instanton number $k_G = c_\alpha k_{SU(2)}$.
    
	Following \cite{Kim:2018gak}, we identify the contribution of an $SU(2)$ instanton of instanton number $k$ to the denominator $\cD$ as
	\be \label{eq:denom_SU2}
	\cD^{A_1}_{k,\alpha} =	 \prod_{ab \le k, a,b >0} \phi_{-1,\frac{1}{2}} ((a-b) \gs + m_\alpha)  \phi_{-1,\frac{1}{2}} ((a-b) \gs - m_\alpha)
	\ee 
    $m_\alpha$ is the contribution of the gauge fugacity or K\"ahler parameter associated to the positive root $\alpha$ of $A_1$ to which we will return below. 
  
    The contribution for a general gauge group $G$ can then be expressed as
    \begin{equation}  \label{eq:denom_G}
        \cD^{\mathfrak{g}}_k=\cD_{k,L}^G\cD_{k/c,S}^\mathfrak{g} \,,
    \end{equation}
    with
    \begin{equation}
        \cD^{\mathfrak{g}}_{k,L}=\prod_{\alpha\in\Delta_L^+}\cD^{A_1}_{k,\alpha}, \quad \cD^{\mathfrak{g}}_{k,S}=\prod_{\alpha\in\Delta_S^+}\cD^{A_1}_{k,\alpha} \,,
    \end{equation}
    where we have indicated the set of positive long and short roots as $\Delta_L^+$ and $\Delta_S^+$ respectively.
   Note that \eqref{eq:denom_G} is invariant under permutations on the sets of long and short positive roots, and $\cD^{A_1}_{k,\alpha}$ is invariant under $\alpha \rightarrow -\alpha$. As the Weyl group $\weyl(\mathfrak{g})$ is a subgroup of the permutation group on all roots that does not mix long and short roots, this establishes the invariance of $\cD^{\mathfrak{g}}_k$ under its action. 
   
   Note further that at $k=1$, \eqref{eq:denom_G} is independent of $\gs$.
    
    The power $n(k)$ of the Dedekind $\eta$ function occurring in \eqref{eq:ZkAnsatz} has been determined from topological string considerations \cite{Huang:2015sta} for the minimal singularities over the base surfaces $\IF_n$ equal to $\IF_0$, $\IF_1$ and $\IF_2$ to be $-12 k C_B \cdot K = 12k(n-2)$. Here, $C_B$ is the base curve of the Hirzebruch surface, $K$ is its canonical divisor. In \cite{DelZotto:2016pvm}, it was given for minimal (i.e. maximally Higgsed) models for $n>2$ as $n(k)= 4k h_G^\vee$, with $h_G^\vee$ the dual Coxeter number of the gauge group of the corresponding model, which happens to be given by $h_G^\vee = 3(n-2)$ for all occurring cases, as noted by \cite{Shimizu:2016lbw}. By matching to Gromov-Witten invariants, we find that expressing $n(k)$ in terms of the dual Coxeter number of the gauge group is misleading. In fact, it is the self-intersection number of the base curve which determines this quantity. The correct expression valid for all rank 1 Higgsing tree geometries is
	\be
	n(k) = 12 k |n-2| \,.
	\ee

    Turning now to the prefactor $c_k(\mathbf{Q})$ which enters in extracting $Z_k$ from $Z_{top}$ in \eqref{eq:Zk_from_Ztop}, it was given in \cite{Huang:2015sta} as $c_k(\mathbf{Q}) = q^{-\frac{k(n-2)}{2}}$ for the minimal models over $\IF_0, \IF_1, \IF_2$. For the minimal models over $\IF_n$, $n>2$, \cite{DelZotto:2016pvm} identified it as $c_k(\mathbf{Q}) = (\sqrt{q}/\prod_i Q_i^{a_i})^{k h_G^\vee/3}$\,. We find that this latter formula should be modified by adding a factor of $Q_0$ to the product in the denominator, and replacing $h_G^\vee$ by $3(n-2)$ for all geometries over a base $\IF_n$. This yields an expression valid for all bases $\IF_n$:
	\be
	c_k(\mathbf{Q}) = \left( \frac{\sqrt{q}}{\prod_i Q_0 Q_i^{a_i}}\right)^{k(n-2)} = q^{-\frac{k(n-2)}{2}} \,.
	\ee    
	
	The final ingredient is the numerator $\cN$ of $Z_k$. Its exact expression depends sensitively on the geometry considered. Upon determining the appropriate ring $J$ of holomorphic Jacobi forms in which it lies, an ansatz can be made in terms of the finite basis of $J$ at appropriate weight and index. The expansion coefficients must then be fixed by imposing appropriate boundary conditions, as we discuss in subsection \ref{ss:bcs}. Beyond the problem of providing sufficient boundary conditions, the sheer number of coefficients to be determined quickly becomes computationally untenable. By imposing the symmetries of the massless spectrum of the 5d theory on all of $Z_k$, the number of coefficients can be sufficiently reduced to render many more calculations feasible. In this work, we provide an a posteriori justification for this procedure by demonstrating that the constrained ansatz is consistent with Gromov-Witten invariants obtained via mirror symmetry.

	\subsection{The map between the K\"ahler cone and elliptic parameters} \label{ss:matching_elliptic_parameters}
	The exceptional fibral curves of the class of elliptic fibrations we are considering organize themselves in terms of representations of the corresponding Lie algebra $\mg$. As such, each curve can be identified with an element of the weight lattice $\LambdaW$ of $\mg$. It is therefore natural to identify the fiber components $m$ of the K\"ahler form with an element of the complexified dual lattice, the coroot lattice, such that the K\"ahler parameter $m_C$ associated to the curve $C$, obtained by integrating the complexified K\"ahler form against the curve class, is given by the pairing
	\be \label{eq:def_m_omega}
	m_\omega = (\omega, m)  \,,
	\ee
	with $\omega \in \LambdaW$ the weight identified with $C$.
	
	$-2$ rational curves in the fiber of the elliptic fibration organize themselves into the adjoint representation. Each such curve thus maps to a root $\alpha$ of $\mg$, which just as any other weight lies in $\LambdaW$. The corresponding K\"ahler parameters
	\be
	m_\alpha = (\alpha, m)
	\ee
	are identified with the gauge fugacities of the elliptic genus. They have already featured in the formula \eqref{eq:denom_SU2} above, while the exponentiated K\"ahler parameters
	\be \label{eq:definition_exponentiated_Kaehler_parameter}
	Q_i = e^{2\pi (\alpha_i,m)} \,,
	\ee
    with $\alpha_i$ a simple root, appear in equation \eqref{eq:generic_fiber} above. Note that the dependence of the elliptic genus on the gauge fugacities will generically contain fractional powers of $e^{2\pi i m_\alpha}$, as the weight lattice is generically finer than the root lattice.
	
	In theories with a Lagrangian description, $m$ can be identified with the VEV of the real scalar field $\phi$ in the 5d gauge multiplet, under identification of the complexified coroot lattice with the Cartan subalgebra of $\mg$. This gives rise to masses for hypermultiplets as follows. The scalar fields $(Q, \tilde{Q})$ of a hypermultiplet transforming in the irreducible representation $\rho$ of the gauge group couple to $\phi$ via 
	\be
	(\tilde{Q} , \rho(\phi) Q) \,.
	\ee 
	Recall that $Q$ and $\tilde{Q}$ transform in dual representations; in the above formula, $(\cdot,\cdot)$ indicates the pairing between the dual spaces. $\phi$ acquiring a VEV gives rise to the mass term
	\be
	(\tilde{Q} , \rho(m) Q) \,.
	\ee 
	Decomposing the vector $Q$ in terms of weight eigenspaces, $Q = \sum_\lambda Q_\lambda$ (assuming non-degenerate eigenspaces for notational simplicity), this yields
	\be \label{hypermultiplet_mass_term}
	(\tilde{Q} , \rho(m) Q) = \sum_{\lambda, \tilde{\lambda}} (\tilde{Q}_{\tilde{\lambda}}, (\lambda,m) Q_\lambda ) = \sum_{\lambda} (\lambda, m) (\tilde{Q}_{\tilde{\lambda}}, Q_\lambda) \,,
	\ee	
	where we have denoted by $\tilde{\lambda}$ the conjugate weight to $\lambda$. Note that if the irreducible representation $\rho$ has highest weight $\lambda_h$, all the weights $\lambda$ occurring in the mass term \eqref{hypermultiplet_mass_term} are of the form
	\be
	\lambda = \lambda_h - \sum_i n_i \alpha_i \,, \quad n_i \in \IN \,,
	\ee
	where the sum is over the simple roots of $\mg$. The hypermultiplet in representation $\rho$ hence introduces dependence on the parameter $(\lambda_h,m)$ in addition to the parameters $(\alpha_i,m)$.
	
	As explained in the previous subsection, the numerator of $Z_k$ as presented in \eqref{eq:ZkAnsatz} is a Weyl invariant Jacobi form with, aside from $\gs$, $m$ featuring as the elliptic parameter. Correctly identifying the dependence on $m$ requires some care. We will mostly take the generators of the ring $J(\mathfrak{g})$ of Weyl invariant Jacobi forms as derived in \cite{Bertola} as our starting point. These depend on $\rk \mg$ parameters $x_i$, which determine a point in a Euclidean lattice $\IE_n$. The action of $\weyl(\mathfrak{g})$, the Weyl group of $\mg$, on these parameters, as well as their behavior under shifts by elements of $\Lambda_r(\mathfrak{g})$, the root lattice of $\mg$, follows from the embedding of the root lattice into $\IE_n$ (note that $n$ can be larger than $\rk(\mg)$; as is the case e.g. for the $A$-series).
	
	From our identification of K\"ahler parameters with \eqref{eq:def_m_omega}, we conclude that it is shifts of $m$ via elements of the coroot lattice which should be symmetries of the theory. When studying the elliptic genus for the Lie algebra $\mg$, we hence need to consider Weyl invariant Jacobi forms of the Lie algebra $\tilde{\mg}$ whose root lattice equals the coroot lattice of $\mg$. For simply laced groups, we can identify roots with the corresponding coroots as elements of the orthogonal lattice, and this distinction is irrelevant.\footnote{Note that in the following, it will be convenient to refer to all the roots of simply laced root systems as long.} For $F_4$ and $G_2$, root and coroot lattice are isomorphic, the map between the two does not however preserve the inner product: short roots are mapped to long coroots and vice versa. E.g., in our conventions, the set of coroots for $G_2$ as embedded in $\IE_3$ is given by 
	\begin{equation*}
	    \pm (e_i-e_j), \quad i\neq j, \quad \quad \pm(2e_i-e_j-e_k), \quad i\neq j\neq k\neq i;
	\end{equation*}
	while the set of roots is given by
	\begin{equation*}
	    \pm (e_i-e_j), \quad i\neq j, \quad \quad \pm\frac{1}{3}(2e_i-e_j-e_k), \quad i\neq j\neq k\neq i.
	\end{equation*}
	Finally, the root lattice of $B_n$ is isomorphic to the coroot lattice of $C_n$, and vice versa. We must hence use the Weyl invariant forms assigned to the Lie algebra $C_n$ in the conventions of \cite{Bertola} to describe $Z_k$ on a background leading to gauge symmetry $B_n$, and vice versa.
	
	\subsection{Determining weight and index}
	
	$Z_k$ is a weight 0 Jacobi form whose index is determined by the anomaly polynomial of the elliptic genus or equivalently by the holomorphic anomaly of the topological string partition function \cite{Gu:2017ccq}. The anomaly polynomial for the elliptic genus of a string carrying charges $Q_i$ (not to be confused with the exponentiated K\"ahler parameters; in the type IIB picture, these charges encode the class of the base curves $C_i$ that the D3 brane giving rise to the string is wrapping) is given by \cite{Ohmori:2014kda,Shimizu:2016lbw} 
    \ba
    I_4 &=& \frac{\interMatrix^{ij} Q_i Q_j}{2} \left(c_2(L) - c_2(R) \right) + \\
    && Q_i \left( \frac{1}{4h^\vee} \interMatrix^{ia} \tr_{adj} F_a^2 - \frac{2- \interMatrix^{ii}}{4} \left(p_1(T) - 2 c_2(L) - 2 c_2(R)\right) + h_{G_i}^\vee c_2(I) \right)\,,  \nn
    \ea
    where $\interMatrix^{ij}=-C_{i}\cdot C_j$ is (minus) the intersection matrix of the curves in the base, $c_2(L)$ and $c_2(R)$ are the second Chern classes associated to the left and right parts of the Poincaré symmetry $SU(2)_L\times SU(2)_R$ of the normal bundle of the string in the 6d spacetime, $c_2(I)$ is the second Chern class for the $SU(2)$ R-symmetry bundle, $p_1(T)$ is the first Pontryagin class of the tangent bundle of the 6d spacetime, and $F_a$ is the field strength associated to the curve $C_a$ in the base. This latter index $a$ includes compact curves associated to gauge fields indexed by $i$ above and non-compact ones associated to global symmetries. 
    
    Specializing to rank 1, i.e. to the case of only one compact cycle in the base $B$, giving rise to one tensor multiplet, making the replacement $c_2(R)=c_2(I)=0,\,c_2(L)=-g_s^2$  for the unrefined string \cite{DelZotto:2016pvm, Gu:2017ccq,DelZotto:2017mee}, and introducing the norm
    \be
    ( \cdot, \cdot) = \frac{1}{2 h^\vee} \tr_{\!\!adj}   
    \ee
    on the Cartan subalgebra of the Lie algebra, which is normalized so that short coroots have norm squared 2, we obtain the index bilinear form for $Z_k$,
    \begin{equation}
    I_Z=i_{Z,top}g_{top}^2+i_{Z,gauge}(m,m)
    \end{equation}
    with
    \begin{align}
    &i_{Z,top}=-\frac{1}{2}(nk^2+(2-n)k) \,,\\
    &i_{Z,gauge}=-kn \,.
    \end{align}
    Here, $-n=\interMatrix^{11}=C_B\cdot C_B$ is the self-intersection number of the base curve, and we have replaced $F$ by $m$, which we will use from now on.
	
	To compute the index of the denominator in the presentation \eqref{eq:ZkAnsatz} for $Z_k$, we add the index bilinear form of each factor. For the universal part $\cD_{univ} $ in \eqref{eq:denom_uni}, it is given by
	\begin{align*}
		\sum_{m=1}^k m^2g_{top}^2 \,.
	\end{align*} 
	For the gauge group contribution $\cD_{k}^G$, the index bilinear form is
	\begin{equation*}
	\left(|\Delta_L|\sum_{\substack{ab\leq k\\ a,b>0}}(a-b)^2+|\Delta_S|\sum_{\substack{ab\leq k/c\\a,b>0}}(a-b)^2\right)g_{top}^2+\abSolLong 2h^\vee(m,m)_l+\abSolShort 2h^\vee(m,m)_s \,,
	\end{equation*}
	where 
	\begin{equation}
	    \abSolLong = \#\{ab\leq k \,|\, a,b>0\} \,, \quad \abSolShort = \#\{ab\leq k/c \,|\,a,b>0\}\,,
	\end{equation}
	and 
	\begin{equation}
	    (m,m)_{S/L}=\frac{1}{2h^\vee}\sum_{\alpha\in\Delta_{S/L}}m_\alpha \,.
	\end{equation}
	As the Weyl group does not mix short and long roots, the two inner products $(\cdot,\cdot)_{S/L}$ are Weyl invariant and therefore proportional to the inner product of the lattice,
	\begin{equation}
	(m,m)_S=\shortNorm_G(m,m), \quad (m,m)_L=(1-\shortNorm_G)(m,m).
	\end{equation}
    The value of $\shortNorm_G$ for all simple Lie algebras is given in table \ref{tab:j}.
    
    \begin{table}[ht]
    \centering
    \def\arraystretch{1.3}
    \begin{tabular}{c|ccccc}
    G& Simply laced & $ B_n$ & $ C_n$ & $ G_2$ & $ F_4$\\ \hline
    $\shortNorm_G$ & --- & $\frac{1}{2n-1}$ & $\frac{n-1}{n+1}$ & $\frac{1}{4}$ & $\frac{1}{3}$\\
    \end{tabular}
    \caption{$(m,m)_s=\shortNorm_G(m,m)$}
    \label{tab:j}
    \end{table}

    Combining these contributions, the index bilinear form of the denominator reads
    \begin{equation}
	I_{\cD}=i_{\cD,top} g_{top}^2+i_{\cD,gauge} (m,m) \,,
    \end{equation}
    where
    \begin{align}
    &i_{\cD,top}=\sum_{m=1}^k m^2+|\Delta_L|\sum_{\substack{ab\leq k\\ a,b>0}}(a-b)^2+|\Delta_S|\sum_{\substack{ab\leq k/c\\a,b>0}}(a-b)^2,\\
    &i_{\cD,gauge}=2h^\vee\left((1-\shortNorm_G)\abSolLong+\shortNorm_G\abSolShort\right).
    \end{align}

    By $I_{\mathcal N}=I_Z+I_\cD$, the index bilinear form of the numerator $\cN$ of \eqref{eq:ZkAnsatz} is then easily determined to be
    \begin{equation}
    I_{\mathcal N}=i_{\mathcal N,top}g_{top}^2+i_{\mathcal N,gauge}(m,m)\,,
    \end{equation}
    with
    \begin{align}
    &i_{\mathcal N,top}=-\frac{1}{2}(nk^2+(2-n)k)+\sum_{m=1}^k m^2+|\Delta_L|\sum_{\substack{ab\leq k\\ a,b>0}}(a-b)^2+|\Delta_S|\sum_{\substack{ab\leq k/c\\a,b>0}}(a-b)^2 \,, \label{eq:topStringIndex}\\
    &i_{\mathcal N ,gauge}=-kn+2h^\vee\left((1-\shortNorm_G)\abSolLong+\shortNorm_G\abSolShort\right)\,.\label{eq:gaugeIndexNumerator}
    \end{align}
    For the case $k=1$, $i_{\mathcal N,top}=0$ and we conclude that the numerator does not depend on $g_{top}$.
	
    To determine the weight $w_\cN$ of the numerator, note that the weights of $\cN$ and $\cD$ must be equal, as $Z_k$ has weight 0. The factor involving the Dedekind $\eta$ function in equation \eqref{eq:ZkAnsatz} contributes the weight $\frac{1}{2}n(k)=6k|n-2|$, the universal contribution $\cD_{univ}$ to the denominator has weight $-2k$, and the gauge contribution $D_k^G$ has weight $-|\Delta_L|\abSolLong-|\Delta_S|\abSolShort$. Adding these contributions, we obtain
    \begin{equation} \label{eq:weightNumerator}
        w_\cN=6k|n-2|-2k-|\Delta_L|\abSolLong-|\Delta_S|\abSolShort\,.
    \end{equation}

	\subsection{Imposing boundary conditions} \label{ss:bcs}
	As the numerator $\cN$ in the ansatz \eqref{eq:ZkAnsatz} for $Z_k$ is a holomorphic Jacobi form, it can be expanded in terms of a finite basis of forms of given weight \eqref{eq:weightNumerator} and index \eqref{eq:topStringIndex} and \eqref{eq:gaugeIndexNumerator}. Most of this work will be concerned with identifying the ring of Jacobi forms best adapted to a given gauge theory, i.e. maximally constrained by the symmetries of the problem. Once the ring is chosen and the expansion of $\cN$ in appropriate generators is performed, appropriate boundary conditions must be imposed on $Z_k$ to determine the expansion coefficients. 
	
	In \cite{Huang:2015sta,Gu:2017ccq,DelZotto:2017mee,Duan:2020cta}, these boundary conditions are imposed in the form of so-called vanishing conditions: the constraint that Gopakumar-Vafa invariants of a given curve class must vanish at sufficiently high genus. In \cite{Huang:2015sta}, it was argued that imposing generic vanishing conditions (i.e. requiring that these invariants vanish eventually) is sufficient to fix $Z_k$ for theories without gauge symmetry; this argument was extended to the refined context in \cite{Gu:2017ccq}. In \cite{DelZotto:2017mee}, it was argued that imposing generic vanishing conditions does not suffice to fix $Z_k$ for theories with gauge symmetry. Imposing sharp vanishing conditions, it was conjectured, does suffice. This was demonstrated in the case of the $A_2$ theory over the base $\IF_3$ up to base wrapping number 3 and the $D_4$ theory over $\IF_4$ for base wrapping 1.
	
	An alternative would be to compute Gopakumar-Vafa invariants for these geometries by imposing elliptic blow-up equations \cite{Gu:2018gmy,Gu:2019dan,Gu:2019pqj,Gu:2020fem}. For the purposes of this work, we rely on the technically less arduous path of mirror symmetry and impose genus 0 Gromov-Witten invariants as boundary conditions. This allows us to fix $Z_k$ at base wrapping $k=1$
	completely, and some coefficients in the expansion of $\cN$ for higher $k$ (only in the case of the E-string (of arbitrary rank) can $Z_k$ for all $k$ be determined solely by imposing genus 0 invariants \cite{Duan:2018sqe}).
	
	To determine the Gromov-Witten invariants for the various nodes of rank 1 Higgsing trees, we construct, where possible, the underlying geometry as hypersurfaces in toric varieties \cite{Candelas:1996su,Bershadsky:1996nh,Perevalov:1997vw,Kashani-Poor:2019jyo} to which we apply well-established mirror symmetry techniques \cite{Candelas:1990rm,Hosono:1993qy}. We organize the invariants in terms of a basis of curve classes adapted to the gauge theory interpretation by identifying the distinguished curves in the geometry as intersections of toric divisors with the hypersurface. We refer to \cite{Kashani-Poor:2019jyo} for a detailed exposition of these techniques.
	
	$C_n$ and $D_n$ singularities cannot be imposed torically on the elliptic fiber over base a Hirzebruch surface $\IF_n$ \cite{Bershadsky:1996nh, Kashani-Poor:2019jyo}. We hence cannot compute the elliptic genus for theories with these gauge groups directly using our techniques. However, the fact that all $D_n$ theories in rank 1 Higgsing trees arise via Higgsing of theories that we can solve provides us with an alternative path to obtaining their elliptic genus, as we explain in section \ref{sec:moving}. Reversing the arrow of dependencies, we can thus compute the genus 0 Gromov-Witten invariants for these geometries, invariants which are not available via traditional mirror symmetry techniques.
	
	\section{Specializing along Higgsing trees} \label{sec:moving}
	
	Descending via Higgsing from a theory with gauge group $\mathfrak{g}$ imposes constraints on the spectrum of charged matter of the resulting theory with gauge group $\mathfrak{g'}$. While the nature of these constraints generically strongly depends on the details of the Higgsing considered, some general statements can be made. E.g., when $\rk(\mathfrak{g}) = \rk(\mathfrak{g'})$, the Weyl symmetry of $\mathfrak{g}$ decomposes into $\weyl(\mathfrak{g'}) \ltimes \dynkinsym(\mathfrak{g'})$, with the Dynkin diagram symmetries continuing to act as an automorphism on the theory. The charged matter spectrum of the Higgsed theory must therefore be invariant under the action of $\dynkinsym(\mathfrak{g'})$ on the representations of $\mathfrak{g'}$. We see many examples of this phenomenon in the rank 1 Higgsing trees:
	\begin{itemize}
	    \item $\weyl(G_2) = \weyl(A_2) \ltimes \dynkinsym(A_2)$: the Dynkin diagram symmetry exchanges the fundamental representations $\mathbf{3}$ and $\mathbf{\bar{3}}$ of $A_2$. Symmetry under this exchange is however already required by CPT invariance, hence does not constrain the $A_2$ gauge theories further.
	    \item $\weyl(F_4) = \weyl(D_4) \ltimes \dynkinsym(D_4)$: the Dynkin diagram symmetry of $D_4$ famously permutes its vector and two spinor representations, and indeed, all $D_4$ theories occurring in rank 1 Higgsing trees (all descending from $F_4$ theories via Higgsing) have a charged matter spectrum invariant under such permutations.
	    \item $\weyl{(B_n)} = \weyl(D_n) \ltimes \dynkinsym(D_n)$: the $\IZ_2$ Dynkin diagram symmetry of $D_n$ ($n>4$) exchanges the two spinor representations. For $n$ odd, these are conjugate to each other, hence the $\IZ_2$ symmetry is imposed by CPT invariance. However, for $n$ even, the spinor representations are self-conjugate, allowing for the presence of half hypermultiplets in the spectrum of these theories. Here, $\dynkinsym(D_n)$ is an additional constraint on the spectrum. In perusing the rank 1 Higgsing trees, we indeed observe that this symmetry is not realized only in the case of the $D_6$ theories of the $\IF_2$ and $\IF_3$ Higgsing trees; these are the only $D_{2n}$ theories not descending from a $B_{2n}$ theory.
	\end{itemize}
	In section \ref{sec:enhanced}, we will see that these symmetries of the massless spectrum of the theory are inherited by the elliptic genus. What is more, at least at the level of the elliptic genus, symmetries of a Higgsing theory can be ``reverse inherited" by the unHiggsed theory. Thus e.g., via the $B_4$ and $F_4$ to $D_4$ branch of a Higgsing tree, $\weyl(F_4)$ has repercussions for the $B_4$ theory, even though it does not descend via Higgsing from a theory with gauge group $F_4$.
	
	In this section, we will study the relation between the elliptic genera of theories related by Higgsing $\mg \rightarrow \mg'$ in detail. We will define for each Higgsing a linear embedding
    \begin{equation} \label{eq:def_iota}
        \iota : \mathfrak{h}' \hookrightarrow \mathfrak{h}
    \end{equation}
    between the Cartan subalgebras. In practice, these maps have simple presentations in terms of the orthogonal coordinates $x_i$ on the Euclidean lattices in which $\LambdaCR(\mg')$ and $\LambdaCR(\mg)$ are embedded. The map $\iota$ induces a map $\iota^*$ between functions on the complexified Cartan algebra and in particular the Jacobi forms associated to the corresponding Lie algebras. Up to some possible change of coordinates, dictated by our explicit embeddings of $\LambdaCR(\mg)$ in $\Eucl_\mg$, the map $\iota^*$ corresponds to restricting the modular forms to the subspace $\iota(\mathfrak h')$. We give these restrictions for all simple Lie algebras, except $E_6,\,E_7$ and $E_8$.
    
    It is natural to ask whether the relation between $Z^{\mathfrak g}$ and $Z^{\mathfrak{g}'}$ is governed by $\iota$. In all the models we consider, we find this to be the case, at least at base wrapping 1:
	\begin{equation}
	    \iota^*(Z^{\mathfrak g}) = Z^{\mathfrak g'}.
	    \label{eq:specializing}
	\end{equation}
	In particular, when the Higgsed gauge algebra has fewer roots, the apparent extra divergences in equation \eqref{eq:denom_G} disappear. 
	
	We will discuss this specialization separately for the numerator and the denominator of the ansatz \eqref{eq:ZkAnsatz} for the elliptic genus.
	
	\subsection{Restriction maps between rings of Jacobi forms} \label{ss:restricting_Jacobi_forms}
	The specialization map $\iota^*$ takes Jacobi forms of the Lie algebra $\mathfrak g$ to Jacobi forms of the Lie algebra $\mathfrak g'$. We begin by studying how to specialize the generators of the ring of Jacobi modular forms along the Higgsing trees. Some technical details are relegated to appendix \ref{app:specializationformulas}.
		
	\subsubsection{$G_2$ to $A_2$} \label{ss:Jacobi_G2_to_A2}
	The (co)root lattices of the Lie algebras $A_2$ and $G_2$ are isomorphic. We can therefore choose $\iota$ to be the identity. The Weyl groups of the two algebras are related via 
	\begin{equation}
	    \weyl(G_2)=\weyl(A_2)\ltimes \text{DynkinSym}(A_2).
	    \label{eq:WeylG2}
	\end{equation}
	The ring $J(G_2)$ of $G_2$ Jacobi forms is thus equal to the subring of the ring $J(A_2)$ of $A_2$ Jacobi forms whose elements are Dynkin diagram symmetric. At the level of the standard basis of the orthogonal lattice $\Eucl_3$, the latter symmetry is realized by exchanging the lattice generators $e_1$ and $e_3$ and flipping the sign of all three generators. Note that $e_1 \leftrightarrow e_3$ is already an element of $\weyl(A_2)$. Of the three generators of $J(A_2)$, only $\phi_{-3,1}^{A_2}$ is not invariant under $x_i \rightarrow - x_i$; it changes by a sign. This observation fixes the relation between the two sets of generators, as summarized in figure \ref{fig:jacobi_G2_to_A2}. 
	
	The relation between the generators of $J(A_2)$ and $J(G_2)$ in particular implies that the numerator $\cN$ of the elliptic genus of all $A_2$ theories obtained via Higgsing from a $G_2$ theory must permit an expansion in an even power of the generator $\phi_{-3,1}^{A_2}$. This however does not impose an independent constraint, as the weight of the numerator in the expression \eqref{eq:ZkAnsatz} for the elliptic genus is even for all $k$, and $\phi_{-3,1}^{A_2}$ is the only generator of odd weight.
\begin{figure}
	    \centering
	    \begin{tikzcd}
	     \phi_{0,1}^{G_2}\arrow[d]& \phi_{-2,1}^{G_2}\arrow[d]&\phi_{-6,2}^{G_2}\arrow[d]\\
	     \phi_{0,1}^{A_2}&\phi_{-2,1}^{A_2} &(\phi_{-3,1}^{A_2})^2
	    \end{tikzcd}
	    \caption{The map between $J(G_2)$ and $J(A_2)$ generators. The vertical arrows denote equality.}
	    \label{fig:jacobi_G2_to_A2}
	\end{figure}

	\subsubsection{$B_3$ to $G_2$} 
	
	All rank 1 models with $G_2$ gauge group arise via Higgsing of theories with gauge group $B_3$. We recall that the relevant ring of Jacobi forms for the latter, in the conventions of \cite{Bertola}, is $J(C_3)$, as its elements are shift symmetric under coroots of $B_3$. 
	We identify the Cartan algebras of $G_2$ and $B_3$ with the subspace $x_1+x_2+x_3=0$ of $\mathbb R^3$ and $
	\mathbb R^3$ itself. The map $\iota$ is then the inclusion
	\begin{equation*}
	    \iota:\{x_1+x_2+x_3=0\}\hookrightarrow \mathbb R^3.
	\end{equation*}
	
	The root lattice of $C_3$, as embedded in the orthogonal lattice $\Eucl_3$, reduces to the root lattice of $G_2$ upon restriction to the subspace $x_1+x_2+x_3=0$; likewise, the Weyl group of $C_3$ (equal to the Weyl group of $B_3$) maps to the Weyl group of $G_2$. The restriction $\iota^*$ hence provides a map from $J(C_3)$ to $J(G_2)$. In fact, this map is surjective: it maps one of the generators of $J(\Bn_3)$ to 0, and the other three to the generators of $J(G_2)$, see figure \ref{fig:jacobiB3}.
	
	\begin{figure}
	    \centering
	    \begin{tikzcd}
	     \phi_{0,1}^{C_3}\arrow[d]& \phi_{-2,1}^{C_3}\arrow[d]&\phi_{-4,1}^{C_3}\arrow[d]&\phi_{-6,2}^{C_3}\arrow[d]\\
	     \phi_{0,1}^{G_2}&\phi_{-2,1}^{G_2} & 0 &\phi_{-6,2}^{G_2}
	    \end{tikzcd}
	    \caption{The specialization of $J(C_3)$ to $J(G_2)$ generators. The vertical lines correspond to setting $x_1+x_2+x_3=0$ and multiplying by a constant.}
	    \label{fig:jacobiB3}
	\end{figure}

    \subsubsection{$F_4$ to $D_4$}
    
    All rank 1 models with $D_4$ gauge group arise via Higgsing of theories with gauge group $F_4$. 

    The root lattices of $D_4$ and $F_4$ are isomorphic, but are embedded differently in $\IR^4$ in the conventions of \cite{Bourbaki}. We give a map $\iota$ between these two realizations in appendix \ref{sec:appF4}.
    The Weyl group of $F_4$ coincides with the semi-direct product of the Weyl group of $D_4$ with the Dynkin diagram symmetry of $D_4$,
    \begin{equation*}
		W(F_4)=O(F_4)=W(D_4)\ltimes \text{DynkinSym}(D_4)=W(D_4)\ltimes S_3 \,.
	\end{equation*}
    Therefore, $\iota^*$ embeds $J(F_4)$ as a subring into $J(D_4)$, its elements consisting of $\weyl(D_4)$ symmetric Jacobi forms which in addition exhibit $D_4$ Dynkin diagram symmetry. Imposing this symmetry on the generic elements of $J(D_4)$ of appropriate weight and index, we arrive at the generators of $J(F_4)$ given in figure \ref{fig:jacobiF4}.
    
    \begin{figure}
	    \centering
	    \begin{tikzcd}[column sep=0.5]
	     \phi_{0,1}^{F_4}\arrow[d]& \phi_{-2,1}^{F_4}\arrow[d]&\phi^{F_4}_{-6,2}\arrow[d]&\phi_{-8,2}^{F_4}\arrow[d]&\phi_{-12,3}^{F_4}\arrow[d]\\
	     \phi_{0,1}^{D_4}-\frac{2}{3}E_4 \phi_{-4,1}^{D_4}  &\phi_{-2,1}^{D_4} & \phi_{-6,2}^{D_4}-\frac{1}{18}\phi_{-2,1}^{D_4}\phi_{-4,1}^{D_4} & \left(\phi_{-4,1}^{D_4}\right)^2+3\left(\omega_{-4,1}^{D_4}\right)^2 &\phi_{-4,1}^{D_4}\left(\omega_{-4,1}^{D_4}\right)^2-\frac{1}{9}\left(\phi_{-4,1}^{D_4}\right)^3
	    \end{tikzcd}
	    \caption{The specialization of $J(F_4)$ to $J(D_4)$ generators. The vertical lines correspond to composition via an isomorphism of the the two lattices which is given explicitly in appendix \ref{sec:appF4}.}
	    \label{fig:jacobiF4}
	\end{figure}

	\subsubsection{$A$-series}
	
	The Higgsing tree over $\IF_1$ and over $\IF_2$ both exhibit a branch of $A_n$ gauge theories for $n$ arbitrarily large, with a sequence of Higgsings $A_{n+1} \rightarrow A_n$ all the way down to a theory with gauge symmetry $A_1$.\footnote{The Higgsing of certain theories over the bases $\IF_1$ and $\IF_2$ is studied from the perspective of the 2D quiver theory living on the BPS strings in \cite{Kim:2015fxa}.}
	
	The root lattice of $A_n$ is embedded in the orthogonal lattice $\Eucl_{n+1}$ via the constraint $\sum_{i=1}^{n+1} x_i = 0$. The Weyl group of $A_n$ is the group $S_{n+1}$ of permutations on the $n+1$ generators of $\Eucl_{n+1}$. The map $\iota$ for the $A_{n+1}\to A_n$ Higgsing is induced by the inclusion of the orthogonal lattices: $\Eucl_{n+1}\hookrightarrow \Eucl_{n+2}$. Restricting to the sublattice $x_{n+2}=0$ thus maps the root lattice and the Weyl group of $A_{n+1}$ to that of $A_{n}$. This restriction maps the lowest weight generator $\phi_{-(n+1),1}$ to 0, and otherwise preserves weight and index, leading to the relation between generators summarized in figure~\ref{fig:jacobiA}.
	
	\begin{figure}
	    \centering
	    	\begin{tikzcd}
	\vdots \arrow[d]&\vdots \arrow[d]&\vdots \arrow[d]&\dots \arrow[d]&\vdots \arrow[d]& \vdots \arrow[d]& \vdots \arrow[d]\\
	\phi_{0,1}^{A_n}\arrow[d]&\phi_{-2,1}^{A_n}\arrow[d]&\phi_{-3,1}^{A_n}\arrow[d]&\dots\arrow[d]&\phi_{-n,1}^{A_n}\arrow[d] &\phi_{-n-1,1}^{A_n}\arrow[d] & 0\\
	\phi_{0,1}^{A_{n-1}}\arrow[d]&\phi_{-2,1}^{A_{n-1}}\arrow[d]&\phi_{-3,1}^{A_{n-1}}\arrow[d]&\dots\arrow[d]&\phi_{-n,1}^{A_{n-1}}&0\\
	\vdots \arrow[d]&\vdots \arrow[d]&\vdots \arrow[d]&\dots \arrow[d]&\vdots \\
	\phi_{0,1}^{A_2}\arrow[d]&\phi_{-2,1}^{A_2}\arrow[d]&\phi_{-3,1}^{A_2}\arrow[d]& 0\\
	\phi_{0,1}&\phi_{-2,1}& 0
	\end{tikzcd}
	    \caption{Restriction of the $A$-series Jacobi forms. The vertical arrows correspond to setting the last coordinate to 0 and multiplying by a constant.}
	    \label{fig:jacobiA}
	\end{figure}
	
   \subsubsection{... $\rightarrow D_{n+1} \rightarrow B_n \rightarrow D_n \rightarrow$ ...}
   The Higgsing trees over $\IF_n$ for $n=0,1,2,3,4$ exhibit branches of $B_n$ and $D_n$ gauge theories for $n$ arbitrarily large. Along these branches, the pattern of Higgsing is $\ldots \rightarrow D_{n+1} \rightarrow B_n \rightarrow D_n \rightarrow \ldots$.\footnote{These branches over the bases $\IF_2$, $\IF_3$ and $\IF_4$ were studied from the perspective of brane-systems in \cite{Kim:2019dqn}.}
	
	Recall that in the conventions of \cite{Bertola}, $J(C_n)$ is the appropriate ring of Jacobi forms for the construction of the elliptic genera for $B_n$ gauge theories, as its elements are shift symmetric under $\LambdaCR(B_n) = \LambdaR(C_n)$.

    The orthogonal lattice for $C_n$ and $D_n$ is $\Eucl_n$. For the Higgising $B_n\to D_n$, the map $\iota$ is simply the identity. The root lattices of $C_n$ and $D_n$ coincide. Furthermore,
	\be
	W(C_n) = S_n \ltimes (\IZ_2)^n = W(D_n) \ltimes \text{DynkinSym}(D_n) \,;
	\ee
	in addition to permutations of the generators of the Euclidean lattice $\Eucl_n$, $W(C_n)$ includes arbitrary sign flips, whereas $W(D_n)$ includes only even numbers of sign flips. The generators of $J(D_n)$ and $J(C_n)$ can be chosen to reflect the close relation between these two groups: $n$ of the generators can be chosen to coincide (i.e. are in particular invariant under arbitrary sign flips). The final generator for $J(D_n)$ is odd under an odd number of sign flips. Its square provides the missing generator for $J(C_n)$.
	
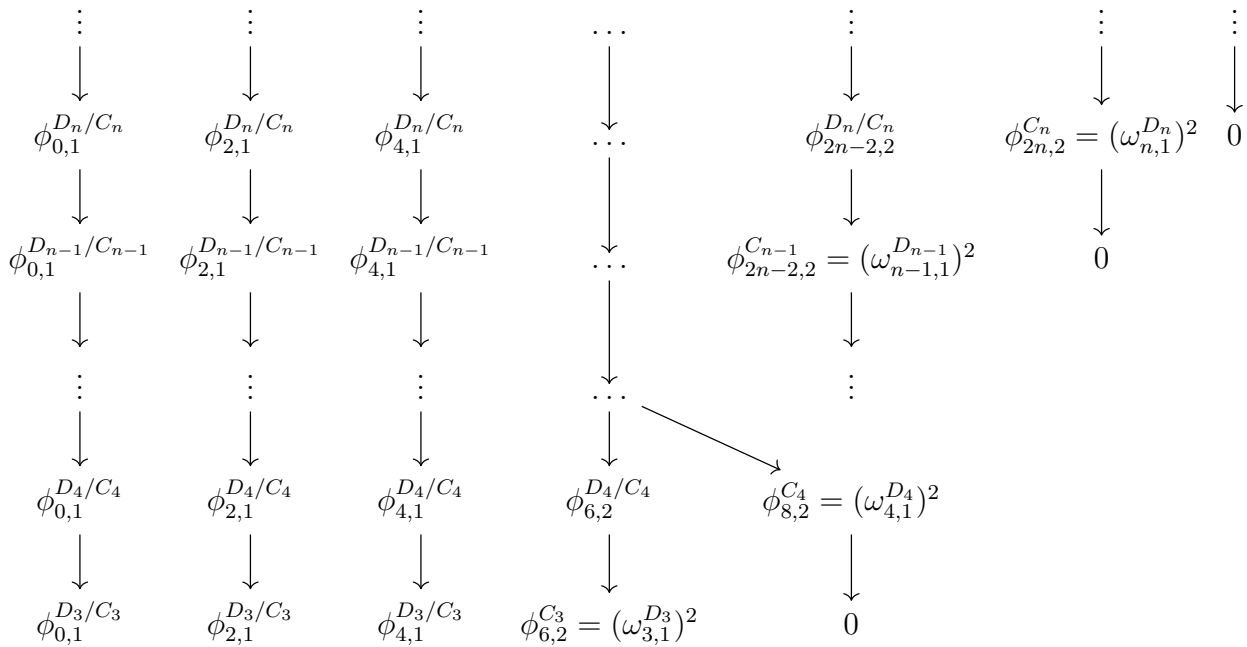
\begin{figure}
	    \centering
	    	\begin{tikzcd}[column sep=0.3]
	\vdots \arrow[d]&\vdots \arrow[d]&\vdots \arrow[d]&\dots \arrow[d]&\vdots \arrow[d]& \vdots \arrow[d]& \vdots \arrow[d]\\
	\phi_{0,1}^{D_n/C_n}\arrow[d]&\phi_{2,1}^{D_n/C_n}\arrow[d]&\phi_{4,1}^{D_n/C_n}\arrow[d]&\dots\arrow[d]&\phi_{2n-2,2}^{D_n/C_n}\arrow[d] &\phi^{C_n}_{2n,2}=(\omega_{n,1}^{D_n})^2\arrow[d] & 0\\
	\phi_{0,1}^{D_{n-1}/C_{n-1}}\arrow[d]&\phi_{2,1}^{D_{n-1}/C_{n-1}}\arrow[d]&\phi_{4,1}^{D_{n-1}/C_{n-1}}\arrow[d]&\dots\arrow[d]&\phi_{2n-2,2}^{C_{n-1}}=(\omega_{n-1,1}^{D_{n-1}})^2\arrow[d] &0\\
	\vdots \arrow[d]&\vdots \arrow[d]&\vdots \arrow[d]&\dots \arrow[d]\arrow[rd]&\vdots \\
	\phi_{0,1}^{D_4/C_4}\arrow[d]&\phi_{2,1}^{D_4/C_4}\arrow[d]&\phi_{4,1}^{D_4/C_4}\arrow[d]&\phi_{6,2}^{D_4/C_4}\arrow[d]&\phi_{8,2}^{C_4}=(\omega_{4,1}^{D_4})^2\arrow[d] \\
	\phi_{0,1}^{D_3/C_3}&\phi_{2,1}^{D_3/C_3}&\phi_{4,1}^{D_3/C_3}&\phi_{6,2}^{C_3}=(\omega_{3,1}^{D_3})^2 & 0
	\end{tikzcd}
	    \caption{Restriction of the $D/C$ series Jacobi forms. The vertical arrows correspond to setting the last coordinate to 0 and multiplying by a constant (for $C_3$ we picked a different basis). This is the same table as in \cite{Adler_2020}.}
	    \label{fig:jacobiBD}
	\end{figure}

	For the $D_n\to B_{n-1}$ Higgsing, the map $\iota$ is induced by the inclusion $\Eucl_{n-1}\hookrightarrow \Eucl_{n}$. By restricting to $x_n=0$, the root lattice and Weyl group of $D_n$ are mapped to $\Lambda_r(C_{n-1})$ and $W(C_{n-1})$, respectively. Consequently, setting $x_n=0$ maps the generators of $J(D_n)$ to those of $J(C_{n-1})$. These relations are summarized in figure \ref{fig:jacobiBD}.
    
    For the case $n=4$, the restriction to $x_n=0$ does not yield the standard basis of $C_3$ Jacobi forms (as given for instance in \cite{Bertola}), which introduces some awkwardness in the reduction. The details are given in appendix \ref{app:specializationformulas}.
	
	\subsubsection{$C$-series}
	
	The numerator of the elliptic genus of models with $C_n$ gauge symmetry, which arise as nodes of the $\mathbb F_1$ Higgsing tree, take value in $J(B_n)$. As the necessary singularity enhancements of the elliptic fiber cannot be obtained torically \cite{Bershadsky:1996nh, Kashani-Poor:2019jyo}, we did not study these models in this work. However, as we will discuss in the next section, due to symmetry enhancement, the numerators of the elliptic genus of theories with gauge symmetry of $D$- and $B$-type over $\mathbb F_4$ are elements of these rings.
	
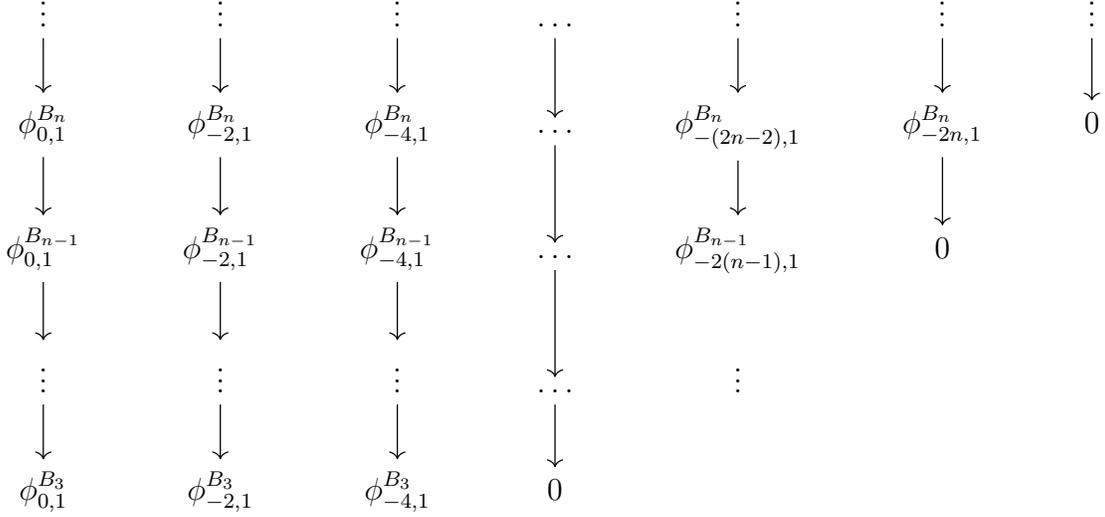
\begin{figure}
	    \centering
	    	\begin{tikzcd}
	\vdots \arrow[d]&\vdots \arrow[d]&\vdots \arrow[d]&\dots \arrow[d]&\vdots \arrow[d]& \vdots \arrow[d]& \vdots \arrow[d]\\
	\phi_{0,1}^{B_n}\arrow[d]&\phi_{-2,1}^{B_n}\arrow[d]&\phi_{-4,1}^{B_n}\arrow[d]&\dots\arrow[d]&\phi_{-(2n-2),1}^{B_n}\arrow[d] &\phi_{-2n,1}^{B_n}\arrow[d] & 0\\
	\phi_{0,1}^{B_{n-1}}\arrow[d]&\phi_{-2,1}^{B_{n-1}}\arrow[d]&\phi_{-4,1}^{B_{n-1}}\arrow[d]&\dots\arrow[d]&\phi_{-2(n-1),1}^{B_{n-1}}&0\\
	\vdots \arrow[d]&\vdots \arrow[d]&\vdots \arrow[d]&\dots \arrow[d]&\vdots \\
	\phi_{0,1}^{B_3}&\phi_{-2,1}^{B_3}&\phi_{-4,1}^{B_3}& 0
	\end{tikzcd}
	    \caption{Restriction of the $B$-series Jacobi forms. The vertical arrows correspond to setting the last coordinate to 0 and multiplying by a constant.}
	    \label{fig:jacobiB}
	\end{figure}

    The map $\iota$ for the $C_{n}\to C_{n-1}$ Higgsing is induced by the inclusion $\Eucl_{n-1}\hookrightarrow 
    \Eucl_n$. Once we set $x_n=0$, the Weyl group and root lattice of $B_n$ map to the Weyl group and root lattice of $B_{n-1}$. So we expect the simple transformation law in figure \ref{fig:jacobiB}.
	
	\subsection{Specialization of the elliptic genus} \label{ss:specDen}
	
	The structure of the denominator as given in \eqref{eq:denom_G} is dictated by the roots of the Lie algebra. As explained in section \ref{ss:matching_elliptic_parameters}, we choose conventions for the roots in the orthogonal basis (i.e. the embedding of roots in a Euclidean lattice) such that the coroots of the Lie algebra $\mathfrak g$ coincide with the expressions as given by \cite{Bourbaki} for the roots of the dual Lie algebra. We recall that this is the natural normalization for us as the argument of the elliptic genus is an element of the (complexified) Cartan algebra $\mathfrak h_{\mathbb C}$ periodic under translation by elements of the coroot lattice (rather than the root lattice).
    
    In subsection \ref{ss:restricting_Jacobi_forms}, we defined for each Higgsing $\mathfrak{g} \rightarrow \mathfrak{g}'$ a restriction map $\iota^*$. Under this mapping, the image of a $\weyl(\mg)$ invariant function is $\weyl(\mg')$ invariant. Furthermore, the positive roots of $\mg'$ are mapped by $\iota$ onto a subset of positive roots of $\mg$. It follows that 
     \begin{equation*}
        \frac{\iota^*(\cD_k^{\mg})}{\cD_k^{\mg'}}
    \end{equation*}
    is an element of the ring $J(\mg') \otimes J(A_1)$ (the second factor having elliptic parameter $\gs$). Our prediction is that this element factors out of $\iota^*(\cN_k^\mg)$, thus establishing \eqref{eq:specializing}.\footnote{Note that the power of the Dedekind $\eta$ function in \eqref{eq:ZkAnsatz} only depends on the base of the elliptic fibration.} Below, we perform this reduction explicitly for various Higgsings.

    Once we have established \eqref{eq:specializing}, we can invoke the specialization mechanism to improve our ansatz for $Z_k$ as we move away from the root of a Higgsing tree. We distinguish between two cases.
    
    The first case is when the map induced by $\iota^*$ between $J(\mg)$ and $J(\mg')$ is one-to-one. In this case, we can impose the reduced denominator $(\iota^*)^{-1}(\cD_k^{\mg'})$ for the theory with $\mg$ gauge symmetry. In fact, the elliptic genera $Z_k^\mg$ and $Z_k^{\mg'}$ as functions on the Euclidean lattice (which coincides for both algebras) coincide. The two differ as functions of gauge fugacities or K\"ahler parameters, as the map between the Euclidean lattice and the K\"ahler cone of the underlying geometries (explained in subsection \ref{ss:matching_elliptic_parameters}) differ.\footnote{This is why \cite{Haghighat:2014vxa,DelZotto:2017mee} could extract Gopakumar-Vafa invariants of the $D_4$ geometry over base $\IF_4$ from the $B_4$ geometry one node up the Higgsing tree, resolving an issue raised in footnote 14 of \cite{Kashani-Poor:2019jyo}.} In section \ref{sec:enhanced}, we will discuss symmetry enhancements which allow to choose $\cN^{\mathfrak{g}}$ to lie in a smaller ring than $J(\mg)$. For such an ansatz to be sufficient, it is necessary to impose the reduced denominator; otherwise, factors not invariant under the enhanced symmetry are required in the numerator to cancel corresponding terms in the denominator. This effect becomes apparent in the base degree 2 example that we study in section \ref{sec:HigherBaseDegree}.

    The second case arises when Higgsing to a theory with smaller rank. In this case, the map $\iota^*:J(\mg)\to J(\mg')$ has a non-trivial kernel. We again have
    \begin{equation*}
        \iota^*(\cN^{\mathfrak{g}})=\cN^{\mathfrak{g}'}\frac{\iota^*(\cD_k^{\mathfrak{g}})}{\cD_k^{\mathfrak{g}'}} \,,
    \end{equation*}
    i.e. 
    \begin{equation*}
        \cN^{\mathfrak{g}}\in (\iota^*)^{-1}\left(\cN^{\mathfrak{g}'}\frac{\iota^*(\cD_k^{\mathfrak{g}})}{\cD_k^{\mathfrak{g}'}}\right) \,.
    \end{equation*}
    Now, any two elements in this preimage differ by an element of the kernel. The kernel of $\iota^*$ is a principal ideal $\psi J(\mg)$ generated by an element $\psi \in J(\mg)$ which can be read off from the figures in section \ref{ss:restricting_Jacobi_forms}. If we pick a particular element in the preimage $\hat \phi\in (\iota^*)^{-1}\left(\cN^{\mathfrak{g}'}\frac{\iota^*(\cD_k^{\mg})}{\cD_k^{\mg'}}\right)$ (for instance by going to section  \ref{ss:restricting_Jacobi_forms} and following the arrows in reverse), we have that 
    \begin{equation*}
        \cN^{\mathfrak{g}}=\hat \phi +\psi\phi.
    \end{equation*}
   If one has already computed $\cN^{\mathfrak{g}'}$, this reduces the calculation of $\cN^{\mathfrak{g}}$ to the determination of $\phi$, whose weight and index are (in absolute value) smaller than those of $\cN^{\mathfrak{g}}$, thus reducing the number of coefficients that must be determined. In practice however, this reduction is not substantial: the weight and index of $\psi$ are small against those of $\cN^\mg$ already at $k=1$; as $\psi$ is a fixed form while the weight and index of the numerator increase rapidly with base degree $k$, the reduction becomes even more marginal as $k$ increases.
    
    \subsubsection{$G_2$ to $A_2$}
    
    As the long roots of $A_2$ and $G_2$ coincide in our conventions (see the discussion at the end of subsection \ref{ss:matching_elliptic_parameters}),  
    \begin{equation*}
        \cD_{k,L}^{G_2}=\cD_{k,L}^{A_2} \,.
    \end{equation*}
    For the short roots, we have for $m \in \LambdaCR(G_2) \otimes \IC$ that by $\sum_{i=1}^3 x_i = 0$,
    \begin{equation*}
        \left(\frac{2}{3}e_i-\frac{1}{3}e_j-\frac{1}{3}e_k,m\right)=\frac{2}{3}x_i-\frac{1}{3}x_j-\frac{1}{3}x_k=x_i,\quad i\neq j\neq k \neq i.
    \end{equation*}
    Thus,
    \begin{equation*}
        \cD_{k,S}^{G_2}=\prod_{i=1}^3\cD^{A_1}_{k,e_i}.
    \end{equation*}
    
    As the map $\iota^*: J(G_2)\to J(A_2)$ is one-to-one, equation \eqref{eq:specializing} actually implies that we can refine our ansatz by taking the $A_2$ denominator in all of the theories with $G_2$ gauge symmetries, as all of them can be Higgsed to $A_2$. 
    
    We have computed $Z_1$ for all $A_2$ and $G_2$ nodes of rank 1 Higgsing trees (these occur over $\IF_n$, $n=0, \ldots, 3$). 
    At $k=1$, only the long roots contribute to the denominator. Hence,
	\be \label{eq:denominators_A2_G2_coincide}
	\cD_1^{A_2}=\cD_1^{G_2} \,.
	\ee
	We have verified that as functions of orthogonal coordinates $x_i$,
	\begin{equation*}
	    Z_1^{G_2}=Z_1^{A_2} \,.
	\end{equation*}

    \subsubsection{$B_3$ to $G_2$}
    The map $\iota^*$ associated to the Higgsing $B_3\to G_2$ is implemented by imposing the condition $x_1+x_2+x_3=0$ on the orthogonal coordinates of the Cartan algebra of $B_3$. The six long roots of $G_2$, given by $\pm(e_i- e_j)$, coincide with 6 of the 12 long roots of $B_3$, while the contribution of the remaining long roots maps under $\iota^*$ to that of the short roots of $G_2$. Thus,
    \begin{equation*}
        \cD_{k,L}^{B_3}|_{\sum x_i=0}=\cD_{k,L}^{G_2}\prod_{\alpha=e_i+e_j}\cD_{k,\alpha}^{A_1}=\cD_{k,L}^{G_2}\cD_{k,S}^{G_2}
    \end{equation*}
    Furthermore, $\iota^*$ maps the contribution of the short roots of $B_3$ to the contribution of the short roots of $G_2$. Hence,
    \begin{equation*}
        \cD_{k,S}^{B_3}|_{x_1+x_2+x_3=0}=\cD_{k,S}^{G_2} \,.
    \end{equation*}
    We thus expect that $\iota^*(\cN_k^{B_3})$ factorizes to cancel the contribution $\cD_{k,S}^{G_2}$.
    
    We have computed $Z_1$ for all $B_3$ nodes of rank 1 Higgsing trees (these occur over the base $\IF_n$ for $n= 0, \ldots, 3$).\footnote{The maximally Higgsed theories over the bases $\IF_0$ and $\IF_2$ are equivalent \cite{Morrison:1996na}. This however is no longer the case once one moves up from the root of the trees \cite{Kashani-Poor:2019jyo}.}

     At $k=1$, $\cD_{k,S}^{G_2}$ evaluates to
    \begin{equation}
      \prod_{\alpha=e_i+e_j}\cD_{1,\alpha}^{A_1}=(-1)^3\prod_i\phi_{-2,1}(x_i)=-\phi_{-6,2}^{G_2} \,,
    \end{equation}
    and we verify that $\iota^*(\cN_1^{B_3})$ is indeed divisible by $-\phi_{-6,2}^{G_2}$, such that
   \begin{equation*}
        \left.Z_1^{B_3}\right|_{\sum x_i=0}=\left.\frac{1}{\eta^{n(1)}}\frac{\mathcal N^{B_3}}{\mathcal D^{B_3}}\right|_{\sum x_i=0}=\frac{1}{\eta^{n(1)}}\frac{\mathcal N^{G_2}}{\mathcal D^{G_2}}=Z_1^{G_2}\,.
    \end{equation*}

	\subsubsection{$F_4$ and $B_4$ to $D_4$} \label{ss:F4 and B4 to D4: denom}
	The respective maps $\iota$ map the long roots of $D_4$ to those of $B_4$, $F_4$ respectively. As $\iota: \mathfrak{h}_{D_4} \rightarrow \mathfrak{h}_{B_4}$ is the identity, we suppress it (as well as the corresponding $\iota^*$) in this subsection to lighten notation. $\iota$ in the following will thus refer to the map \eqref{eq: F4 to D4}. With this understanding, 
	\begin{equation*}
	    \iota^*\cD_{k,L}^{F_4}=\cD_{k,L}^{B_4}=\cD_{k,L}^{D_4}=\cD_k^{D_4} \,.
	\end{equation*}
    For the specializations $\iota^*(Z_k^{F_4}) = Z_k^{D_4}$ and $Z_k^{B_4} = Z_k^{D_4}$ (the latter as an identity of functions of the orthogonal lattice coordinates) to be correct, we thus expect 
	\be \label{eq:F4/B4toD4Factorization}
	\iota^*(\cN_k^{F_4}) = \iota^*(\cD_{k/c,S}^{F_4}) \cN_k^{D_4} \,, \quad \cN_k^{B_4} = \cD_{k/c,S}^{B_4} \cN_k^{D_4} \,.
	\ee
    We have computed $Z_1$ for the $B_4$ and $F_4$ nodes of rank 1 Higgsing trees over the base $\IF_n$ for $n=0, \ldots, 4$ and checked the specialization equations by verifying
    \be \label{eq:z1F4=z1B4}
    \iota^*(Z_1^{F_4}) = Z_1^{B_4} \,.
    \ee
    As explained in section \ref{ss:bcs}, we could not directly compute the elliptic genera for $D_4$, as we cannot compute the required Gromov-Witten invariants using toric methods. But having demonstrated that specialization holds by verifying \eqref{eq:z1F4=z1B4}, we can use our results to compute $Z_1^{D_4}$ and extract the associated Gromov-Witten invariants.
    A non-trivial check on our results is that they specialize correctly (for $\IF_n$, $n=0,\ldots,3$) to the appropriate $B_3$ gauge theories, whose elliptic genera we compute independently.
    
    In section \ref{sec:HigherBaseDegree}, we test the factorization \eqref{eq:F4/B4toD4Factorization} at base degree 2 and genus 0.

	\subsubsection{$A$-series}
	The map $\iota^*$ for the Higgsing $A_n \rightarrow A_{n-1}$ is given by
	\be
	\iota^* = \left. \cdot \right|_{x_{n+1}=0} \,.
	\ee
	The roots of $A_n$ are given by the roots of $A_{n-1}$ together with the set $\pm(e_i-e_{n+1})$, $i=1,\dots,n$. Therefore, 
	\begin{equation*}
 \left.\cD_{k}^{A_n}\right|_{x_{n+1}=0}=\cD_k^{A_{n-1}}\prod_{i=1}^{n}D_{k,e_i}^{A_1} \,.
	\end{equation*}
	 For the specialization $\iota^*( Z_k^{A_n}) = Z_k^{A_{n-1}}$ to be correct, we thus expect 
	 \be  \label{eq:factorization A-series}
	\iota^*(\cN_k^{A_n}) = \iota^*(\prod_{i=1}^{n}D_{k,e_i}^{A_1})\cN_k^{A_{n-1}} \,.
	\ee
	At base degree 1, the prefactor of $\cN_k^{A_{n-1}}$ in the above equation evaluates to
	\begin{equation*}
    \left.\prod_{i=1}^{n}D_{1,e_i}^{A_1}\right|_{x_{n+1}=0}= \prod_{i=1}^{n}\left(-\phi_{-2,1}(x_i)\right)=(-1)^{n}(\phi_{-n,1}^{A_{n-1}})^2 \,.
	\end{equation*}
	We have checked the factorization \eqref{eq:factorization A-series} for the gaugings $A_3 \rightarrow A_2$ over the bases $F_1$ and $F_2$.

    \subsubsection{... $\rightarrow D_{n+1} \rightarrow B_n \rightarrow D_n \rightarrow$ ...}
    The map $\iota^*$ for the Higgsing $D_n \rightarrow B_{n-1}$ is given by
	\be
	\iota^* = \left. \cdot \right|_{x_{n}=0} \,.
	\ee
    The long roots of $D_n$ are given by the long roots of $B_{n-1}$ together with the set $\pm(e_i\pm e_n)$, $i=1,\dots,n-1$, which gives rise to the second factor on the RHS of the following equality,
    \begin{equation*}
        \left.\mathcal D_k^{D_n}\right|_{x_n=0}=\mathcal D_{k,L}^{B_{n-1}}\prod_{i=1}^{n-1}(D_{k,e_i}^{A_1})^2
    \end{equation*}
    For the specialization $\iota^*( Z_k^{D_n}) = Z_k^{B_{n-1}}$ to be correct, we thus expect 
	 \be  \label{eq:factorization D to B}
	\iota^*(\cN_k^{D_n}) = \iota^*\left(\prod_{i=1}^{n}D_{k,e_i}^{A_1}\right)^2\cN_k^{B_{n-1}} \,.
	\ee
    At base degree 1, we have
    \begin{equation*}
        \left.\mathcal D_1^{D_n}\right|_{x_n=0}=\mathcal D_1^{B_{n-1}}\left(\prod_{i=1}^{n-1} \phi_{-2,1}(x_i)\right)^2 = \mathcal D_1^{B_{n-1}}(\phi_{-2(n-1),2}^{C_{n-1}})^2.
    \end{equation*}
    For $n=4$, we checked the factorization \eqref{eq:factorization D to B} and thus the specialization
    \be
    \left. Z_1^{D_4}\right|_{x_4=0}=Z_1^{B_3}
    \ee
    for the corresponding nodes of the rank 1 Higgsing trees over bases $\mathbb F_0,\mathbb F_1,$ and $\mathbb F_2$, with $Z_1^{D_4}$ computed as explained in subsection \ref{ss:F4 and B4 to D4: denom}
	
	The map $\iota$ associated to the Higgsing $B_n \rightarrow D_n$ is the identity, the induced map $\iota^*:J(C_n)\hookrightarrow J(D_n)$ is injective. The long roots of both gauge algebras coincide, hence 
	\begin{equation*}
	    \mathcal D_{k,L}^{B_n}=\mathcal D_k^{D_n}.
	\end{equation*}
	For the specialization $\iota^*( Z_k^{B_n}) = Z_k^{D_{n}}$ to be correct, we thus expect 
	 \be  \label{eq:factorization B to D}
	\iota^*(\cN_k^{B_n}) = \iota^*(D_{k/c,S}^{B_n})\cN_k^{D_{n}} \,.
	\ee
	Furthermore, as $\iota^*$ is injective, we expect
	\be \label{eq:ZB = ZD}
	Z_k^{B_n} = Z_k^{D_n}
	\ee
	as functions of orthogonal coordinates. 
	
	Since we cannot compute $Z_k^{D_n}$ directly (see the discussion in subsection \ref{ss:bcs}), we cannot check \eqref{eq:factorization B to D} and \eqref{eq:ZB = ZD} directly. However, given the numerous checks that our specialization formulae have passed, we feel confident in invoking \ref{eq:ZB = ZD} to identify the elliptic genera of $D_n$ gauge theories.
	
   \subsubsection{$C$-series}
   The map $\iota^*$ for the Higgsing $C_n \rightarrow C_{n-1}$ is given by
	\be
	\iota^* = \left. \cdot \right|_{x_{n}=0} \,.
	\ee
    The denominators $\cD_k^{C_n}$ are mapped to zero under $\iota^*$, as the root $e_n$ leads to a factor $\phi(2x_n)$ which vanishes at $x_n=0$. However, one of the generators of $J(B_n)$ also lies in the kernel of $\iota^*$, and must factor out of the numerators $\cN_k^{C_n}$ to cancel these apparent poles.
   
   The long roots of $C_n$ are given by the long roots of $C_{n-1}$ together with the roots $\pm e_n$\footnote{Note that $(m,e_n)=2x_n$ as the inner product here is $2 dx^2$, see the $B$-series column in table \ref{tab:rootSystems1}.} together with the $C_{n-1}$ long roots. Then 
   \begin{equation*}
        \cD^{C_n}_{k,L} =\cD_{e_n,L}^{A_1}\cD^{C_{n-1}}_{k,L}=-\phi_{-2,0}(2x_n)\prod_{\substack{ab \le k, a,b >0 \\ ab\neq 0}} \phi_{-2,0}[(a-b)g_{top}]\cD^{C_{n-1}}_{k,L} + \cO(x_n^2)\,.
   \end{equation*}
    The short roots of $C_n$ are given by the short roots of $C_{n-1}$ together with $\pm e_i\pm e_n$. Therefore,
    \begin{equation*}
        \left. \cD_{k,S}^{C_n} \right|_{x_n=0}=\prod_{i=1}^{n-1}(D_{k,e_i}^{A_1})^2\cD_{k,S}^{C_{n-1}} \,.
    \end{equation*}
   At base degree 1, these relations reduce to
   \begin{equation*}
       \cD^{C_n}_{1}=-\phi_{-2,0}(2x_n)\cD^{C_{n-1}}_{1}
   \end{equation*}
   
   We did not study models with $C_n$ gauge symmetry, as the Gromov-Witten invariants are not accessible torically, see the discussion in subsection \ref{ss:bcs}. They could presumably be solved by imposing vanishing conditions on Gopakumar-Vafa invariants, but we did not pursue this approach here.
    
    \subsubsection{$E_6$, $E_7$, $E_8$}
    Rank 6 puts already the smallest of the Lie algebras $E_6,E_7,E_8$ out of computational reach. We do however want to briefly discuss the specialization of the denominator for these theories. Similar to the $C$-series, the denominator for a theory with $E_{n+1}$ gauge symmetry vanishes once we restrict to the Cartan algebra of $E_n$.
    
    In somewhat more detail, the contribution of the roots $\pm(e_7+e_8)$ of $E_8$ to  the denominator cause it to vanish once we impose the $E_7$ constraint $x_7+x_8=0$:
    \begin{equation*}
        \cD_1^{E_8}\propto \alpha^2(x_7+x_8)\xmapsto{x_7+x_8=0}\alpha^2(0)=0 \,.
    \end{equation*}
    We expect the $E_8$ numerator to be proportional to $\alpha^2(0)$ once we restrict to $x_7+x_8=0$. Checking this claim however lies beyond the scope of this work: the ring $J(E_8)$ being the only ring of $\weyl(\mg)$ invariant Jacobi forms for $\mg$ a simple Lie algebra which is not polynomially generated \cite{Wirthmuller:Jacobi,Wang:2018fil}, the strategy employed in this paper does not extend straightforwardly to this case.
    
    For the $E_7\to E_6$ Higgsing, the story is very similar. The roots $\pm(e_7-e_6)$ lead to a divergence once we impose the $E_6$ constraint $x_7-x_6=0$: 
    \begin{equation*}
        \cD_1^{E_8}\propto \alpha^2(x_7-x_6)\xmapsto{x_7-x_6=0}\alpha^2(0)=0.
    \end{equation*}
    Again, we expect this divergence to be cancelled by the numerator. Contrary to the $E_8$ case, theories with $E_6$ and $E_7$ gauge symmeetry could be solved with our techniques, given more computational power, or more patience, and this claim could thus be checked explicitly.
    
    \subsection{Specializing Gromov-Witten invariants} \label{ss:GWinv}
    The specialization in equation $\eqref{eq:specializing}$ implies that the Gromov-Witten invariants of theories related by Higgsing $\mg \rightarrow \mg'$ are closely related. Naively,
    \begin{equation} \label{eq:GW_expansion}
    F^{\mg'}=\sum_{\kappa\in H_2(X^{\mg'})} \gs^{2g-2}r^{\mg'}_{g,\kappa}Q^\kappa=\log Z^{\mg'}=\iota^*\log Z^{\mg}=\iota^* F^{\mg}=\sum_{\kappa\in H_2(X^{\mg})}\gs^{2g-2}r^{\mg}_{g,\kappa}\iota^*Q^\kappa\,,
    \end{equation}
    such that
    \begin{equation}
    \label{eq:gwinv}
    r^{\mg'}_{g,\kappa'}=\sum_{\substack{\kappa\in H_2(X^\mg)\\ \iota^*(Q^\kappa)=Q^{\kappa'} }}r^{\mg}_{g,\kappa}\,.
    \end{equation}
    This reasoning must however be applied with care: the map $\iota$ was introduced in \eqref{eq:def_iota} to relate Weyl invariant expressions. As such, composition with any Weyl transformation yields an equally suitable map. However, Weyl transformations on the parameter $m$ introduced in section \ref{ss:matching_elliptic_parameters} map the set $\{Q_i=\exp(2\pi i (\alpha_i, m))\}$ of exponentiated fibral K\"ahler parameters to a new set $\{\tilde{Q}_i\}$ involving negative powers of the $Q_i$ (a consequence of the fact that any Weyl transformation maps a Weyl chamber to a distinct Weyl chamber). They hence do not commute with the Taylor expansion, indicated in \eqref{eq:GW_expansion}, required to extract Gromov-Witten invariants from $Z_g$.  When comparing Gromov-Witten invariants, the appropriate representative in the Weyl orbit of the map $\iota$ needs to be chosen in order to relate the Taylor expansions in terms of positive powers of the respective exponentiated K\"ahler parameters. Note that $\iota^*$ only acts on the exponentiated fibral K\"ahler parameters, not on those of the base curve $Q_b$ nor of the general fiber $q$. The action on the affine parameter $Q_0$ follows from $Q_0=q/\prod Q_i^{a_i}$. 

    How to determine the appropriate representative?
    \begin{enumerate}
    \item  \label{how_to_1} When a choice exists such that that $\iota^* Q^{\mathfrak g'}$ is a monomial in only positive powers of $Q^{\mathfrak g}$, this uniquely determines the map $\iota$ which commutes with the Taylor expansion.
    
    \item \label{how_to_2} Else, if in terms of orthogonal coordinates on the coroot lattice, $\iota^*$ identifies the denominator of $Z^\mathfrak{g}$, in the presentation \eqref{eq:ZkAnsatz}, with that of $Z^\mathfrak{g'}$, the appropriate representative is the one which allows the identification (up to a constant) of the prefactors $p_k^\mg$ and $p_k^{\mg'}$ in the presentation 
    \be
    \cD_k = p_k(Q) (1 + \sum_{\kappa>0} a_\kappa Q^\kappa)
    \ee
    of the denominator, as functions of orthogonal coordinates. 
    
    For $k=1$, the prefactor takes the simple form
    \begin{equation}
        p_1(Q) = \exp(-(2i\pi)m_{\rho_L}) \,,
    \end{equation}
    where $\rho_L$ is the sum of all positive long roots.
    
    \item Finally, only in the case of specializations $A_n \rightarrow A_{n-1}$, neither condition \ref{how_to_1} nor condition \ref{how_to_2} apply. It turns out that the composition of the Weyl transformation $x_n\leftrightarrow x_{n+1}$ with the map $\iota$ which sets $x_{n+1}=0$ gives the correct transformation in this case.
    
    \end{enumerate}

    \subsubsection{$G_2$ to $A_2$}
    For the specialization from $G_2$ to $A_2$, no choice of $\iota$ exists which maps the fibral K\"ahler parameters for the $G_2$ geometry to monomials of the K\"ahler parameters for the $A_2$ geometry. However, the denominators of the associated $Z_1$ coincide, see the discussion around equation \eqref{eq:denominators_A2_G2_coincide}. We are hence in case \ref{how_to_2}. With 
    \begin{equation}
        m_{\alpha_1}^{G_2} = -x_1 \,, \quad m_{\alpha_2}^{G_2} = x_1 - x_2
    \end{equation}
    and 
    \begin{equation}
        m_{\alpha_1}^{A_2} = x_1 - x_2 \,, \quad m_{\alpha_2}^{A_2} = x_1 + 2 x_2 \,,
    \end{equation}
    we obtain
    \begin{equation}
        m_{\rho_L}^{G_2}=-2x_1-2x_4\,, m_{\rho}^{A_2}=4x_1+2x_2\,.
    \end{equation}
    The map $\iota$ proposed in section \ref{ss:Jacobi_G2_to_A2} is the identity. To match the quantities $m_{\rho_L}^{G_2}$ and $m_{\rho}^{A_2}$, we thus perform the Weyl transformation $x_1\mapsto -x_2\,, x_2\mapsto -x_1$. This yields the map
    \begin{align}
        Q_1^{G_2}&\mapsto\left(\frac{Q_2^{A_2}}{Q_1^{A_2}}\right)^{1/3}\,,\nonumber\\
        Q_2^{G_2}&\mapsto Q_{1}^{A_2} \,,
    \end{align}
    so that 
    \begin{equation}
        Q_0^{G_2}=\frac{q}{(Q_1^{G_2})^3(Q_2^{G_2})^2} \mapsto \frac{q}{Q_1^{A_2}Q_2^{A_2}}=Q_0^{A_2}\,.
    \end{equation}
    
    We conclude that the $G_2$ Gromov-Witten invariant associated to the curve $a_BC_B+a_0C_0+a_1C_1+a_2C_2$ coincides with the $A_2$ Gromov-Witten invariant associated to the curve $a_BC_B+a_0C_0+(a_2-\frac{a_1}{3})C_1+\frac{a_2}{3} C_2$:
    \begin{equation}
        r^{G_2}_{a_B,a_0,a_1,a_2}=r^{A_2}_{a_B,a_0,a_2-\frac{a_1}{3},\frac{a_1}{3}}\,.
    \end{equation}
    
    \subsubsection{$B_3$ to $G_2$}
    For the specialization $B_3$ to $G_2$, we are in case \ref{how_to_1}. Performing the Weyl transformation
    \begin{align*}
        x_1&\mapsto x_3 \,,\\
        x_2& \mapsto -x_2 \,, \\
        x_3&\mapsto -x_1\,,
    \end{align*}
    we obtain the map
    \begin{align}
        Q_1^{B_3}&\mapsto Q_1^{G_2} \nonumber\\
        Q_2^{B_3}&\mapsto Q_2^{G_2} \nonumber\\
        Q_3^{B_3}&\mapsto Q_1^{G_2}\,,
    \end{align}
    and
    \begin{equation}
        Q_0^{B_3}=\frac{q}{Q_1^{B_3}(Q_2^{B3})^2(Q_3^{B_3})^2}\mapsto\frac{q}{(Q_1^{G_2})^3(Q_2^{G_2})^2}=Q_0^{G_2}\,,
    \end{equation}
    which involves only positive powers of the exponentiated K\"ahler parameters of the $G_2$ geometry. The relation among Gromov-Witten invariants which follows from this map is
    \begin{equation} \label{eq:gwB3G2}
        r^{G_2}_{a_B,a_0,a_1,a_2}=\sum_{a_3}r^{B_3}_{a_B,a_0,a_1-a_3,a_2,a_3} \,.
    \end{equation}
    Table \ref{tab:gwB3G2} in the appendix exemplifies this relation.
    
    \subsubsection{$F_4$ and $B_4$ to $D_4$}
    
    Both specializations $F_4 \rightarrow D_4$ and $B_4 \rightarrow D_4$ fall under case 2. For $D_4$ and $B_4$,
    \begin{equation*}
    m_{\rho}^{D_4}=6x_1+4x_2+2x_3=m_{\rho_L}^{B_4}\,,
    \end{equation*}
    while for $F_4$, the choice for $\iota$ corresponding to \eqref{eq: F4 to D4} yields
    \begin{equation}
        m_{\rho_L}^{F_4}\xmapsto{ \iota}6x_1+4x_2+2x_3\,.
    \end{equation}
    It follows that no additional Weyl transformations are needed. We obtain the maps
    \begin{equation}
        Q_1^{B_4} \mapsto Q_1^{D_4} \,,\,\, Q_2^{B_4} \mapsto Q_2^{D_4} \,,\,\, Q_3^{B_4} \mapsto Q_3^{D_4} \,,\,\, Q_4^{B_4} \mapsto \sqrt{\frac{Q_4^{D_4}}{Q_3^{D_4}}} \,, 
    \end{equation}
    and
    \begin{equation}
        Q_1^{F_4} \mapsto Q_2^{D_4} \,,\,\, Q_2^{F_4} \mapsto Q_1^{D_4} \,,\,\, Q_3^{F_4} \mapsto \sqrt{\frac{Q_3^{D_4}}{Q_1^{D_4}}}  \,,\,\, Q_4^{F_4} \mapsto \sqrt{\frac{Q_4^{D_4}}{Q_3^{D_4}}}  \,, 
    \end{equation}
    such that
    \begin{equation}
    \begin{array}{ccccc}
    \frac{q}{Q_1^{B_4}(Q_2^{B_4})^2(Q_3^{B_4})^2(Q_4^{B_4})^2}&=&\frac{q}{Q_1^{D_4}(Q_2^{D_4})^2Q_3^{D_4}Q_4^{D_4}}&=&\frac{q}{(Q_1^{F_4})^2(Q_2^{F_4})^3(Q_3^{F_4})^4(Q_4^{F_4})^2}\\
    Q_0^{B_4}&=&Q_0^{D_4}&=&Q_0^{F_4}
    \end{array}
    \end{equation}
    
    This implies the following relation between Gromov-Witten invariants:
    \begin{equation}
       r^{B_4}_{a_B,a_0,a_1,a_2,a_3,a_4}=r^{D_4}_{a_B,a_0,a_1,a_2,a_3-\frac{a_4}{2},\frac{a_4}{2}}=r^{F_4}_{a_B,a_0,a_2,a_1+a_3,2a_3,a_4} \,.
        \label{eq:GVB4D4F4}
    \end{equation}
    Table \ref{tab:B4F4gw} in the appendix exemplifies these relations.

    \subsubsection{$A$-series}
    
    For the $A_3\to A_2$ specialization, we find that composing $\iota$ with the Weyl transformation $x_3\leftrightarrow x_4$ yields the correct map between K\"ahler parameters:
    \begin{align}
        Q^{A_3}_1&\mapsto Q^{A_2}_1\,,\nonumber\\
        Q^{A_3}_2&\mapsto \frac{(Q^{A_2}_2)^{1/3}}{(Q^{A_2}_1)^{1/3}}\,,\nonumber\\
        Q^{A_3}_3&\mapsto (Q^{A_2}_1)^{1/3}(Q^{A_2}_2)^{2/3}
        \,,
    \end{align}
    so that,
    \begin{equation}
        Q_0^{A_3}=Q_0^{A_2}\,.
    \end{equation}
    
    This gives the relation among the invariants 
    \begin{equation}
        r^{A_2}_{a_B,a_0,a_1,a_2}=\sum_{a_3}r^{A_3}_{a_B,a_0,a_1+a_2-a_3,3a_2-2a_3,a_3}\,.
    \end{equation}
    
    We conjecture that this generalizes to the general case $A_{n+1} \rightarrow A_n$, yielding the map 
    \begin{align}
        Q_i^{A_n}& \mapsto Q_i^{A_{n-1}}, \quad i=1,\dots,n-2\,, \nonumber\\
        Q_{n-1}^{A_n}& \mapsto \frac{(Q_{n-1}^{A_{n-1}})^{1/n}}{\prod_{i=1}^{n-2}(Q_{i}^{A_{n-1}})^{i/n}}\,,\nonumber\\
        Q_n^{A_n}&\mapsto\prod_{i=1}^{n-1}(Q_{i}^{A_{n-1}})^{i/n}\,,
    \end{align}
    so that 
    \begin{equation}
        Q_0^{A_n}=Q_0^{A_{n-1}}\,.
    \end{equation}
    
    The relation among the invariants is then
    \begin{equation*}
        r^{A_{n-1}}_{a_B,a_0,a_1,\dots,a_{n-1}}=\sum_{a_n}r^{A_{n}}_{a_B,a_0,a_1+a_{n-1}-a_n,\dots,a_i+i(a_{n-1}-a_{n}),\dots,a_{n-2}+(n-2)(a_{n-1}-a_n),na_{n+1}-(n-1)a_n,a_n} \,. 
    \end{equation*}

    \subsubsection{... $\rightarrow D_{n+1} \rightarrow B_n \rightarrow D_n \rightarrow$ ...}
    For the $B_n \rightarrow D_n$ specialization, we are in case \ref{how_to_2}. As $m_{\rho_L}^{D_n}$ and $m_{\rho_L}^{B_n}$ coincide, no extra Weyl transformation is required. The map between K\"ahler parameters is
    \begin{align}
        Q_i^{B_n}&\mapsto Q_i^{D_n}\,, \quad\quad i=1,\dots,n-1\,, \nonumber \\
        Q_n^{B_n}&\mapsto \frac{(Q_n^{D_n})^{1/2}}{(Q_{n-1}^{D_{n}})^{1/2}}\,, \label{eq:gwBnDn}
    \end{align}
    so that 
    \begin{equation}
        Q_0^{B_n}=Q_0^{D_n}\,,
    \end{equation}
    implying the relation 
    \begin{equation}
        r^{B_n}_{a_B,a_0,a_1,\dots,a_n}=r^{D_n}_{a_B,a_0,a_1,\dots,a_{n-2},a_{n-1}-\frac{a_n}{2},\frac{a_n}{2}}
    \end{equation}
    among the Gromov-Witten invariants.
    
    Regarding the $D_n\to B_{n-1}$ Higgsing, the naive transformation reveals this to be a case 1 specialization, with map between K\"ahler parameters
    \begin{align}
        Q_i^{D_n}&\mapsto Q_i^{B_{n-1}}, \quad\quad i=1,\dots,n-1\,, \nonumber\\
        Q_n^{D_n}&\mapsto Q_{n-1}^{B_{n-1}} \,. \label{eq:gwDnBn-1}
    \end{align}
    
    The relation among the Gromov-Witten invariants that follows is
    \begin{equation}
         r^{B_{n-1}}_{a_B,a_0,a_1,\dots,a_{n-1}}=\sum_{a_n}r^{D_n}_{a_B,a_0,a_1,\dots,a_{n-2},a_{n-1}-a_n,a_n}=\sum_{a_n}r^{B_n}_{a_B,a_0,a_1,\dots,a_{n-2},a_{n-1},a_n} \,.
    \end{equation}
    We check this relation for the specialization $B_4 \rightarrow B_3$ in table \ref{tab:B4B3gw}.

	\section{Enhanced symmetries} 
	\label{sec:enhanced}
	The methods developed up to this point in this paper allow us to express the elliptic genus for any of the rank 1 models in terms of Weyl invariant Jacobi forms. However, depending on the matter representations present, some theories exhibit a higher symmetry than that implied merely by the gauge symmetry. Invoking this higher symmetry, the elliptic genus can be expanded in terms of a set of more restrictive Jacobi forms, in some cases drastically reducing the number of expansion coefficients which need to be determined.
	
	Symmetry enhancement due to the absence of certain matter representations is reminiscent of center symmetry in gauge theory: a given Lie algebra $\mg$ is compatible with a maximal center symmetry (the center of the simply connected Lie group among all Lie groups associated to $\mg$), but part or all of this symmetry is ruled out by certain matter representations. We will show that the symmetry enhancement we are seeing can indeed in some cases be interpreted in terms of center symmetry, though we also have examples in which the symmetry is enhanced beyond what full center symmetry would suggest.
	
	What is remarkable is that the symmetry enhancement that we see at the level of the massless matter spectrum, or equivalently, at the level of fibral curves, turns out to persist at the level of the entire elliptic genus. We present abundant evidence for this enhancement at base wrapping degree 1, and further evidence at base wrapping degree 2 in section \ref{sec:HigherBaseDegree}. To solve for the elliptic genus completely at higher base wrapping degree requires an alternative to genus 0 Gromov-Witten invariants as a source of boundary conditions. The elliptic genus for the $A_2$ theory over base $\IF_3$ was computed in \cite{DelZotto:2017mee} up to base degree 3 by invoking exact vanishing conditions on Gopakumar-Vafa invariants. Those results further corroborate the claim that symmetry enhancement persists beyond the massless spectrum. It would be important to establish this fact independently from explicit computations, both by arguing within the framework of gauge theory and, independently, geometrically.
	
	In terms of explicit computation, imposing the additional constraints dictated by enhanced symmetries puts otherwise burdensome models within computational reach. The extent of these simplifications is tabulated in table \ref{tab:reduction}.
	
	The symmetry enhancement in our computations rests on two pillars: the enhancement of the Weyl group, and the enhancement of the shift symmetry. We will discuss these two mechanisms in turn.
	
	\subsection{Enhancement of the shift symmetry by elements of the coweight lattice and center symmetry}
	Let $\omega$ lie in the subspace of $H_2(X,\IZ)$ spanned by the classes of exceptional fibral curves of $X$. The elliptic genus depends on the exponential of parameters
	\be
	(\omega, m) 
	\ee
	whenever the class $\omega$ is represented by a holomorphic curve. If a generating set of the entire weight lattice $\LambdaW$ is represented by such curves, the theory will possess the minimal amount of shift symmetry of $m$ compatible with the gauge symmetry, namely shift symmetry by the dual lattice to $\LambdaW$, the coroot lattice $\LambdaCR$. Conversely, the presence of enhanced gauge symmetry with Lie algebra $\mg$ implies that at least the classes $\alpha$ corresponding to roots of $\mg$ (in particular the simple roots $\alpha_i$) are represented by holomorphic curves. If only these classes are represented, the shift symmetry of $m$ will be enhanced from $\LambdaCR$ to the dual of the root lattice $\LambdaR$, the coweight lattice $\LambdaCW$ of $\mg$. The cases in between are when only some weights in $\LambdaW/\LambdaR$ are represented by holomorphic curves. Let $\lambda$ represent such a class. The shift symmetry of $m$ compatible with the presence of $\lambda$ will then be given by elements of the sublattice of $\LambdaCW$ for which
	\be
	\Lambda_{\mathrm{shift}}(\lambda) = \{ \omega^\vee \in \LambdaCW \,|\, (\lambda, \omega^\vee) \in \IZ \} \,. 
	\ee
	Now let us connect this discussion to center symmetry. Recall that the center of a Lie group equals the intersection of all possible choices of maximal tori, hence lies in the image of any choice $\mathfrak{h}$ of Cartan subalgebra of the Lie algebra under the exponential map. Recall also that for given Lie algebra $\mg$, the exponential map depends on the particular Lie group $G$ associated to $\mg$. For any $G$,
	\be
	Z = \exp(\LambdaCW) \,, \quad \LambdaCR \subset \ker(\exp) \,,
	\ee
    the identification $\mathfrak{h} \cong \LambdaCR \otimes \IC$ being understood. The kernel of the exponential map however depends on the choice of $G$. For $G$ simply connected, $\ker(\exp) = \LambdaCR$, hence
    \be
    Z \cong \LambdaCW/ \LambdaCR \,.
    \ee
	The question of the amount of center symmetry preserved by a representation $\boldsymbol{\lambda}$ of the group (we are using bold faced symbols to distinguish between representations of the group and of the algebra) hence amounts to deciding when an element of the center is represented trivially. The following lemma answers this question:
	\be
	\boldsymbol{\lambda}(\exp(\omega^\vee)) = 1 \quad \Leftrightarrow \quad (\lambda,\omega^\vee) \in \IZ \,.
	\ee
	We conclude that the lattice of shift symmetries of the elliptic genus due to the absence of certain matter representations is equal to $\exp^{-1}(Z)$, with the $Z$ the largest center compatible with the matter representations present.

	\subsection{Enhancement of the Weyl group by Dynkin diagram symmetries}
    At the level of the gauge theory, the Weyl symmetry acting on the Coulomb moduli of the theory is a remnant of the full gauge symmetry not fixed by the choice of which maximal torus the gauge symmetry is broken to. In contrast, symmetries of the Dynkin diagram of the gauge algebra, if respected by the matter representations present in the theory, give rise to an additional automorphism of the theory.

	\subsection{Further symmetry enhancements}
	We also encounter examples where the shift symmetry is enhanced even beyond the maximal amount of center symmetry compatible with the Lie algebra $\mg$, i.e. beyond $\LambdaCW(\mg)$, and the Weyl symmetry beyond the extension by the Dynkin diagram symmetry. All examples of such symmetry enhancements that we encounter arise when the specialization maps exhibited in the previous section are invertible (e.g. in the case $G_2 \rightarrow A_2$). In these cases, the theory with larger gauge group can "reverse inherit" the enhanced symmetry of the Higgsed theory. In such cases, the enhancement can be explained intrinsically (i.e. without reference to the Higgsed theory) by noting that certain weights associated to the matter representations present in the theory coincide with roots of $\mg$, and their contribution to the elliptic genus cancels. The cancellation is best described geometrically \cite{Kashani-Poor:2019jyo}: a root contributes the value -2 to the Gromov-Witten invariant of the associated curve class, and each half-hypermultiplet associated to a weight contributes +1. It is not at all evident that the enhanced symmetries of the lattice obtained due to this cancellation should be inherited by the full elliptic genus. In all of the examples we study, this however turns out to be the case. It would be interesting to study whether this symmetry enhancement has repercussions in the gauge theory beyond its effect on the elliptic genus.

	\subsection{Examples}

	\subsubsection{$A_2$ and $G_2$ over $\mathbb F_3$}
	The model with $A_2$ gauge symmetry over $\mathbb F_3$ has no matter. Hence, the shift symmetry is enhanced from $\LambdaCR(\mathfrak{a}_2)$ to $\LambdaCW(\mathfrak{a}_2)$, and the Weyl group symmetry is enhanced by the Dynkin diagram symmetry $\IZ_2$, yielding the Weyl group of $G_2$, see \eqref{eq:WeylG2}. Our task is thus to construct Jacobi forms invariant under the action of $W(G_2)$ and shifts by $\LambdaCW(\mathfrak{a}_2)$.
	
    Conveniently, the $A_2$ coweight lattice is isomorphic to the coroot lattice of $G_2$, upon multiplying the inner product of the former by a factor of 3.\footnote{We thank Haowu Wang for pointing this out to us, as well as further very useful comments both regarding the $A_2$ theory over base $\IF_3$ and the $D_4$ theory over base $\IF_4$ discussed further below.} With regard to the realizations of these two lattices in the conventions of Bourbaki \cite{Bourbaki}, an explicit isomorphism is given by 

    \begin{align*}
    \psi:&\left\{
    \begin{array}{rl}
         \frac{1}{3} (2 e_1^{A_2}-e_2^{A_2}-e_3^{A_2})&\mapsto e_1^{G_2}-e_2^{G_2} \\
         \frac{1}{3} (- e_1^{A_2}+2e_2^{A_2}-e_3^{A_2})&\mapsto e_2^{G_2}-e_3^{G_2}
    \end{array}
        \right.\\
         &\qquad \qquad \quad \quad  \big\Updownarrow \nonumber\\
    \psi:    x_1e_1^{A_2}+x_2e_2^{A_2}&+x_3e_3^{A_2} \mapsto (x_2-x_3)e_1^{G_2}+(x_1-x_2)e_2^{G_2}+(x_3-x_1)e_3^{G_2} \,.
    \end{align*}
	Note that this isomorphism pulls back the metric on $\LambdaCR(\mathfrak{g}_2)$ to 3 times the metric on $\LambdaCW(\mathfrak{a}_2)$, as announced: 
	\begin{equation*}
	    \psi^*\left(ds^2_{G_2}=\sum_i (dx_i)^2\right)=3\sum_i (dx_i)^2 \,.
	\end{equation*}
	The map $\psi$ thus induces via composition a map from $J(G_2)$ (whose elements for the purposes of the composition are considered as functions on $\LambdaCR(\mathfrak{g}_2)$) to $J(A_2)$ which triples the index. The image of this map is a subring of $J(A_2)$ generated by
    \begin{align} \label{eq:subring_of_W(G_2)}
    \phi_{0,3}^{A_2}&=\phi^{G_2}_{0,1}\circ \psi \,, \nn\\
    \phi_{-2,3}^{A_2}&=\phi^{G_2}_{-2,1}\circ \psi \,, \nn \\
    \phi_{-6,6}^{A_2}&=\phi^{G_2}_{-6,2}\circ \psi \,.
    \end{align}
    By the underlying assumption of this section that symmetries of the massless spectrum are inherited by the elliptic genus, the numerator $\cN$ in the ansatz \eqref{eq:ZkAnsatz} of the elliptic genus of the massless $A_2$ theory should be an element of this subring. The subring of Dynkin diagram and $\LambdaCW(\mathfrak{a}_2)$ shift symmetric forms was already identified in \cite{DelZotto:2017mee} as the appropriate ring in which $\cN$ should lie, and it was conjectured that this ring should be generated by the set of generators \eqref{eq:subring_of_W(G_2)}. We have now integrated the former observation into the larger context of enhanced symmetries of elliptic genera of rank 1 models, and provided a proof of the latter statement.

    In \cite{DelZotto:2017mee}, it was furthermore noticed that the function $\psi$ has a very simple expansion in terms of the coefficients $m_i=(m,\alpha_i)$, $m=m_1\omega_1^\vee+m_2\omega_2^\vee$:
    \begin{align*}
    m=&m_1\omega_1^\vee+m_2\omega_2^\vee=\frac{2m_1+m_2}{3}e_1^{A_2}+\frac{-m_1+m_2}{3}e_2^{A_2}+\frac{-m_1-2m_2}{3}e_2^{A_2}\\&\xmapsto{\psi} m_1 e_1^{G_2}+m_2 e_2^{G_2}+(-m_1-m_2)e_3^{G_2} \,.
    \end{align*}
	
	Now let us climb up one node in the Higgsing tree, and consider the theory with gauge group $G_2$ arising over the base $\IF_3$. As we discussed in detail in section \ref{sec:moving}, the elliptic genus specializes when moving from the branches towards the roots of the Higgsing tree. The specialization from a $G_2$ node to an $A_2$ node is particularly simple, see figure \ref{fig:jacobi_G2_to_A2}; in particular, it is invertible: the numerator $\cN$ of the elliptic genus of an $A_2$ theory obtained via Higgsing from a $G_2$ theory necessarily involves only even powers of the generator $\phi_{-3,1}^{A_2}$. Replacing the square of this generator by $\phi_{-6,2}^{G_2}$ and retaining the two other generators which coincide between the two gauge groups yields the numerator of the $G_2$ theory. This however poses a conundrum: the symmetry enhancement of the matterless $A_2$ theory just described must be reverse inherited by the $G_2$ theory with matter in the $\mathbf{7}$ representation. How does the enhancement of the shift symmetry to $\LambdaCW(\mathfrak{a_2})$ arise in this theory? The answer to this puzzle was already explained in general terms above: the weights of the $\mathbf{7}$ representation coincide with the short roots of $G_2$. The curves giving rise to the 2 half hypermultiplets in the $\mathbf{7}$ hence contribute to the same Gromov-Witten  invariants as those giving rise to the gauge fields associated to these roots, and in fact cancel the latter contribution. The remaining contributions are those due to the long roots. The lattice spanned by these is precisely the $A_2$ root lattice, with dual lattice $\LambdaCW(\mathfrak{a_2})$.
	
	Note that as the two fundamental representations of $A_2$ are complex conjugates, the spectrum of any $A_2$ gauge theory with $8$ supercharges will always be Dynkin diagram symmetric, hence cannot present an obstruction to the theory descending via Higgsing from a $G_2$ theory.

	\subsubsection{The $B_4$ and $F_4$ to $D_4$ branch}\label{ss:B_4_and_F_4_to_D_4_branch}
	
	All rank 1 Higgsing trees branch above a $D_4$ node, to a branch with adjacent node $B_4$, and another with adjacent node $F_4$. The matter content of the $D_4$ theory over the base $\IF_k$ is $ V^{\oplus (k-4)} \oplus S_+^{\oplus (k-4)} \oplus S_-^{\oplus (k-4)}$. In particular, it is invariant under the Dynkin diagram symmetry of $D_4$, consistent with the fact that the theory is obtained upon Higgsing an $F_4$ theory, given that the Weyl group of $D_4$ enhanced by the Dynkin diagram symmetry yields the Weyl group of $F_4$.
	
	By the hypothesis underlying this section, the numerator $\cN$ of the elliptic genus of the $D_4$ theory should therefore lie in the subring of $J(D_4)$ whose elements exhibit Dynkin diagram symmetry. As the root lattices of $D_4$ and $F_4$ are isomorphic, this subring is isomorphic to $J(F_4)$, as explained further in appendix \ref{sec:appF4}, hence spanned by $\{\phi_{-n,k}^{F_4} \circ i^{-1} \}$, the generators of $J(F_4)$ composed with the isomorphism $i^{-1} : \LambdaR(D_4) \rightarrow \LambdaR(F_4)$.
	
	The specialization of the elliptic genus for $B_n$ theories to that of $D_n$ theories is equally simple as that from $G_2$ to $A_2$ theories: all but one generator of the two rings $J(C_n)$ and $J(D_n)$ coincide.\footnote{Recall that $J(C_n)$ is the appropriate ring to expand the elliptic genus of $B_n$ theories in.} The remaining generator of the former, $\phi_{-2n,2}^{C_n}$, is the square of that of the latter, $\omega_{-n,1}^{D_n}$. This implies in particular that the expansion of the  numerator $\cN$ of the elliptic genus of the $D_4$ theory in $J(D_4)$ generators should exhibit only even powers of $\omega_{-4,1}^{D_4}$. However, as only even powers of $\omega_{-4,1}^{D_4}$ occur in the subring of $J(D_4)$ compatible with the $F_4$ Higgsing, this provides no further restriction on $\cN$.
	
	Inverting the specialization map from $B_4$ to $D_4$, we can conclude that $B_4$ must exhibit the same symmetry enhancement as the $D_4$ theory, i.e. from $W(B_4)$ to $W(F_4)$. To explain this enhancement intrinsically (i.e. without reference to the $D_4$ theory), we can invoke a more intricate realization of the mechanism at play in the transition from $G_2$ to $A_2$: the $B_4$ theories have matter content $V^{\oplus (4-k)+1}\oplus S^{\oplus (4-k)}$ over $\IF_k$. Two half hypermultiplets in the vector representation cancel the contribution of the short roots of $B_4$ to the Gromov-Witten invariants of the theory. The long roots of $B_4$ coincide with the roots of $D_4$, hence are invariant under $W(F_4)$. Furthermore, having cancelled one vector representation against the short roots,
	the remaining matter content, $(V \oplus S)^{\oplus (4-k)}$, is also invariant under $W(F_4)$, thus explaining the occurrence of this symmetry for these $B_4$ theories.

    \subsubsection{$D_4$, $B_4$ and $F_4$ over $\mathbb{F}_4$} \label{ss:enhanced symmetry D4 B4 F4}
    The symmetry enhancement of the $D_4$ theory over $\IF_4$ was already noted in \cite{DelZotto:2017mee}. The absence of matter means, once again, that the numerator $\cN$ of the elliptic genus is invariant under translations by the coweight lattice $\LambdaCW(D_4)$, and that the expansion is Dynkin-diagram symmetric. 

    The $D_4$ coweight lattice is isomorphic to the coroot lattice of $F_4$ upon multiplying the inner product of the former by a factor of 2. The $D_4$ Weyl group is enhanced by the Dynkin diagram symmetry of $D_4$ to the $F_4$ Weyl group. We conclude that the ring $J(F_4)$ can be mapped to the subring of $J(D_4)$ with precisely the enhancement of shift and Weyl symmetry that we require. Note that unlike the situation in section \ref{ss:B_4_and_F_4_to_D_4_branch}, we do not need to invoke the map $\iota^*$, as the two lattices $\LambdaCW(D_4)$ and $\LambdaCR(F_4)$ coincide as embedded in the Euclidean space $\IR^4$. In particular, this implies that due to the factor of 2 relating the inner products, elements of $J(F_4)$ interpreted as elements of $J(D_4)$ with enhanced shift symmetry have twice the index. Specifically, the subring of $J(D_4)$ in question is spanned by the generators
    \begin{align}
    \phi_{0,2}^{D_4}&=\phi_{0,1}^{F_4} \,, \label{eq:J(D_4)_with_enhanced_symmetry_generators}\\
    \phi_{-2,2}^{D_4}&=\phi_{-2,1}^{F_4}\,,\nn\\
    \phi_{-6,4}^{D_4}&=\phi_{-6,2}^{F_4}\,,\nn\\
    \phi_{-8,4}^{D_4}&=\phi_{-8,2}^{F_4}\,,\nn\\
    \phi_{-12,6}^{D_4}&=\phi_{-12,3}^{F_4}\,.\nn
    \end{align}

    We can move up the Higgsing tree to the $F_4$ node as in the previous subsection via the map $(\iota^*)^{-1}$. As only the generators \eqref{eq:J(D_4)_with_enhanced_symmetry_generators} arise, the $F_4$ theory will inverse inherit the enhanced symmetry of the $D_4$ theory. In particular, the numerator of the corresponding elliptic genus will permit an expansion in terms of the generators
    \begin{align*}
    \phi_{0,2}^{F_4}&= (\iota^*)^{-1}\phi_{0,2}^{D_4}   \,,\\
    \phi_{-2,2}^{F_4}&= (\iota^*)^{-1}\phi_{-2,2}^{D_4} \,,\\
    \phi_{-6,4}^{F_4}&= (\iota^*)^{-1}\phi_{-6,4}^{D_4} \,,\\
    \phi_{-8,4}^{F_4}&= (\iota^*)^{-1}\phi_{-8,4}^{D_4}\,,\\
    \phi_{-12,6}^{F_4}&=(\iota^*)^{-1}\phi_{-12,6}^{D_4} \,.
    \end{align*}
    To argue for this enhancement intrinsically, note that the contribution of the 2 half hypermultiplets in the $\mathbf{26}$ representation of $F_4$ cancel the contribution of the gauge fields associated to the short roots of $F_4$ to the Gromov-Witten invariants. Hence, following the logic of this section, we expect the numerator $\cN$ of the elliptic genus of this $F_4$ theory to have an expansion in the subring of $J(F_4)$ whose elements have a shift symmetry under the dual lattice to the sublattice of $\LambdaR(F_4)$ spanned by the long roots. This dual lattice is isomorphic to $\LambdaCW(D_4)$, as we can verify explicitly by considering the image of the sublattice of long roots of $F_4$ under the map $i$. Recall that we are identifying the embedding of the lattice $\LambdaR(F_4)$ in $\IZ^4$ as given in table \ref{tab:rootSystems2} (which reflects  the conventions of Bourbaki \cite{Bourbaki}) with the isomorphic lattice $\LambdaCR(F_4)$. We can obtain the dual lattice $\LambdaR(F_4)$ via the map $\alpha=\frac{2}{(\alpha^\vee,\alpha^\vee)}\alpha^\vee$, which in particular maps short coroots to long roots.
    The image of the long roots under the map $i$ is thus equal to the image of the short coroots, which we readily identify with $\LambdaR(D_4) = \LambdaCR(D_4)$. The dual lattice is hence $\LambdaCW(D_4)$, as we wished to argue.

    Choosing the other branch of the Higgsing tree, the next node up from the minimal $D_4$ theory is a theory with $B_4$ gauge symmetry and 2 half hypermultiplets in the vector representation. In addition to the symmetry enhancement discussed in section \ref{ss:B_4_and_F_4_to_D_4_branch}, this theory must also reverse inherit the enhanced shift symmetry under $\LambdaCW(D_4)$ which is particular to the massless $D_4$ theory. This enhancement follows the general pattern exhibited in this section: the contribution of the two half hypermultiplets to the Gromov-Witten invariants of the geometry cancels that of the gauge fields associated to the short roots of $B_4$. The remaining lattice of long roots of $B_4$ is the $D_4$ root lattice, with dual the $D_4$ co-weight lattice.

    \subsubsection{$D_n$ and $B_n$ over $\IF_4$}
    As already pointed out in \cite{Morrison:2020ool}, the only rank 1 models with matter present which does not break all of the center symmetry are the $D_n$ and $B_n$ models over $\IF_4$. These exhibit matter in the vector representation.
    
    We first consider the theories with $D_n$ gauge symmetry. The $\IZ_2$ Dynkin diagram symmetry (for $n>2$) exchanges the two spinor representations. As the only matter present in the $D_n$ series over $\IF_4$ is in the vector representation, this is a symmetry of these theories, enhancing the Weyl group symmetry to that of $B_n$:
    \be
    W(D_n) \ltimes \text{DynkinSym}(D_n) = W(B_n) \,.
    \ee
	Furthermore, the shift symmetry consists of all element of the coweight lattice of $D_n$ which have integral pairing with the fundamental weight $\omega_1$ associated to the vector representation: in our conventions, this yields the lattice $\IZ^n$, which is the root lattice of $B_n$. The subring of $J(D_n)$ which exhibits the enhanced symmetry of the $D_n$ theory over $\IF_4$, is thus the ring $J(B_n)$. Note however that the inner product on the enhancement of the lattice $\LambdaCR(D_4)$ is the conventionally normalized Euclidean inner product $dx^2$. With regard to this norm, the indices of the generators of $J(B_n)$ (defined with regard to the inner product $2dx^2$) are doubled.
	
	Following the general strategy of this section, we expect the numerator of the elliptic genus for the $D_n$ theory over $\IF_4$ to lie in this subring. 
	
	We have computed the elliptic genus explicitly at base wrapping 1 for the model $D_5$. The ansatz in terms of the ring $J(B_5)$ with indices doubled indeed allows us to match all Gromov-Witten invariants as obtained via mirror symmetry. 
	
	For theories with $B_n$ gauge symmetry, the shift symmetry is enhanced from $\LambdaCR(B_n)$ (identified with $\LambdaR(C_n)$, as explained above) to the sublattice of $\LambdaCW(B_n)$ whose elements have integer pairing with $\omega_1$, the fundamental weight associated to the vector representation of $B_n$. As $\omega_1$ is also a root of $B_n$, this sublattice is indeed all of $\LambdaCW(B_n)$, i.e. $\IZ^n$ with inner product $dx^2$. We are hence seeking a subring of $J(C_n)$ with this shift symmetry. The ring $J(B_n)$ provides this subring, as $W(B_n) = W(C_n)$, and $\LambdaR(B_n) = \IZ^n$. The indices of the canonical generators are defined however with regard to the inner product $2dx^2$. As a subring of $J(C_n)$, they need to be doubled.

	In table \ref{tab:reduction} we summarize the reduction in the number of coefficients. 
	
	\begin{table}[ht]
    \centering
    \begin{tabular}{c|c|c|c}
         Base &Gauge algebra & Naive number of coefficients & Improved number of coefficients  \\\hline\hline
         
         \multirow{ 3}{*}{$\mathbb{F}_1$}
& $D_4$ & 902 &  \\
& $B_4$ & 495 & 197 \\
& $F_4$ & 197 &  \\\hline
         \multirow{ 4}{*}{$\mathbb{F}_2$} & $A_2/G_2$ & 4 & 1 \\\cline{2-4}
& $D_4$ & 295 &  \\
& $B_4$ & 161 & 64 \\
& $F_4$ & 64 &  \\\hline
        
         \multirow{ 3}{*}{$\mathbb{F}_3$} 
& $D_4$ & 310 &  \\
& $B_4$ & 171 & 69  \\
& $F_4$ & 69 &  \\\hline

 \multirow{ 6}{*}{$\mathbb{F}_4$} & $D_4$ & 287 &  \\
& $F_4$ & 69 & 4 \\
& $B_4$ & 163 &  \\\cline{2-4}
& $D_5/B_5$ & 1088 & 280 \\\cline{2-4}
& $B_6$ & 7086 & \multirow{ 2}{*}{1950}\\
& $D_6$ & 13248 &\\
    \end{tabular}
    \caption{By looking carefully at the symmetries of the low energy theory and assuming they hold for the full theory, we could reduce the number of coefficients for some models. The naive number is the number of coefficients one would have to fix using an arbitrary ansatz in $J(\tilde \mg)$ and the improved number of coefficients is the number of coefficients one would have to fix if one refines the ansatz to the subring of $J(\tilde \mg)$ invariant under the previously mentioned symmetries.}
    \label{tab:reduction}
\end{table}

	\subsection{A result at base wrapping 2}
	\label{sec:HigherBaseDegree}
	
	Several new features arise when we consider $Z_k$ at base wrapping $k>1$. Most importantly, the index of the numerator in the topological string coupling \eqref{eq:topStringIndex} is no longer 0, so the numerator depends on $g_{top}$. This dependence is through the Jacobi forms $\phi_{0,1}$ and $\phi_{-2,1}$. Expanding these in $x=\left(2\sin\left(\frac{g_{top}}{2}\right)\right)^2\sim \gs^2$ yields $\phi_{0,1}(g_{top})=2+o(x)$ and $\phi_{-2,0}=x+o(x^2)$.
	
	The universal part of the denominator \eqref{eq:denom_uni} at base degree $k$ scales as $x^k$. Therefore, a term of order $x^m$ in the numerator will contribute only to the Gopakumar-Vafa invariants of curves of genus $g_{m}$ or higher, with 
	\begin{equation*}
	    g_{m}-1=m-k.
	\end{equation*}
	Consequently, expanding the numerator as
	\begin{equation}
	    \cN_{i_{top},i_G,w}=\sum_{m=1}^{i_{top}}\cN^{(m)}_{i_G,w+2m}\phi_{0,1}(g_{top})^{i_{top}-m}\phi_{-2,1}(g_{top})^m,
	\end{equation}
	where $\cN^{(m)}$ is an element of $J(\mg)$ and we have indicated weights and indices by subscripts, the $\cN^{(m)}$ will contribute at genus $g_m$ and higher. For $m<k-1$, $\cN^{(m)}$ is completely fixed by requiring a cancellation of the contribution of order $x^{1+m-k}$ in the free energy $F=\sum_g F_g \gs^{2g-2}=\log Z_{top}$. $\cN^{(k-1)}$ can be fixed by imposing genus 0 Gromov-Witten invariants.
	
	As the dimension of the space of Jacobi forms of which $\cN$ is an element increases rapidly with $k$, results beyond $k>1$ are computationally expensive. We will here only discuss one example, the $F_4$ gauge theory over the base $\mathbb F_4$. For this model, invoking the enhanced shift symmetry discussed in subsection \ref{ss:enhanced symmetry D4 B4 F4} allows us to compute $\cN^{(1)}$ at $k=2$.
	
	From the expansions
	\begin{align*}
	   F&=\sum_g \gs^{2g-2} F_g=\log Z_{top}= \log Z_0 +Q_b \hat Z_1+Q_b^2\left(\hat Z_2-\frac{1}{2}\hat Z_1^2\right)+ \dots,\\
	    F_0&=F_0^{(0)}+F_0^{(1)}Q_b+F_0^{(2)}Q_b^2 + o(Q_b^3),
	\end{align*}
	where we have denoted $\hat Z_k=c_k(\mathbf{Q})Z_k$, we conclude that 
	\begin{equation}
	   \frac{F_0^{(2)}}{g_s^2}=\frac{F_0^{(2)}}{x}+o(x^0)=\hat Z_2-\frac{1}{2}\hat Z_1^2\,.
	   \label{eq:basedeg2}
	\end{equation}
	The universal contribution of the denominator for $Z_1$, $Z_2$ scale as $x$, $x^2$ respectively. The corresponding contributions must vanish in the linear combination appearing on the RHS of equation \eqref{eq:basedeg2}. This fixes $\cN^{(0)}$. 
	
	To fix $\cN^{(1)}$, we need to impose Gromov-Witten invariants obtained by mirror symmetry. The naive ansatz requires fixing 13189 coefficients. As explained in subsection \ref{ss:specDen}, we can impose the $D_4$ denominator in the ansatz \eqref{eq:ZkAnsatz}, i.e. we expect only divergences corresponding to the long roots of $F_4$. This reduces the number of coefficients to be fixed to 8620. Imposing the enhanced shift symmetry of this theory finally reduces the number of coefficients to be fixed to determine $\cN^{(1)}$ to 21. 
	
	We find that the ansatz with 21 coefficients is sufficient to match a large number of genus Gromov-Witten invariants at base degree $k=1$. We consider this strong evidence that our considerations in this and the previous section are also valid at higher base wrapping.

    \section*{Acknowledgements}
    We are grateful to Albrecht Klemm, Kimyeong Lee, Guglielmo Lockhart, Kaiwen Sun, Haowu Wang, Xin Wang, Timo Weigand for helpful discussions. ZD is supported by KIAS Individual Grant PG076901.

	\appendix
	
	\section{The numerator of the $F_4$, $B_4$, and $D_4$ theory over $\IF_4$}
	\label{app:explicit}
    To give a flavor for the form of our results, we give one explicit example in this appendix. Further explicit results are available upon request.

    The numerator $\cN_1$ for the $B_4$ and $D_4$ theories over the base $\IF_4$ is given by
    \begin{equation*}
        \cN=\frac{1}{4} \phi _{-12,3}^{F_4} \phi _{-2,1}^{F_4}-\frac{1}{2} \phi _{-8,2}^{F_4} \phi _{-6,2}^{F_4} \,.
    \end{equation*}
    The numerator for the $F_4$ theory can be obtained from this result by application of the map $(\iota^*)^{-1}$.
    
    To exemplify the power of imposing enhanced symmetries, we also give the result in term of standard $D_4$ forms:
    
    \hspace{1cm}
    
    \resizebox{\linewidth}{!}{\setstretch{1.5}
  $
   \begin{array}{l}
 \frac{E_4 \phi_{-6,2}^{D_4} \phi_{-4,1}^{D_4} \phi_{0,1}^{D_4} \left(\phi_{-2,1}^{D_4}\right){}^4}{1492992}-\frac{E_6 \phi_{-6,2}^{D_4} \phi_{-4,1}^{D_4} \left(\phi_{-2,1}^{D_4}\right){}^5}{6718464}-\frac{\phi_{-6,2}^{D_4} \phi_{-4,1}^{D_4} \left(\phi_{0,1}^{D_4}\right){}^3 \left(\phi_{-2,1}^{D_4}\right){}^2}{497664}+\frac{\phi_{-6,2}^{D_4} \left(\phi_{-4,1}^{D_4}\right){}^2 \left(\phi_{0,1}^{D_4}\right){}^4}{82944} \\
 -\frac{5 E_4^2 \phi_{-6,2}^{D_4} \left(\phi_{-4,1}^{D_4}\right){}^2 \left(\phi_{-2,1}^{D_4}\right){}^4}{8957952}-\frac{E_4 \phi_{-6,2}^{D_4} \left(\phi_{-4,1}^{D_4}\right){}^2 \left(\phi_{0,1}^{D_4}\right){}^2 \left(\phi_{-2,1}^{D_4}\right){}^2}{995328}-\frac{7 E_4 \phi_{-6,2}^{D_4} \left(\phi_{-4,1}^{D_4}\right){}^3 \left(\phi_{0,1}^{D_4}\right){}^3}{248832}+\frac{7 E_6 \phi_{-6,2}^{D_4} \left(\phi_{-4,1}^{D_4}\right){}^2 \phi_{0,1}^{D_4} \left(\phi_{-2,1}^{D_4}\right){}^3}{4478976} \\+
 \frac{23 E_4^2 \phi_{-6,2}^{D_4} \left(\phi_{0,1}^{D_4}\right){}^2 \left(\phi_{-4,1}^{D_4}\right){}^4}{995328}+\frac{E_4^2 \phi_{-6,2}^{D_4} \left(\phi_{-2,1}^{D_4}\right){}^2 \phi_{0,1}^{D_4} \left(\phi_{-4,1}^{D_4}\right){}^3}{186624}-\frac{E_4 E_6 \phi_{-6,2}^{D_4} \left(\phi_{-2,1}^{D_4}\right){}^3 \left(\phi_{-4,1}^{D_4}\right){}^3}{839808}-\frac{E_6 \phi_{-6,2}^{D_4} \phi_{-2,1}^{D_4} \left(\phi_{0,1}^{D_4}\right){}^2 \left(\phi_{-4,1}^{D_4}\right){}^3}{248832} \\
 -\frac{35 E_4^3 \phi_{-6,2}^{D_4} \phi_{0,1}^{D_4} \left(\phi_{-4,1}^{D_4}\right){}^5}{4478976}-\frac{35 E_4^3 \phi_{-6,2}^{D_4} \left(\phi_{-2,1}^{D_4}\right){}^2 \left(\phi_{-4,1}^{D_4}\right){}^4}{13436928}+\frac{E_4 E_6 \phi_{-6,2}^{D_4} \phi_{-2,1}^{D_4} \phi_{0,1}^{D_4} \left(\phi_{-4,1}^{D_4}\right){}^4}{165888}-\frac{E_6^2 \phi_{-6,2}^{D_4} \left(\phi_{-2,1}^{D_4}\right){}^2 \left(\phi_{-4,1}^{D_4}\right){}^4}{26873856} \\
 -\frac{5 E_4^2 E_6 \phi_{-6,2}^{D_4} \phi_{-2,1}^{D_4} \left(\phi_{-4,1}^{D_4}\right){}^5}{2239488}+\frac{25 E_4^4 \phi_{-6,2}^{D_4} \left(\phi_{-4,1}^{D_4}\right){}^6}{26873856}-\frac{5 E_4 E_6^2 \phi_{-6,2}^{D_4} \left(\phi_{-4,1}^{D_4}\right){}^6}{26873856}+\frac{E_6^2 \phi_{-6,2}^{D_4} \phi_{0,1}^{D_4} \left(\phi_{-4,1}^{D_4}\right){}^5}{4478976} \\
 -\frac{E_4 \left(\phi_{-6,2}^{D_4}\right){}^2 \phi_{0,1}^{D_4} \left(\phi_{-2,1}^{D_4}\right){}^3}{165888}+\frac{E_6 \left(\phi_{-6,2}^{D_4}\right){}^2 \left(\phi_{-2,1}^{D_4}\right){}^4}{746496}-\frac{E_6 \left(\phi_{-6,2}^{D_4}\right){}^2 \phi_{-4,1}^{D_4} \phi_{0,1}^{D_4} \left(\phi_{-2,1}^{D_4}\right){}^2}{41472}+\frac{\left(\phi_{-6,2}^{D_4}\right){}^2 \left(\phi_{0,1}^{D_4}\right){}^3 \phi_{-2,1}^{D_4}}{55296} \\+
 \frac{E_4^2 \left(\phi_{-6,2}^{D_4}\right){}^2 \phi_{-4,1}^{D_4} \left(\phi_{-2,1}^{D_4}\right){}^3}{124416}-\frac{5 E_4^2 \left(\phi_{-6,2}^{D_4}\right){}^2 \left(\phi_{-4,1}^{D_4}\right){}^2 \phi_{0,1}^{D_4} \phi_{-2,1}^{D_4}}{82944}+\frac{11 E_4 E_6 \left(\phi_{-6,2}^{D_4}\right){}^2 \left(\phi_{-4,1}^{D_4}\right){}^2 \left(\phi_{-2,1}^{D_4}\right){}^2}{497664}+\frac{E_6 \left(\phi_{-6,2}^{D_4}\right){}^2 \left(\phi_{-4,1}^{D_4}\right){}^2 \left(\phi_{0,1}^{D_4}\right){}^2}{18432} \\+
 \frac{E_4^2 E_6 \left(\phi_{-6,2}^{D_4}\right){}^2 \left(\phi_{-4,1}^{D_4}\right){}^4}{41472}+\frac{7 E_4^3 \left(\phi_{-6,2}^{D_4}\right){}^2 \phi_{-2,1}^{D_4} \left(\phi_{-4,1}^{D_4}\right){}^3}{186624}-\frac{E_4 E_6 \left(\phi_{-6,2}^{D_4}\right){}^2 \phi_{0,1}^{D_4} \left(\phi_{-4,1}^{D_4}\right){}^3}{13824}-\frac{E_6^2 \left(\phi_{-6,2}^{D_4}\right){}^2 \phi_{-2,1}^{D_4} \left(\phi_{-4,1}^{D_4}\right){}^3}{373248} \\
 -\frac{E_4^2 \left(\phi_{-2,1}^{D_4}\right){}^2 \left(\phi_{-6,2}^{D_4}\right){}^3}{41472}+\frac{E_4^2 \phi_{-4,1}^{D_4} \phi_{0,1}^{D_4} \left(\phi_{-6,2}^{D_4}\right){}^3}{3456}-\frac{E_4 \left(\phi_{0,1}^{D_4}\right){}^2 \left(\phi_{-6,2}^{D_4}\right){}^3}{4608}+\frac{E_6 \phi_{-2,1}^{D_4} \phi_{0,1}^{D_4} \left(\phi_{-6,2}^{D_4}\right){}^3}{6912} \\
 -\frac{\left(\phi_{-2,1}^{D_4}\right){}^3 \left(\phi_{0,1}^{D_4}\right){}^3 \left(\omega_{-4,1}^{D_4}\right){}^2}{5971968}-\frac{E_4^3 \left(\phi_{-4,1}^{D_4}\right){}^2 \left(\phi_{-6,2}^{D_4}\right){}^3}{6912}-\frac{E_4 E_6 \phi_{-4,1}^{D_4} \phi_{-2,1}^{D_4} \left(\phi_{-6,2}^{D_4}\right){}^3}{10368}+\frac{E_6^2 \left(\phi_{-4,1}^{D_4}\right){}^2 \left(\phi_{-6,2}^{D_4}\right){}^3}{20736} \\+
 \frac{E_4 \phi_{0,1}^{D_4} \left(\phi_{-2,1}^{D_4}\right){}^5 \left(\omega_{-4,1}^{D_4}\right){}^2}{17915904}+\frac{E_4 \phi_{-4,1}^{D_4} \left(\phi_{0,1}^{D_4}\right){}^2 \left(\phi_{-2,1}^{D_4}\right){}^3 \left(\omega_{-4,1}^{D_4}\right){}^2}{373248}-\frac{E_6 \left(\phi_{-2,1}^{D_4}\right){}^6 \left(\omega_{-4,1}^{D_4}\right){}^2}{80621568}-\frac{\phi_{-4,1}^{D_4} \left(\phi_{0,1}^{D_4}\right){}^4 \phi_{-2,1}^{D_4} \left(\omega_{-4,1}^{D_4}\right){}^2}{124416} \\
 -\frac{E_4^2 \phi_{-4,1}^{D_4} \left(\phi_{-2,1}^{D_4}\right){}^5 \left(\omega_{-4,1}^{D_4}\right){}^2}{13436928}+\frac{37 E_4 \left(\phi_{-4,1}^{D_4}\right){}^2 \left(\phi_{0,1}^{D_4}\right){}^3 \phi_{-2,1}^{D_4} \left(\omega_{-4,1}^{D_4}\right){}^2}{1492992}-\frac{5 E_6 \phi_{-4,1}^{D_4} \phi_{0,1}^{D_4} \left(\phi_{-2,1}^{D_4}\right){}^4 \left(\omega_{-4,1}^{D_4}\right){}^2}{13436928}+\frac{7 E_6 \left(\phi_{-4,1}^{D_4}\right){}^2 \left(\phi_{0,1}^{D_4}\right){}^2 \left(\phi_{-2,1}^{D_4}\right){}^2 \left(\omega_{-4,1}^{D_4}\right){}^2}{5971968} \\
 -\frac{47 E_4^2 \left(\phi_{-4,1}^{D_4}\right){}^2 \phi_{0,1}^{D_4} \left(\phi_{-2,1}^{D_4}\right){}^3 \left(\omega_{-4,1}^{D_4}\right){}^2}{8957952}-\frac{41 E_4^2 \left(\phi_{-4,1}^{D_4}\right){}^3 \left(\phi_{0,1}^{D_4}\right){}^2 \phi_{-2,1}^{D_4} \left(\omega_{-4,1}^{D_4}\right){}^2}{1492992}+\frac{101 E_4 E_6 \left(\phi_{-4,1}^{D_4}\right){}^2 \left(\phi_{-2,1}^{D_4}\right){}^4 \left(\omega_{-4,1}^{D_4}\right){}^2}{161243136}-\frac{E_4 E_6 \left(\phi_{-4,1}^{D_4}\right){}^3 \phi_{0,1}^{D_4} \left(\phi_{-2,1}^{D_4}\right){}^2 \left(\omega_{-4,1}^{D_4}\right){}^2}{497664} \\+
 \frac{739 E_4^3 \phi_{-2,1}^{D_4} \phi_{0,1}^{D_4} \left(\phi_{-4,1}^{D_4}\right){}^4 \left(\omega_{-4,1}^{D_4}\right){}^2}{53747712}+\frac{53 E_4^3 \left(\phi_{-2,1}^{D_4}\right){}^3 \left(\phi_{-4,1}^{D_4}\right){}^3 \left(\omega_{-4,1}^{D_4}\right){}^2}{20155392}-\frac{E_4 E_6 \left(\phi_{0,1}^{D_4}\right){}^2 \left(\phi_{-4,1}^{D_4}\right){}^4 \left(\omega_{-4,1}^{D_4}\right){}^2}{497664}-\frac{E_6^2 \left(\phi_{-2,1}^{D_4}\right){}^3 \left(\phi_{-4,1}^{D_4}\right){}^3 \left(\omega_{-4,1}^{D_4}\right){}^2}{5038848} \\+
 \frac{E_4^2 E_6 \phi_{0,1}^{D_4} \left(\phi_{-4,1}^{D_4}\right){}^5 \left(\omega_{-4,1}^{D_4}\right){}^2}{373248}+\frac{E_4^2 E_6 \left(\phi_{-2,1}^{D_4}\right){}^2 \left(\phi_{-4,1}^{D_4}\right){}^4 \left(\omega_{-4,1}^{D_4}\right){}^2}{1679616}-\frac{115 E_4^4 \phi_{-2,1}^{D_4} \left(\phi_{-4,1}^{D_4}\right){}^5 \left(\omega_{-4,1}^{D_4}\right){}^2}{40310784}+\frac{37 E_6^2 \phi_{-2,1}^{D_4} \phi_{0,1}^{D_4} \left(\phi_{-4,1}^{D_4}\right){}^4 \left(\omega_{-4,1}^{D_4}\right){}^2}{53747712} \\
 -\frac{5 E_4^3 E_6 \left(\phi_{-4,1}^{D_4}\right){}^6 \left(\omega_{-4,1}^{D_4}\right){}^2}{5971968}-\frac{13 E_4 E_6^2 \phi_{-2,1}^{D_4} \left(\phi_{-4,1}^{D_4}\right){}^5 \left(\omega_{-4,1}^{D_4}\right){}^2}{40310784}-\frac{E_6^3 \left(\phi_{-4,1}^{D_4}\right){}^6 \left(\omega_{-4,1}^{D_4}\right){}^2}{17915904}+\frac{\phi_{-6,2}^{D_4} \left(\phi_{0,1}^{D_4}\right){}^4 \left(\omega_{-4,1}^{D_4}\right){}^2}{27648} \\+
 \frac{E_4^2 \phi_{-6,2}^{D_4} \left(\phi_{-2,1}^{D_4}\right){}^4 \left(\omega_{-4,1}^{D_4}\right){}^2}{2985984}-\frac{E_4 \phi_{-6,2}^{D_4} \left(\phi_{0,1}^{D_4}\right){}^2 \left(\phi_{-2,1}^{D_4}\right){}^2 \left(\omega_{-4,1}^{D_4}\right){}^2}{110592}-\frac{11 E_4 \phi_{-6,2}^{D_4} \phi_{-4,1}^{D_4} \left(\phi_{0,1}^{D_4}\right){}^3 \left(\omega_{-4,1}^{D_4}\right){}^2}{82944}+\frac{E_6 \phi_{-6,2}^{D_4} \phi_{0,1}^{D_4} \left(\phi_{-2,1}^{D_4}\right){}^3 \left(\omega_{-4,1}^{D_4}\right){}^2}{1492992} \\+
 \frac{E_4^2 \phi_{-6,2}^{D_4} \phi_{-4,1}^{D_4} \phi_{0,1}^{D_4} \left(\phi_{-2,1}^{D_4}\right){}^2 \left(\omega_{-4,1}^{D_4}\right){}^2}{20736}+\frac{E_4^2 \phi_{-6,2}^{D_4} \left(\phi_{-4,1}^{D_4}\right){}^2 \left(\phi_{0,1}^{D_4}\right){}^2 \left(\omega_{-4,1}^{D_4}\right){}^2}{6144}-\frac{E_4 E_6 \phi_{-6,2}^{D_4} \phi_{-4,1}^{D_4} \left(\phi_{-2,1}^{D_4}\right){}^3 \left(\omega_{-4,1}^{D_4}\right){}^2}{139968}-\frac{E_6 \phi_{-6,2}^{D_4} \phi_{-4,1}^{D_4} \left(\phi_{0,1}^{D_4}\right){}^2 \phi_{-2,1}^{D_4} \left(\omega_{-4,1}^{D_4}\right){}^2}{27648} \\
 -\frac{E_4^3 \phi_{-6,2}^{D_4} \left(\phi_{-4,1}^{D_4}\right){}^2 \left(\phi_{-2,1}^{D_4}\right){}^2 \left(\omega_{-4,1}^{D_4}\right){}^2}{27648}-\frac{61 E_4^3 \phi_{-6,2}^{D_4} \left(\phi_{-4,1}^{D_4}\right){}^3 \phi_{0,1}^{D_4} \left(\omega_{-4,1}^{D_4}\right){}^2}{746496}+\frac{13 E_4 E_6 \phi_{-6,2}^{D_4} \left(\phi_{-4,1}^{D_4}\right){}^2 \phi_{-2,1}^{D_4} \phi_{0,1}^{D_4} \left(\omega_{-4,1}^{D_4}\right){}^2}{248832}+\frac{11 E_6^2 \phi_{-6,2}^{D_4} \left(\phi_{-4,1}^{D_4}\right){}^2 \left(\phi_{-2,1}^{D_4}\right){}^2 \left(\omega_{-4,1}^{D_4}\right){}^2}{1492992} \\
 -\frac{7 E_4^2 E_6 \phi_{-6,2}^{D_4} \phi_{-2,1}^{D_4} \left(\phi_{-4,1}^{D_4}\right){}^3 \left(\omega_{-4,1}^{D_4}\right){}^2}{373248}+\frac{149 E_4^4 \phi_{-6,2}^{D_4} \left(\phi_{-4,1}^{D_4}\right){}^4 \left(\omega_{-4,1}^{D_4}\right){}^2}{8957952}-\frac{13 E_4 E_6^2 \phi_{-6,2}^{D_4} \left(\phi_{-4,1}^{D_4}\right){}^4 \left(\omega_{-4,1}^{D_4}\right){}^2}{8957952}-\frac{E_6^2 \phi_{-6,2}^{D_4} \phi_{0,1}^{D_4} \left(\phi_{-4,1}^{D_4}\right){}^3 \left(\omega_{-4,1}^{D_4}\right){}^2}{746496} \\
 -\frac{E_4^2 \left(\phi_{-6,2}^{D_4}\right){}^2 \phi_{-2,1}^{D_4} \phi_{0,1}^{D_4} \left(\omega_{-4,1}^{D_4}\right){}^2}{9216}+\frac{E_4 E_6 \left(\phi_{-6,2}^{D_4}\right){}^2 \left(\phi_{-2,1}^{D_4}\right){}^2 \left(\omega_{-4,1}^{D_4}\right){}^2}{55296}-\frac{E_4 E_6 \left(\phi_{-6,2}^{D_4}\right){}^2 \phi_{-4,1}^{D_4} \phi_{0,1}^{D_4} \left(\omega_{-4,1}^{D_4}\right){}^2}{4608}+\frac{E_6 \left(\phi_{-6,2}^{D_4}\right){}^2 \left(\phi_{0,1}^{D_4}\right){}^2 \left(\omega_{-4,1}^{D_4}\right){}^2}{6144} \\+
 \frac{E_4^2 E_6 \left(\phi_{-6,2}^{D_4}\right){}^2 \left(\phi_{-4,1}^{D_4}\right){}^2 \left(\omega_{-4,1}^{D_4}\right){}^2}{13824}-\frac{E_4^3 \left(\phi_{-6,2}^{D_4}\right){}^3 \left(\omega_{-4,1}^{D_4}\right){}^2}{6912}+\frac{E_4^3 \left(\phi_{-6,2}^{D_4}\right){}^2 \phi_{-4,1}^{D_4} \phi_{-2,1}^{D_4} \left(\omega_{-4,1}^{D_4}\right){}^2}{6912}-\frac{E_6^2 \left(\phi_{-6,2}^{D_4}\right){}^2 \phi_{-4,1}^{D_4} \phi_{-2,1}^{D_4} \left(\omega_{-4,1}^{D_4}\right){}^2}{13824} \\
 -\frac{E_4^2 \left(\phi_{-2,1}^{D_4}\right){}^3 \phi_{0,1}^{D_4} \left(\omega_{-4,1}^{D_4}\right){}^4}{2985984}+\frac{E_4 \phi_{-2,1}^{D_4} \left(\phi_{0,1}^{D_4}\right){}^3 \left(\omega_{-4,1}^{D_4}\right){}^4}{497664}+\frac{E_6^2 \left(\phi_{-6,2}^{D_4}\right){}^3 \left(\omega_{-4,1}^{D_4}\right){}^2}{6912}-\frac{E_6 \left(\phi_{-2,1}^{D_4}\right){}^2 \left(\phi_{0,1}^{D_4}\right){}^2 \left(\omega_{-4,1}^{D_4}\right){}^4}{1990656} \\+
 \frac{E_4^3 \phi_{-4,1}^{D_4} \left(\phi_{-2,1}^{D_4}\right){}^3 \left(\omega_{-4,1}^{D_4}\right){}^4}{2239488}-\frac{E_4^2 \phi_{-4,1}^{D_4} \phi_{-2,1}^{D_4} \left(\phi_{0,1}^{D_4}\right){}^2 \left(\omega_{-4,1}^{D_4}\right){}^4}{497664}+\frac{5 E_4 E_6 \left(\phi_{-2,1}^{D_4}\right){}^4 \left(\omega_{-4,1}^{D_4}\right){}^4}{53747712}-\frac{E_4 E_6 \phi_{-4,1}^{D_4} \left(\phi_{-2,1}^{D_4}\right){}^2 \phi_{0,1}^{D_4} \left(\omega_{-4,1}^{D_4}\right){}^4}{1492992} \\+
 \frac{5 E_4^2 E_6 \left(\phi_{-4,1}^{D_4}\right){}^2 \left(\phi_{-2,1}^{D_4}\right){}^2 \left(\omega_{-4,1}^{D_4}\right){}^4}{4478976}-\frac{7 E_4^3 \left(\phi_{-4,1}^{D_4}\right){}^2 \phi_{-2,1}^{D_4} \phi_{0,1}^{D_4} \left(\omega_{-4,1}^{D_4}\right){}^4}{2985984}+\frac{E_4 E_6 \left(\phi_{-4,1}^{D_4}\right){}^2 \left(\phi_{0,1}^{D_4}\right){}^2 \left(\omega_{-4,1}^{D_4}\right){}^4}{248832}-\frac{E_6^2 \left(\phi_{-4,1}^{D_4}\right){}^2 \phi_{-2,1}^{D_4} \phi_{0,1}^{D_4} \left(\omega_{-4,1}^{D_4}\right){}^4}{2985984} \\+
 \frac{11 E_4^3 E_6 \left(\phi_{-4,1}^{D_4}\right){}^4 \left(\omega_{-4,1}^{D_4}\right){}^4}{5971968}-\frac{E_4^2 E_6 \left(\phi_{-4,1}^{D_4}\right){}^3 \phi_{0,1}^{D_4} \left(\omega_{-4,1}^{D_4}\right){}^4}{186624}+\frac{13 E_4^4 \left(\phi_{-4,1}^{D_4}\right){}^3 \phi_{-2,1}^{D_4} \left(\omega_{-4,1}^{D_4}\right){}^4}{6718464}+\frac{E_4 E_6^2 \left(\phi_{-4,1}^{D_4}\right){}^3 \phi_{-2,1}^{D_4} \left(\omega_{-4,1}^{D_4}\right){}^4}{6718464} \\
 -\frac{E_4^3 \phi_{-6,2}^{D_4} \left(\phi_{-2,1}^{D_4}\right){}^2 \left(\omega_{-4,1}^{D_4}\right){}^4}{497664}-\frac{E_4^2 \phi_{-6,2}^{D_4} \left(\phi_{0,1}^{D_4}\right){}^2 \left(\omega_{-4,1}^{D_4}\right){}^4}{110592}+\frac{E_4 E_6 \phi_{-6,2}^{D_4} \phi_{-2,1}^{D_4} \phi_{0,1}^{D_4} \left(\omega_{-4,1}^{D_4}\right){}^4}{165888}-\frac{E_6^3 \left(\phi_{-4,1}^{D_4}\right){}^4 \left(\omega_{-4,1}^{D_4}\right){}^4}{17915904} \\
 -\frac{E_4^2 E_6 \phi_{-6,2}^{D_4} \phi_{-4,1}^{D_4} \phi_{-2,1}^{D_4} \left(\omega_{-4,1}^{D_4}\right){}^4}{248832}+\frac{E_4^3 \phi_{-6,2}^{D_4} \phi_{-4,1}^{D_4} \phi_{0,1}^{D_4} \left(\omega_{-4,1}^{D_4}\right){}^4}{55296}+\frac{E_6^2 \phi_{-6,2}^{D_4} \left(\phi_{-2,1}^{D_4}\right){}^2 \left(\omega_{-4,1}^{D_4}\right){}^4}{995328}-\frac{E_6^2 \phi_{-6,2}^{D_4} \phi_{-4,1}^{D_4} \phi_{0,1}^{D_4} \left(\omega_{-4,1}^{D_4}\right){}^4}{165888} \\+
 \frac{E_4^3 \phi_{-2,1}^{D_4} \phi_{0,1}^{D_4} \left(\omega_{-4,1}^{D_4}\right){}^6}{1990656}-\frac{11 E_4^4 \phi_{-6,2}^{D_4} \left(\phi_{-4,1}^{D_4}\right){}^2 \left(\omega_{-4,1}^{D_4}\right){}^4}{995328}+\frac{7 E_4 E_6^2 \phi_{-6,2}^{D_4} \left(\phi_{-4,1}^{D_4}\right){}^2 \left(\omega_{-4,1}^{D_4}\right){}^4}{995328}-\frac{E_4 E_6 \left(\phi_{0,1}^{D_4}\right){}^2 \left(\omega_{-4,1}^{D_4}\right){}^6}{497664} \\
 -\frac{E_4^2 E_6 \left(\phi_{-2,1}^{D_4}\right){}^2 \left(\omega_{-4,1}^{D_4}\right){}^6}{4478976}+\frac{E_4^2 E_6 \phi_{-4,1}^{D_4} \phi_{0,1}^{D_4} \left(\omega_{-4,1}^{D_4}\right){}^6}{373248}-\frac{E_4^4 \phi_{-4,1}^{D_4} \phi_{-2,1}^{D_4} \left(\omega_{-4,1}^{D_4}\right){}^6}{1492992}+\frac{5 E_6^2 \phi_{-2,1}^{D_4} \phi_{0,1}^{D_4} \left(\omega_{-4,1}^{D_4}\right){}^6}{5971968} \\
 -\frac{7 E_4^3 E_6 \left(\phi_{-4,1}^{D_4}\right){}^2 \left(\omega_{-4,1}^{D_4}\right){}^6}{5971968}+\frac{E_4^4 \phi_{-6,2}^{D_4} \left(\omega_{-4,1}^{D_4}\right){}^6}{331776}-\frac{E_4 E_6^2 \phi_{-4,1}^{D_4} \phi_{-2,1}^{D_4} \left(\omega_{-4,1}^{D_4}\right){}^6}{4478976}+\frac{5 E_6^3 \left(\phi_{-4,1}^{D_4}\right){}^2 \left(\omega_{-4,1}^{D_4}\right){}^6}{17915904} \\+
 \frac{E_4^3 E_6 \left(\omega_{-4,1}^{D_4}\right){}^8}{5971968}-\frac{E_4 E_6^2 \phi_{-6,2}^{D_4} \left(\omega_{-4,1}^{D_4}\right){}^6}{331776}-\frac{E_6^3 \left(\omega_{-4,1}^{D_4}\right){}^8}{5971968} \\
\end{array}$}

\section{Tables of Gromov-Witten invariants and their specializations}
\label{app:GWinv}

In tables \ref{tab:B4F4gw},\ref{tab:B4B3gw},\ref{tab:gwB3G2} we provide some data to check the relations in section \ref{ss:GWinv}.

\begin{table}[]
    \centering
    \footnotesize
    \setlength{\tabcolsep}{2pt}
    \begin{tabular}{c|c|ccccccccccc||ccccccccccc}
         $(a_0,a_3^{B_4},a_4^{B_4})$&\multicolumn{12}{c||}{$B_4$ and $D_4$ invariants}&  \multicolumn{11}{c}{$F_4$ invariants}\\\hline\hline
         \multirow{11}{*}{(0,0,0)}& \diagbox{$a_1$}{$a_2$}
          & 0 & $\frac{1}{2}$ & 1 & $\frac{3}{2}$ & 2 & $\frac{5}{2}$ & 3 & $\frac{7}{2}$ & 4 & $\frac{9}{2}$ & 5 & 0 & $\frac{1}{2}$ & 1 & $\frac{3}{2}$ & 2 & $\frac{5}{2}$ & 3 &
   $\frac{7}{2}$ & 4 & $\frac{9}{2}$ & 5 \\\cline{2-24}&
 0 & $-$2 & \zero & $-$2 & \zero & $-$4 & \zero & $-$6 & \zero & $-$8 & \zero & $-$10 & $-$2 & \zero & \zero & \zero & \zero
   & \zero & \zero & \zero & \zero & \zero & \zero \\&
 $\frac{1}{2}$ & \zero & \zero & \zero & \zero & \zero & \zero & \zero & \zero & \zero & \zero &
   \zero & \zero & \zero & \zero & \zero & \zero & \zero & \zero & \zero & \zero & \zero &
   \zero \\&
 1 & \zero & \zero & $-$2 & \zero & $-$6 & \zero & $-$10 & \zero & $-$14 & \zero & $-$18 & $-$2 & \zero & $-$2 & \zero &
   \zero & \zero & \zero & \zero & \zero & \zero & \zero \\&
 $\frac{3}{2}$ & \zero & \zero & \zero & \zero & \zero & \zero & \zero & \zero & \zero & \zero &
   \zero & \zero & \zero & \zero & \zero & \zero & \zero & \zero & \zero & \zero & \zero &
   \zero \\&
 2 & \zero & \zero & \zero & \zero & $-$6 & \zero & $-$12 & \zero & $-$18 & \zero & $-$24 & $-$4 & \zero & $-$6 & \zero &
   $-$6 & \zero & $-$4 & \zero & $-$6 & \zero & $-$8 \\&
 $\frac{5}{2}$ & \zero & \zero & \zero & \zero & \zero & \zero & \zero & \zero & \zero & \zero &
   \zero & \zero & \zero & \zero & \zero & \zero & \zero & \zero & \zero & \zero & \zero &
   \zero \\&
 3 & \zero & \zero & \zero & \zero & $-$4 & \zero & $-$12 & \zero & $-$20 & \zero & $-$28 & $-$6 & \zero & $-$10 & \zero &
   $-$12 & \zero & $-$12 & \zero & $-$10 & \zero & $-$14 \\&
 $\frac{7}{2}$ & \zero & \zero & \zero & \zero & \zero & \zero & \zero & \zero & \zero & \zero &
   \zero & \zero & \zero & \zero & \zero & \zero & \zero & \zero & \zero & \zero & \zero &
   \zero \\&
 4 & \zero & \zero & \zero & \zero & $-$6 & \zero & $-$10 & \zero & $-$20 & \zero & $-$30 & $-$8 & \zero & $-$14 & \zero &
   $-$18 & \zero & $-$20 & \zero & $-$20 & \zero & $-$18 \\&
 $\frac{9}{2} $& \zero & \zero & \zero & \zero & \zero & \zero & \zero & \zero & \zero & \zero &
   \zero & \zero & \zero & \zero & \zero & \zero & \zero & \zero & \zero & \zero & \zero &
   \zero \\&
 5 & \zero & \zero & \zero & \zero & $-$8 & \zero & $-$14 & \zero & $-$18 & \zero & $-$30 & $-$10 & \zero & $-$18 & \zero &
   $-$24 & \zero & $-$28 & \zero & $-$30 & \zero & $-$30 \\\hline\hline
   
    \multirow{11}{*}{$(1,\frac{1}{2},0)$}& \diagbox{$a_1$}{$a_2$}&  0 & $\frac{1}{2}$ & 1 & $\frac{3}{2}$ & 2 & $\frac{5}{2}$ & 3 & $\frac{7}{2}$ & 4 & $\frac{9}{2}$ & 5 & 0 & $\frac{1}{2}$ & 1 & $\frac{3}{2}$ & 2 & $\frac{5}{2}$ & 3 & $\frac{7}{2}$
   & 4 & $\frac{9}{2}$ & 5 \\\cline{2-24}&
 0 & \zero & \zero & \zero & \zero & \zero & \zero & \zero & \zero & \zero & \zero & \zero &
   \zero & \zero & \zero & \zero & \zero & \zero & \zero & \zero & \zero & \zero & \zero \\&
 $\frac{1}{2}$ & \zero & \zero & 12 & \zero & 16 & \zero & 32 & \zero & 48 & \zero & 64 & \zero & \zero & \zero &
   \zero & \zero & \zero & \zero & \zero & \zero & \zero & \zero \\&
 1 & \zero & \zero & \zero & \zero & \zero & \zero & \zero & \zero & \zero & \zero & \zero &
   \zero & \zero & 12 & \zero & \zero & \zero & \zero & \zero & \zero & \zero & \zero \\&
 $\frac{3}{2}$ & \zero & \zero & \zero & \zero & 16 & \zero & 48 & \zero & 80 & \zero & 112 & \zero & \zero &
   \zero & \zero & \zero & \zero & \zero & \zero & \zero & \zero & \zero \\&
 2 & \zero & \zero & \zero & \zero & \zero & \zero & \zero & \zero & \zero & \zero & \zero &
   \zero & \zero & 16 & \zero & 16 & \zero & \zero & \zero & \zero & \zero & \zero \\&
 $\frac{5}{2}$ & \zero & \zero & \zero & \zero & \zero & \zero & 48 & \zero & 96 & \zero & 144 & \zero &
   \zero & \zero & \zero & \zero & \zero & \zero & \zero & \zero & \zero & \zero \\&
 3 & \zero & \zero & \zero & \zero & \zero & \zero & \zero & \zero & \zero & \zero & \zero &
   \zero & \zero & 32 & \zero & 48 & \zero & 48 & \zero & 32 & \zero & 48 \\&
 $\frac{7}{2}$ & \zero & \zero & \zero & \zero & \zero & \zero & 32 & \zero & 96 & \zero & 160 & \zero &
   \zero & \zero & \zero & \zero & \zero & \zero & \zero & \zero & \zero & \zero \\&
 4 & \zero & \zero & \zero & \zero & \zero & \zero & \zero & \zero & \zero & \zero & \zero &
   \zero & \zero & 48 & \zero & 80 & \zero & 96 & \zero & 96 & \zero & 80 \\&
 $\frac{9}{2}$ & \zero & \zero & \zero & \zero & \zero & \zero & 48 & \zero & 80 & \zero & 160 & \zero &
   \zero & \zero & \zero & \zero & \zero & \zero & \zero & \zero & \zero & \zero \\&
 5 & \zero & \zero & \zero & \zero & \zero & \zero & \zero & \zero & \zero & \zero & \zero &
   \zero & \zero & 64 & \zero & 112 & \zero & 144 & \zero & 160 & \zero & 160 \\\hline\hline
   
   \multirow{11}{*}{$(0,\frac{1}{2},1)$}& \diagbox{$a_1$}{$a_2$} & 0 & $\frac{1}{2}$ & 1 & $\frac{3}{2}$ & 2 & $\frac{5}{2}$ & 3 & $\frac{7}{2}$ & 4 & $\frac{9}{2}$ & 5 & 0 & $\frac{1}{2}$ & 1 & $\frac{3}{2}$ & 2 & $\frac{5}{2}$ & 3 & $\frac{7}{2}$
   & 4 & $\frac{9}{2}$ & 5 \\\cline{2-24}&
 0 & \zero & \zero & \zero & \zero & \zero & \zero & \zero & \zero & \zero & \zero & \zero &
   \zero & \zero & \zero & \zero & \zero & \zero & \zero & \zero & \zero & \zero & \zero \\&
 $\frac{1}{2}$ & \zero & \zero & 4 & \zero & 12 & \zero & 20 & \zero & 28 & \zero & 36 & \zero & \zero & \zero &
   \zero & \zero & \zero & \zero & \zero & \zero & \zero & \zero \\&
 1 & \zero & \zero & \zero & \zero & \zero & \zero & \zero & \zero & \zero & \zero & \zero &
   \zero & \zero & 4 & \zero & \zero & \zero & \zero & \zero & \zero & \zero & \zero \\&
 $\frac{3}{2}$ & \zero & \zero & \zero & \zero & 16 & \zero & 32 & \zero & 48 & \zero & 64 & \zero & \zero &
   \zero & \zero & \zero & \zero & \zero & \zero & \zero & \zero & \zero \\&
 2 & \zero & \zero & \zero & \zero & \zero & \zero & \zero & \zero & \zero & \zero & \zero &
   \zero & \zero & 12 & \zero & 16 & \zero & 12 & \zero & 20 & \zero & 28 \\&
 $\frac{5}{2}$ & \zero & \zero & \zero & \zero & 12 & \zero & 36 & \zero & 60 & \zero & 84 & \zero & \zero &
   \zero & \zero & \zero & \zero & \zero & \zero & \zero & \zero & \zero \\&
 3 & \zero & \zero & \zero & \zero & \zero & \zero & \zero & \zero & \zero & \zero & \zero &
   \zero & \zero & 20 & \zero & 32 & \zero & 36 & \zero & 32 & \zero & 48 \\&
 $\frac{7}{2}$ & \zero & \zero & \zero & \zero & 20 & \zero & 32 & \zero & 64 & \zero & 96 & \zero & \zero &
   \zero & \zero & \zero & \zero & \zero & \zero & \zero & \zero & \zero \\&
 4 & \zero & \zero & \zero & \zero & \zero & \zero & \zero & \zero & \zero & \zero & \zero &
   \zero & \zero & 28 & \zero & 48 & \zero & 60 & \zero & 64 & \zero & 60 \\&
 $\frac{9}{2}$ & \zero & \zero & \zero & \zero & 28 & \zero & 48 & \zero & 60 & \zero & 100 & \zero & \zero &
   \zero & \zero & \zero & \zero & \zero & \zero & \zero & \zero & \zero \\&
 5 & \zero & \zero & \zero & \zero & \zero & \zero & \zero & \zero & \zero & \zero & \zero &
   \zero & \zero & 36 & \zero & 64 & \zero & 84 & \zero & 96 & \zero & 100 \\
    \end{tabular}
    \caption{Some genus 0 Gromov-Witten invariants of the $B_4,\,D_4$ and $F_4$ theories over $\mathbb F_2$. On the left, we give the values of $a_0,a_3^{B_4},a_4^{B_4}$, which determine the values of $a_3,a_4$ for $F_4$ and $D_4$ via  $2a_3^{B_4},a_4^{B_4}$ and $a_3^{B_4}-\frac{a_4^{B_4}}{2},\frac{a_4^{B_4}}{2}$ respectively. Inside each block, $a_1$ (resp. $a_2$) grows from $0$ to $5$ in steps of $1/2$ from top to bottom (resp. left to right); this growth in steps of $1/2$ is artificial for $F_4$, as all the possible weights have integer indices, but it makes it simpler to visually compare the invariants. In accordance with \eqref{eq:GVB4D4F4}, this presentation allows us to record the values of the Gromov-Witten invariants for the $B_4$ and $D_4$ geometries in a single table. Regarding the $F_4$ geometry, we note that if we reflect the $B_4$ invariants through the diagonal and shift them horizontally by the value of $a_3$, we recover the $F_4$ invariants, again in agreement with \eqref{eq:GVB4D4F4}.}
    \label{tab:B4F4gw}
\end{table}

\begin{table}[]
    \centering
\begin{tabular}{c|c|ccccccccccc||c}
    $(a_0,a_1,a_3)$&\multicolumn{12}{c||}{$B_4$ invariants}&$B_3$ invariants\\\hline\hline
        \multirow{12}{*}{$(0,0,0)$}& \diagbox{$a_2$}{$a_4$} & 0 & $\frac{1}{2}$
        & 1 & $\frac{3}{2}$ & 2 & $\frac{5}{2}$ & 3 & $\frac{7}{2}$ & 4 & $\frac{9}{2}$ & 5 &\\\cline{2-14}&
 0 & $-$2 & \zero & \zero & \zero & \zero & \zero & \zero & \zero & \zero & \zero & \zero &$-$2\\&
 $\frac{1}{2}$ & \zero & \zero & \zero & \zero & \zero & \zero & \zero & \zero & \zero & \zero &
   \zero &\zero\\&
 1 & $-$2 & \zero & \zero & \zero & \zero & \zero & \zero & \zero & \zero & \zero & \zero &$-$2\\&
 $\frac{3}{2}$ & \zero & \zero & \zero & \zero & \zero & \zero & \zero & \zero & \zero & \zero &
   \zero &\zero\\&
 2 & $-$4 & \zero & \zero & \zero & \zero & \zero & \zero & \zero & \zero & \zero & \zero &$-$4\\&
 $\frac{5}{2}$ & \zero & \zero & \zero & \zero & \zero & \zero & \zero & \zero & \zero & \zero &
   \zero &\zero\\&
 3 & $-$6 & \zero & \zero & \zero & \zero & \zero & \zero & \zero & \zero & \zero & \zero &$-$6\\&
 $\frac{7}{2}$ & \zero & \zero & \zero & \zero & \zero & \zero & \zero & \zero & \zero & \zero &
   \zero &\zero\\&
 4 & $-$8 & \zero & \zero & \zero & \zero & \zero & \zero & \zero & \zero & \zero & \zero&$-$8 \\&
 $\frac{9}{2}$ & \zero & \zero & \zero & \zero & \zero & \zero & \zero & \zero & \zero & \zero &
   \zero &\zero\\&
 5 & $-$10 & \zero & \zero & \zero & \zero & \zero & \zero & \zero & \zero & \zero & \zero &$-$10\\\hline\hline
 
 \multirow{12}{*}{$(0,1,1)$}&\diagbox{$a_2$}{$a_4$} & 0 & $\frac{1}{2}$ & 1 & $\frac{3}{2}$ & 2 & $\frac{5}{2}$ & 3 & $\frac{7}{2}$ & 4 & $\frac{9}{2}$ & 5 \\\cline{2-14}&
 0 & \zero & \zero & \zero & \zero & \zero & \zero & \zero & \zero & \zero & \zero & \zero&\zero \\&
 $\frac{1}{2}$ & \zero & \zero & \zero & \zero & \zero & \zero & \zero & \zero & \zero & \zero &
   \zero &\zero \\&
 1 & $-$2 & \zero & 4 & \zero & $-$2 & \zero & \zero & \zero & \zero & \zero & \zero&\zero \\&
 $\frac{3}{2}$ & \zero & \zero & \zero & \zero & \zero & \zero & \zero & \zero & \zero & \zero &
   \zero &\zero\\&
 2 & $-$20 & \zero & $-$16 & \zero & $-$20 & \zero & \zero & \zero & \zero & \zero & \zero&$-56$\\&
 $\frac{5}{2}$ & \zero & \zero & \zero & \zero & \zero & \zero & \zero & \zero & \zero & \zero &
   \zero&\zero \\&
 3 & $-$40 & \zero & $-$32 & \zero & $-$40 & \zero & \zero & \zero & \zero & \zero & \zero &$-112$\\&
 $\frac{7}{2}$ & \zero & \zero & \zero & \zero & \zero & \zero & \zero & \zero & \zero & \zero &
   \zero&\zero \\&
 4 & $-$60 & \zero & $-$48 & \zero & $-$60 & \zero & \zero & \zero & \zero & \zero & \zero &$-168$\\&
 $\frac{9}{2}$ & \zero & \zero & \zero & \zero & \zero & \zero & \zero & \zero & \zero & \zero &
   \zero &\zero\\&
 5 & $-$80 & \zero & $-$64 & \zero & $-$80 & \zero & \zero & \zero & \zero & \zero & \zero&$-224$ \\\hline\hline
 
   \multirow{12}{*}{$(1,\frac{1}{2},\frac{3}{2})$}&\diagbox{$a_2$}{$a_4$}& 0 & $\frac{1}{2}$ & 1 & $\frac{3}{2}$ & 2 & $\frac{5}{2}$ & 3 & $\frac{7}{2}$ & 4 & $\frac{9}{2}$ & 5 \\\cline{2-14}&
 0 & \zero & \zero & \zero & \zero & \zero & \zero & \zero & \zero & \zero & \zero & \zero &\zero\\&
 $\frac{1}{2}$ & \zero & \zero & \zero & \zero & \zero & \zero & \zero & \zero & \zero & \zero &
   \zero &\zero\\&
 1 & \zero & \zero & 12 & \zero & 12 & \zero & \zero & \zero & \zero & \zero & \zero &24\\&
 $\frac{3}{2}$ & \zero & \zero & \zero & \zero & \zero & \zero & \zero & \zero & \zero & \zero &
   \zero &\zero\\&
 2 & 16 & \zero & \zero & \zero & \zero & \zero & 16 & \zero & \zero & \zero & \zero&32 \\&
 $\frac{5}{2}$ & \zero & \zero & \zero & \zero & \zero & \zero & \zero & \zero & \zero & \zero &
   \zero &\zero\\&
 3 & 48 & \zero & $-$36 & \zero & $-$36 & \zero & 48 & \zero & \zero & \zero & \zero&24\\&
 $\frac{7}{2}$ & \zero & \zero & \zero & \zero & \zero & \zero & \zero & \zero & \zero & \zero &
   \zero &\zero \\&
 4 & 80 & \zero & $-$60 & \zero & $-$60 & \zero & 80 & \zero & \zero & \zero & \zero&40 \\&
 $\frac{9}{2}$ & \zero & \zero & \zero & \zero & \zero & \zero & \zero & \zero & \zero & \zero &
   \zero&\zero \\&
 5 & 112 & \zero & $-$84 & \zero & $-$84 & \zero & 112 & \zero & \zero & \zero & \zero &56\\
    \end{tabular}
    \caption{Genus 0 Gromov-Witten invariants for $B_4$ and $B_3$ theories over $\mathbb F_2$. We see that adding the rows of $B_4$, we get the $B_3$ invariants in agreement with equations \eqref{eq:gwBnDn} and  \eqref{eq:gwDnBn-1}.}
    \label{tab:B4B3gw}
\end{table}

\begin{table}[]
    \footnotesize
        \setlength{\tabcolsep}{2pt}
    \centering
    \begin{tabular}{c|c|ccccccccccc||c}
    $(a_0,a_2)$&\multicolumn{12}{c||}{$B_3$ invariants}&$G_2$ invariants\\\hline\hline

     \multirow{12}{*}{$(0,0)$}&\diagbox{$a_1$}{$a_3$}& 0 & $\frac{1}{2}$ & 1 & $\frac{3}{2}$ & 2 & $\frac{5}{2}$ & 3 & $\frac{7}{2}$ & 4 & $\frac{9}{2}$ & 5 &  \\\hline&
 0 & $-$2 & \zero & \zero & \zero & \zero & \zero & \zero & \zero & \zero & \zero & \zero & $-$2 \\&
 $\frac{1}{2}$ & \zero & \zero & \zero & \zero & \zero & \zero & \zero & \zero & \zero & \zero & \zero
   & \zero \\&
 1 & \zero & \zero & \zero & \zero & \zero & \zero & \zero & \zero & \zero & \zero & \zero &
   \zero \\&
 $\frac{3}{2}$ & \zero & \zero & \zero & \zero & \zero & \zero & \zero & \zero & \zero & \zero & \zero
   & \zero \\&
 2 & \zero & \zero & \zero & \zero & \zero & \zero & \zero & \zero & \zero & \zero & \zero &
   \zero \\&
 $\frac{5}{2}$ & \zero & \zero & \zero & \zero & \zero & \zero & \zero & \zero & \zero & \zero & \zero
   & \zero \\&
 3 & \zero & \zero & \zero & \zero & \zero & \zero & \zero & \zero & \zero & \zero & \zero &
   \zero \\&
 $\frac{7}{2}$ & \zero & \zero & \zero & \zero & \zero & \zero & \zero & \zero & \zero & \zero & \zero
   & \zero \\&
 4 & \zero & \zero & \zero & \zero & \zero & \zero & \zero & \zero & \zero & \zero & \zero &
   \zero \\&
 $\frac{9}{2}$ & \zero & \zero & \zero & \zero & \zero & \zero & \zero & \zero & \zero & \zero & \zero
   & \zero \\&
 5 & \zero & \zero & \zero & \zero & \zero & \zero & \zero & \zero & \zero & \zero & \zero &
   \zero \\\hline\hline

  \multirow{12}{*}{$(1,2)$}&\diagbox{$a_1$}{$a_3$}& 0 & $\frac{1}{2}$ & 1 & $\frac{3}{2}$ & 2 & $\frac{5}{2}$ & 3 & $\frac{7}{2}$ & 4 & $\frac{9}{2}$ & 5 \\\hline &
 0 & $-$6 & \zero & \zero & \zero & $-$8 & \zero & \zero & \zero & $-$6 & \zero & \zero &$-$6\\&
 $\frac{1}{2}$ & \zero & 32 & \zero & 32 & \zero & 32 & \zero & 32 & \zero & \zero & \zero &\zero\\&
 1 & $-$8 & \zero & $-$56 & \zero & $-$160 & \zero & $-$56 & \zero & $-$8 & \zero & \zero &24\\&
 $\frac{3}{2}$ & \zero & 32 & \zero & 32 & \zero & 32 & \zero & 32 & \zero & \zero & \zero &\zero\\&
 2 & $-$6 & \zero & \zero & \zero & $-$8 & \zero & \zero & \zero & $-$6 & \zero & \zero&$-$6 \\&
 $\frac{5}{2}$ & \zero & \zero & \zero & \zero & \zero & \zero & \zero & \zero & \zero & \zero &
   \zero & \zero\\&
 3 & \zero & \zero & \zero & \zero & \zero & \zero & \zero & \zero & \zero & \zero & \zero&$-$96 \\&
 $\frac{7}{2}$ & \zero & \zero & \zero & \zero & \zero & \zero & \zero & \zero & \zero & \zero &
   \zero&\zero \\&
 $4$ & \zero & \zero & \zero & \zero & \zero & \zero & \zero & \zero & \zero & \zero & \zero &$-$6\\&
 $\frac{9}{2}$& \zero & \zero & \zero & \zero & \zero & \zero & \zero & \zero & \zero & \zero &
   \zero &\zero\\&
 5 & \zero & \zero & \zero & \zero & \zero & \zero & \zero & \zero & \zero & \zero & \zero &24\\\hline\hline
 
   \multirow{12}{*}{$(1,3)$}&\diagbox{$a_1$}{$a_3$}& 0 & $\frac{1}{2}$ & 1 & $\frac{3}{2}$ & 2 & $\frac{5}{2}$ & 3 & $\frac{7}{2}$ & 4 & $\frac{9}{2}$ & 5 &  \\\hline&
 0 & $-$10 & \zero & \zero & \zero & $-$6 & \zero & \zero & \zero & $-$6 & \zero & \tikz[remember picture]\node[inner sep=0pt](f){\zero}; & $-$10 \\&
 $\frac{1}{2}$ & \zero & 64 & \zero & 24 & \zero & 32 & \zero & 32 & \zero & 24 & \zero & \zero \\&
 1 & $-$16 & \zero & $-$168 & \zero & $-$166 & \zero & $-$112 & \zero & $-$166 & \zero & $-$168 & 48 \\&
 $\frac{3}{2}$ & \zero & 96 & \zero & 256 & \zero & 392 & \zero & 392 & \zero & 256 & \zero & \zero \\&
 2 & $-$18 & \zero & $-$224 & \zero & $-$308 & \zero & $-$224 & \zero & $-$308 & \zero & $-$224 & $-$72 \\&
 $\frac{5}{2}$ & \zero & 96 & \zero & 248 & \zero & 288 & \zero & 288 & \zero & 248 & \zero & \zero \\&
 3 & $-$16 & \zero & $-$168 & \zero & $-$432 & \zero & $-$336 & \zero & $-$432 & \zero & $-$168 & $-$22 \\&
 $\frac{7}{2}$ & \zero & 64 & \zero & 448 & \zero & 512 & \zero & 512 & \zero & 448 & \zero & \zero \\&
 4 & $-$10 & \zero & $-$280 & \zero & $-$710 & \zero & $-$560 & \zero & $-$710 & \zero & $-$280 & 132 \\&
 $\frac{9}{2}$ & \zero & 96 & \zero & 672 & \zero & 768 & \zero & 768 & \zero & 672 & \zero & \zero \\&
 5 & $-$\tikz[remember picture]\node[inner sep=0pt](i){14}; & \zero & $-$392 & \zero & $-$994 & \zero & $-$784 & \zero & $-$994 & \zero & $-$392 &\tikz[remember picture]\node[inner sep=0pt](c){132}; \\
\end{tabular}

\begin{tikzpicture}[remember picture,overlay]
\draw[line width=4mm,myblue,opacity=0.2] (i.west)--(f.east);
\draw[fill=myblue,opacity=0.2] (c) circle (0.3);
\end{tikzpicture}

    \caption{Genus 0 Gromov-Witten invariants of $B_3$ and $G_2$ over $\mathbb F_2$. Adding the entries in the diagonals given by $a_1+a_3$ constant gives the $G_2$ invariant with $a_1^{G_2}=a_1^{B_3}+a_3^{B_3}$, in accordance with equation \eqref{eq:gwB3G2}. For example, the sum of the $B_3$ invariants in the blue box gives the $G_2$ invariant in the blue circle.}
    \label{tab:gwB3G2}
\end{table}

\section{Root systems} \label{app:root_systems}
Root systems, and consequently finite simple Lie algebras, enjoy a classification into four infinite families, the $A$, $B$, $C$, and $D$ series, and five exceptional cases $G_2$, $F_4$, $E_6$, $E_7$, $E_8$. In this section, we will summarize the properties of these root systems, following the conventions of \cite{Bourbaki}. Note in particular that in these conventions, short roots are normalized to have norm squared 2; roots of simply laced Lie algebras are considered as short.

It is natural to embed the root systems in a Euclidean space $\mathbb{R}^n$. We will we denote the standard basis of this space by $e_i,\, i=1,\dots n$ and the coordinates in this basis by $x_i$. We refer to this coordinate system as the orthogonal basis. In most cases, $n$ equals the rank of the gauge group, i.e. the dimension of the root system. When it does not, the root system lies in a subspace of $\mathbb{R}^n$ and the last $n-r$ coordinates of this space can be expressed in terms of the first $r$ of them.

In tables \ref{tab:rootSystems1}, \ref{tab:rootSystems2}, and \ref{tab:rootSystems3} we summarize the properties of all root systems.

As we explain in section \ref{ss:matching_elliptic_parameters}, the elliptic genus of a gauge theory with gauge algebra $\mathfrak{g}$ is a meromorphic Jacobi forms whose elliptic parameters (aside from $\gs$) take values in the complexified coroot lattice $\LambdaCR(\mathfrak{g}) \otimes \IC$. The Weyl invariant Jacobi forms defined in the literature \cite{Bertola, Adler_2020,adler2020structure} have elliptic parameters taking values in  $\LambdaR(\mathfrak{g})\otimes \IC$ (in the conventions of \cite{Bourbaki}). Rather than redefine these forms, we use conventions in this paper in which the root systems as defined by \cite{Bourbaki} are interpreted as the coroots of the dual algebra $\tilde{\mathfrak{g}}$. In these conventions, short coroots have norm squared 2. The roots that enter in the definition of the denominator of the elliptic genus \eqref{eq:ZkAnsatz} are obtained from these coroots, hence are normalized such that all long roots have norm squared 2; with regard to these normalization conventions, roots of simply laced algebras are considered long. The roots of the Lie algebra $G_2$ in our conventions are worked out as an example in section \ref{ss:matching_elliptic_parameters}.

\begin{table}
    \centering
    \small
    
\newcolumntype{A}{ >{\centering\arraybackslash} m{1.75cm} }

\newcolumntype{B}{ >{\centering\arraybackslash} m{3.3cm} }
    
\renewcommand{\arraystretch}{2}
    \begin{tabular}{A||B|B|B|B}
        $\mg$& $A_n$& $B_n$& $C_n$ &$D_n$ \\\hline
        $\tilde{\mg}$& $A_n$& $C_n$&$B_n$& $D_n$\\\hline
        Embedding space &$V=\{\sum_{i=1}^{n+1}x_i=0\}\subset \mathbb{R}^{n+1}$& $\mathbb{R}^n$& $\mathbb{R}^n$& $\mathbb{R}^n$ \\\hline
        Roots& $e_i-e_j, i\neq j$& $\pm e_i $ \newline $\pm e_i\pm e_j,\, i\neq j$&$\pm 2e_i $ \newline $\pm e_i\pm e_j,\, i\neq j$&$\pm e_i\pm e_j,\, i\neq j$\\\hline
        Simple roots& $\alpha_i=e_i-e_{i+1},\, i=1,\dots,n$&$\alpha_i=e_i-e_{i+1}$\newline$i=1,\dots,n-1$ \newline $\alpha_n=e_n$&$\alpha_i=e_i-e_{i+1}$\newline$i=1,\dots,n-1$ \newline $\alpha_n=2 e_n$&$\alpha_i=e_i-e_{i+1}$\newline$i=1,\dots,n-1$ \newline $\alpha_n=e_{n-1}+e_n$
        \\\hline
        $\Lambda_r$&$\Z^{n+1}\cap V$&$\Z^n$&$x\in \Z^n$ such that $\sum x_i\in 2\Z$&$x\in \Z^n$ such that $\sum x_i\in 2\Z$\\\hline
        Fund. Weights & $\omega_i=e_1+\cdots+e_i-\frac{i}{n+1}\sum e_i$& $\omega_i=e_1+\cdots+e_i,\,i\neq n$\newline$\omega_n=\frac{1}{2}\sum e_i$&$\omega_i=e_1+\cdots+e_i$&$\omega_i=e_1+\cdots+e_i, \, i\neq n-1,n$\newline
        $\omega_{n-1}=\frac{1}{2}(e_1+\dots+e_{n-1}-e_n)$\newline
        $\omega_{n}=\frac{1}{2}(e_1+\dots+e_{n-1}+e_n)$\\\hline
        $\Lambda_w$&$\left<\Lambda_r,e_1-\frac{1}{n+1}\sum e_i\right>$&$\Z^{n}+\left(\frac{1}{2}\sum e_i\right)\IZ$& $\Z^n$&$\Z^{n}+\left(\frac{1}{2}\sum e_i\right)\IZ$\\\hline
        $ds^2$&$dx^2$&$2dx^2$& $dx^2$&$dx^2$\\\hline
        $W$& $S_n$ permuting the $x_i$&$S_{n}\ltimes \Z_2^n$ permuting $x_i$
        and multiplying each coordinate by $\pm 1$&$S_{n}\ltimes \Z_2^n$ permuting $x_i$
        and multiplying each coordinate by $\pm 1$&$S_{n}\ltimes \Z_2^{n-1}$ permuting $x_i$
        and multiplying an even number of coordinates by $-1$\\\hline

    \end{tabular}
    \caption{Description and properties of the A, B, C, and D type root systems}
    \label{tab:rootSystems1}

\end{table}
\begin{table}
    \centering
    \small
    
\newcolumntype{A}{ >{\centering\arraybackslash} m{1.75cm} }

\newcolumntype{B}{ >{\centering\arraybackslash} m{5cm} }
    
\renewcommand{\arraystretch}{2}
    \begin{tabular}{A||B|B}
        $\mg$& $G_2$& $F_4$ \\\hline
        $\tilde{\mg}$& $G_2$& $F_4$\\\hline
        Embedding space &$V=\{x_1+x_2+x_3=0\}\subset \mathbb{R}^{3}$& $\mathbb{R}^4$ \\\hline
        Roots& $e_i-e_j, i\neq j$ \newline
        $ \pm(2e_i-e_j-e_k),\,i\neq j\neq k\neq i$& $\pm e_i$\newline
        $\pm e_i \pm e_j$\newline
        $\frac{1}{2}(\pm e_1\pm e_2\pm e_3\pm e_4)$
    \\\hline
        Simple roots& $\alpha_1=e_1-e_{2}$ \newline
        $\alpha_2=-2e_1+e_2+e_3$& $\alpha_1=e_2-e_3$\newline
        $\alpha_2=e_3-e_4$\newline
        $\alpha_3=e_4$\newline
        $\alpha_4=\frac{1}{2}(e_1-e_2-e_3-e_4)$
        \\\hline
        $\Lambda_r$&$\Z^{3}\cap V$&$\Z^4+\left(\frac{1}{2}\sum e_i\right)\Z$\\\hline
        Fund. Weights & $\omega_1=-e_2+e_3$\newline
        $\omega_2=-e_1-e_2+2e_3$& $\omega_1=e_1+e_2$\newline
        $\omega_2=2e_1+e_2+e_3$\newline$\omega_3=\frac{1}{2}(3e_1+e_2+e_3+e_4)$\newline$\omega_4=e_4$\\\hline
        $\Lambda_w$&$\Z^{3}\cap V$&$\Z^4+\left(\frac{1}{2}\sum e_i\right)\Z$\\\hline
        $ds^2$&$dx^2$&$2dx^2$\\\hline
        $W$& $D_4=S_2\ltimes \Z_2$ permuting $x_1,x_2$ and taking $x\to -x$ &$S_3\ltimes(S_4\ltimes \Z_2^3)$ where $S_4$ permutes the coordinates $\Z_2^3$ changes the sign of an even number of coordinates and $S_3$ is generated by the reflection along $\frac{1}{2}\sum e_i$\\\hline

    \end{tabular}
    \caption{Description and properties of the $G_2$ and $F_4$ root systems}
    \label{tab:rootSystems2}

\end{table}

\begin{table}
    \centering
    \small
    
\newcolumntype{A}{ >{\centering\arraybackslash} m{2.0cm} }

\newcolumntype{B}{ >{\centering\arraybackslash} m{4cm} }
    
\renewcommand{\arraystretch}{2}
    \begin{tabular}{A||B|B|B}
        $\mg$& $E_6$& $E_7$& $E_8$\\\hline
        $\tilde{\mg}$& $E_6$& $E_7$&$E_8$\\\hline
        Embedding space & $V = \{x_6 = x_7 = -x_8\} \subset \mathbb{R}^8$ & $V = \{x_7 = -x_8\} \subset \mathbb{R}^8$ & $\mathbb{R}^8$ \\\hline
        Roots& $\{\pm e_i \pm e_j | i < j \leq 5\} \cup \{\frac{1}{2} \sum_{i=1}^8 (-1)^{n(i)} e_i|\newline \sum_{i=1}^8 n(i) \in 2\mathbb{Z}\}$   & $\{\pm e_i \pm e_j | i < j \leq 6\} \cup \{\pm(e_7-e_8)\}\cup \{\frac{1}{2} \sum_{i=1}^8 (-1)^{n(i)} e_i|\newline \sum_{i=1}^8 n(i) \in 2\mathbb{Z}\} $ &$\{\pm e_i \pm e_j | i < j\} \cup \{\frac{1}{2} \sum_{i=1}^8 (-1)^{n(i)} e_i|\newline \sum_{i=1}^8 n(i) \in 2\mathbb{Z}\}$ \\\hline
        Simple roots& $\alpha_1 = \frac{1}{2}(e_8 - e_7 -e_6-e_5-e_4-e_3-e_2+e_1)$\newline $\alpha_2 = e_2 + e_1$ \newline $\alpha_3 = e_2 - e_1 $\newline $\alpha_4 = e_3 - e_2$\newline $\alpha_5 = e_4 - e_3$\newline $\alpha_6 = e_5 -e_4$\newline  &$\alpha_1 = \frac{1}{2}(e_8 - e_7 -e_6-e_5-e_4-e_3-e_2+e_1)$\newline $\alpha_2 = e_2 + e_1$ \newline $\alpha_3 = e_2 - e_1 $\newline $\alpha_4 = e_3 - e_2$\newline $\alpha_5 = e_4 - e_3$\newline $\alpha_6 = e_5 -e_4$\newline $\alpha_7 = e_6-e_5$\newline &$\alpha_1 = \frac{1}{2}(e_8 - e_7 -e_6-e_5-e_4-e_3-e_2+e_1)$\newline $\alpha_2 = e_2 + e_1$ \newline $\alpha_3 = e_2 - e_1 $\newline $\alpha_4 = e_3 - e_2$\newline $\alpha_5 = e_4 - e_3$\newline $\alpha_6 = e_5 -e_4$\newline $\alpha_7 = e_6-e_5$\newline $\alpha_8 = e_7-e_6$\newline 
        \\\hline
        $\Lambda_r$&$\Lambda_r(E_8)\cap V$& $\Lambda_r(E_8)\cap V$ & $x\in \mathbb R^8$ such that $2x_i\in \Z, \,x_i-x_j\in \Z, \,\sum x_i\in 2\Z$  \\\hline
        Fund. Weights & $\omega_1= \frac{2}{3}(-e_6-e_7+e_8)$\newline $\omega_2 = \frac{1}{2}(e_1 + e_2 + e_3 + e_4 +e_5 -e_6 -e_7 +e_8)$\newline $\omega_3 =  \frac{1}{2}(-e_1+e_2+e_3+e_4+e_5) + \frac{5}{6}(-e_6-e_7 +e_8)$\newline $\omega_4 = e_3+e_4+e_5 -e_6 -e_7+e_8$\newline $\omega_5 =  e_4 + e_5 + \frac{2}{3}(-e_6 - e_7 +e_8)$\newline $\omega_6 = e_5 + \frac{1}{3}(-e_6 -e_7 +e_8)$ &
        $\omega_1= -e_7 +e_8$\newline $\omega_2 = \frac{1}{2}(e_1 + e_2 + e_3 + e_4 +e_5 +e_6 - 2e_7 +2e_8)$\newline $\omega_3 = \frac{1}{2}(-e_1+e_2+e_3+e_4+e_5 + e_6-3e_7+3e_8)$\newline $\omega_4 = e_3+e_4+e_5 +e_6 +2(-e_7 + e_8)$\newline $\omega_5 =  e_4 + e_5 + e_6 + \frac{3}{2}(-e_7 + e_8)$\newline $\omega_6 = e_5 +e_6 -e_7 + e_8$\newline $\omega_7 = e_6 + \frac{1}{2}(-e_7 + e_8)$  & $\omega_1= 2e_8$\newline $\omega_2 = \frac{1}{2}(e_1 + e_2 + e_3 + e_4 +e_5 +e_6 +e_7 +5e_8)$\newline $\omega_3 = \frac{1}{2}(-e_1+e_2+e_3+e_4+e_5 + e_6+e_7+7e_8)$\newline $\omega_4 = e_3+e_4+e_5 +e_6 + e_7 + 5e_8$\newline $\omega_5 = e_4 + e_5 + e_6 + e_7 + 4e_8$\newline $\omega_6 = e_5 +e_6 +e_7 + 3e_8$\newline $\omega_7 = e_6 + e_7 + 2e_8$\newline $\omega_8 = e_7+e_8$  \\\hline
        $\Lambda_w$ &$<\Lambda_r,\frac{1}{3}(e_1 + e_2 +e_3$ \newline $+ e_4 -2e_5-2e_6)> $ & $<\Lambda_r,\frac{1}{4}(e_1 + e_2 +e_3$ \newline $+ e_4 +e_5+e_6-3e_7-3e_8)>$ & $\Z^8$ \\\hline
        $ds^2$&$dx^2$& $dx^2$ & $dx^2$\\\hline

    \end{tabular}
    \caption{Description and properties of the E type root systems}
    \label{tab:rootSystems3}

\end{table}

\section{The rings of Jacobi forms} \label{app:Jacobi_forms}

To simplify notation, we set 
\begin{equation*}
    e(z)=\exp(2\pi i z).
\end{equation*}

We use the following conventions for the Jacobi $\theta$ functions,

\begin{equation}
    \Theta\left[
    \begin{array}{c}
         a\\b
    \end{array}\right](\tau,z)=\sum_{n\in \Z}e^{2\pi i b n}q^{\frac{1}{2}(n+a)^2}Z^{n+a},\quad \text{where } q=e(\tau), \, Z=e(z) \,,
\end{equation}

\begin{equation}
    \theta_1=i\Theta\left[
    \begin{array}{c}
         1/2\\1/2
    \end{array}\right],\quad\theta_2=\Theta\left[
    \begin{array}{c}
         1/2\\0
    \end{array}\right],\quad\theta_3=\Theta\left[
    \begin{array}{c}
         0\\0
    \end{array}\right],\quad\theta_4=\Theta\left[
    \begin{array}{c}
         0\\1/2
    \end{array}\right],
\end{equation}

for the Dedekind $\eta$ function,

\begin{equation}
    \eta(\tau)=q^{1/24}\prod_n (1-q^n), \quad  \text{where }  q=e(\tau),
\end{equation}

and for the Eisenstein series,
\begin{equation}
    E_n(\tau)=1+\frac{2}{\zeta(1-k)}\sum_{n\in \mathbb N^*}\frac{n^{k-1}q^n}{1-q^n}, \quad \text{where } \mathbb N^*=\{1,2,\dots\} \text{ and } q=e(\tau).
\end{equation}

A Jacobi modular form of weight $w$ and index $n$ of the Lie algebra $\mg$ is a holomorphic function $\phi_{w,n}:\mathbb H\times \mathfrak{h}^*_{\mathbb{C}}\to \mathbb C$, where $\mathbb H$ is the upper-half plane, and $\mathfrak{h}^*_{\mathbb C}$ is the complexified root system, satisfying the following conditions:
\begin{itemize}
    \item Modularity:
    \begin{equation*}
        \phi_{w,n}\left(\frac{a\tau+b}{c\tau+d},\frac{z}{c\tau+d}\right)=(c\tau+d)^w e\left[\frac{nc}{2(c\tau+d)}(z,z)\right]\phi_{w,n}(\tau,z), \quad \forall \left(\begin{array}{cc}
             a& b \\
            c & d
        \end{array}\right)\in SL(2,\Z)
    \end{equation*}
    \item Quasi-periodicity:
    \begin{equation*}
        \phi_{w,n}(\tau,z+\lambda \tau +\mu)= e\left[-n\left(\frac{(\lambda,\lambda)}{2}+(\lambda,z)\right)\right]\phi_{w,n}(\tau,z), \quad \forall \lambda,\mu \in \Lambda_r
    \end{equation*}
    \item Weyl Symmetry:
    \begin{equation*}
    \phi_{w,n}(\tau,wz)=\phi_{w,n}(\tau,z),\quad \forall w \in W 
    \end{equation*}
    \item Fourier expansion: $\phi_{w,n}$ can be expanded as
    \begin{equation*}
        \phi_{w,n}=\sum_{l\in \mathbb{N},\gamma \in \Lambda_{cw}=(\Lambda_r)^*}c(l,\gamma)q^l\zeta^\gamma, \quad \text{where } z^\gamma=e[(z,\gamma)]
    \end{equation*}
\end{itemize}

For every finite simple Lie algebra, except $E_8$, the space of Jacobi modular forms is a freely generated algebra over $\mathbb{C}[E_4,E_6]$ (the space of holomorphic modular forms) \cite{Wirthmuller:Jacobi,wang2020weyl}. Below, we will give explicit expressions for the generators of the rings of Jacobi forms we consider in the text. The forms are compactly written in terms of 
\begin{equation}
    \alpha:=\phi_{-1,1/2}=i\frac{\theta_1}{\eta^3} \,.
\end{equation}

\subsection{$A_n$ Jacobi forms}
We can construct the Jacobi forms of the Lie algebras $A_n$ by repeated use of the differential operator \cite{Bertola}
\begin{equation}
    \cZ=\frac{1}{2\pi i}\left(\sum_{i=1}^{n+1}\frac{\partial}{\partial x_i}+\frac{\pi^2}{3}E_2\sum_{i=1}^{n+1}x_i\right)
    \label{eq:Zoperator}
\end{equation}
on the lowest weight form. Let
\begin{equation}
    \Phi^{A_n}=\prod_{i=1}^{n+1}\alpha(x_i).
\end{equation}
then 
\begin{equation}
    \phi^{A_n}_{-k,1}=\left.\left(\cZ^{n+1-k}\Phi^{A_n}\right)\right|_{\sum x_i=0}, \quad k=0,2,3,4,\dots,n+1
\end{equation}
where the $\sum x_i=0$ condition is imposed after acting by the operator $\cZ$.

\subsection{$B_n$ Jacobi forms}
The Jacobi forms were given in \cite{Bertola}. We follow the conventions in \cite{Kim:2018gak}:

In terms of the Weierstrass function 
\begin{equation}
    \wp(z)=\frac{\theta_3(0)\theta_2(0)^2}{4}\frac{\theta_4(z)}{\theta_1(z)}-\frac{1}{12}\left(\theta_3(0)^4+\theta_2(0)^4\right),
\end{equation}
we have 
\begin{equation}
    \sum_{i=0}^n\wp^{(2i-2)}(v)\phi_{-2i,1}(x)=-\frac{1}{2^{n-2}(n-1)!}
    \frac{
    \left|\begin{array}{cccc}
         1& \wp(v)&\cdots&\wp^{(2n-2)}(v) \\
         1& \wp(x_1)&\cdots&\wp^{(2n-2)}(x_1) \\
         \vdots&\vdots &&\vdots\\
         1& \wp(x_n)&\cdots&\wp^{(2n-2)}(x_n) \\
    \end{array}\right|
    }{
    \left|\begin{array}{cccc}
         1& \wp(x_1)&\cdots&\wp^{(2n-4)}(x_1) \\
         \vdots&\vdots &&\vdots\\
         1& \wp(x_n)&\cdots&\wp^{(2n-4)}(x_n) \\
    \end{array}\right|
    }\prod_{i=1}^n\alpha^2(x_i),
    \label{eq:Bforms}
\end{equation}
where $\wp^{(-2)}$ is to be understood as $1$.
\subsection{$D_n$ Jacobi Forms}
	
	The Jacobi forms for $D_n$ ($n\leq 8$) were built in \cite{Adler_2020}.
	 
	 In $D_n$ there is a form of weight $-n$ and index 1 given by
	 \begin{equation}
	 	\omega_{-n,1}^{D_n}
	 	=\prod_{i=1}^{n}\alpha(x_i).
	 \end{equation}
	 The remaining forms of index 1 can be obtained from the lowest weight form by the repeated use of Hecke operator. On forms of weight $k$ and index $m$, the Hecke operator is given by
	 \begin{equation}
 H_{k}=q\partial_q -\frac{1}{2m}\left(X_i\partial_{X_i}+X_i^2\partial_{X_i}^2\right)-\frac{2k-8}{24}E_2,\quad X_i=e(x_i).
 \end{equation}
 The forms of index 2 follow from the inclusion $D_n(2)\leq nA_1$.
	 
	 Explicitly, all the $D_8$ forms are given by 
	 \begin{align}
	   	\omega_{-8,1}^{D_8}&=\prod_{i=1}^{8}\alpha(x_i),\\
	   	\phi_{-4,1}^{D_8}&=\frac{1}{\eta^{24}}\left(E_4\sum_{j=1}^4\prod_{i=1}^8 \theta_j(x_i)-\sum_{j=2}^4\theta_j(0)^8\prod_{i=1}^8\theta_j(x_i)\right)-E_4\omega_{8,1}^{D_8},\\
	   	 \phi_{-2,1}^{D_8}&=3 H_4 \phi_{4,1}^{D_8},\\
 \phi_{0,1}^{D_8}&=\frac{1}{32}\left(2 H_2 \phi_{-2,1}^{D_8} -E_4 \phi_{-4,1}^{D_8}\right),\\
	   	 \phi_{-2k,2}^{D_8}&=\frac{1}{k!(n-k)!}\sum_{\sigma\in S_n}\phi_{-2,1}^{A_1}(\tau,z_{\sigma(1)})\cdots \phi_{-2,1}^{A_1}(\tau,z_{\sigma(k)})\phi_{0,1}^{A_1}(\tau,z_{\sigma(k+1)})\cdots \phi_{0,1}^{A_1}(\tau,z_{\sigma(1)})\\
	   	&\quad k=3,4,5,6,7.
	 \end{align}
	
	Besides $\omega^{D_n}_{n,1}$, the $D_n$ forms with $n\leq 8$ can be obtained by setting $x_i=0, \text{ for } i>n$. 
For $D_4$, to agree with previous conventions \cite{Bertola,DelZotto:2017mee}, we use the forms $\omega_{-4,1}^{D_4}$ and
\begin{align}
\phi_{-4,1}^{D_4}&=\left.-\frac{1}{16}\phi_{-4,1}^{D_8}\right|_{x_{i>4}=0}\,,\\
\phi_{-2,1}^{D_4}&=\left.-\frac{1}{8}\phi_{-2,1}^{D_8}\right|_{x_{i>4}=0}\,,\\
\phi_{0,1}^{D_4}&=\left.2\phi_{0,1}^{D_8}\right|_{x_{i>4}=0}\,,\\
\phi_{-6,2}^{D_4}&=\left.\frac{1}{32}\phi_{-6,2}^{D_8}\right|_{x_{i>4}=0}\,.
\end{align}

\subsection{$C_n$ Jacobi forms }

For $n\geq 4$, we have 
\begin{equation}
    \text{Weyl}(C_n)=\text{Weyl}(D_n)\ltimes \mathbb {Z}_2.
\end{equation}

From the basis of $D_n$ Jacobi forms, the only one that is not invariant under the extra involution is $\omega_{n,1}^{D_n}$ which switches sign. We thus get a basis of $C_n$ Jacobi forms by squaring this last form
\begin{align*}
    \phi_{-k,1}^{C_n}&=\phi^{D_n}_{-k,1},\quad k=0,2,4\\
    \phi_{-k,2}^{C_n}&=\phi^{D_n}_{-k,2},\quad k=6,8,\dots,2n-2\\
    \phi_{-2n,2}^{C_n}&=(\omega_{-n,1}^{D_n})^2.
\end{align*}

 For $C_3$, we can get a basis of the forms by going down from $D_4$ as sketched in figure \ref{fig:jacobiBD} but, we decided to use the convention from \cite{Bertola} that uses the fact that the $C_3$ and $A_3$ root lattices are isomorphic and the Weyl group of $C_3$ compared to the one of $A_3$ just has an extra involution \cite{Bertola}.

Consider the map 
\begin{align}
		j:\left\{\begin{array}{ccc}
		C_3&\to& A_3\\
		\left(
		\begin{array}{c}
		x_1\\x_2\\x_3
		\end{array}\right)&\mapsto&\left(
		\begin{array}{c}
		\frac{x_1-x_2-x_3}{2}\\\frac{-x_1+x_2-x_3}{2}\\\frac{-x_1-x_2+x_3}{2}\\\frac{x_1+x_2+x_3}{2}
		\end{array}\right)
		\end{array}\right.
		\label{eq: C3 to A3}
	\end{align}
we then set 
\begin{align}
    \phi^{C_3}_{-k,1}&=\phi^{A_3}_{-k,1}\circ j,\quad k=0,2,4\\
    \phi^{C_3}_{-6,2}&=(\phi^{A_3}_{-3,1})^2\circ j
\end{align}

	\subsection{$F_4$ Jacobi forms}
	\label{sec:appF4}
	The generators of the ring $J(F_4)$ can be built from the results in \cite{Bertola,Wirthmuller:Jacobi} and were already given explicitly in \cite{adler2020structure}.
	
	The $F_4$ and $D_4$ lattices are isomorphic\footnote{If one takes the Euclidean norm, the exact statement is that $F_4(2)$, the $F_4$ lattice with norm scaled by 2 ( $(\cdot,\cdot)_{F_4(2)}=2(\cdot,\cdot)$), is isomorphic to $D_4$. In the present paper, we fix the norm by imposing the condition that short coroots have norm squared 2.}. An explicit isomorphism is given by\footnote{We choose this particular isomorphism as it maps the vector $\rho_L=\sum_{\alpha \in \Delta_L^+}\alpha$ of $F_4$ to the corresponding vector of $D_4$. This is only relevant for the considerations of section \ref{ss:GWinv}.}
	\begin{align}
		i:\left\{\begin{array}{ccc}
		F_4&\to& D_4\\
		\left(
		\begin{array}{c}
		x_1\\x_2\\x_3\\x_4
		\end{array}\right)&\mapsto&\left(
		\begin{array}{c}
		x_3-x_4\\-x_3-x_4\\x_1+x_2\\x_1-x_2
		\end{array}\right)
		\end{array}\right..
		\label{eq: F4 to D4}
	\end{align}
	
	The Weyl groups of the two lattices are different, however. The $F_4$ Weyl group is the full orthogonal group of the lattice which is given by the semi-direct product of the $D_4$ Weyl group and the $D_4$ Dynkin diagram symmetry, $S_3$:
	\begin{equation*}
		W(F_4)=O(F_4)=W(D_4)\ltimes \text{DynkinSym}(D_4)=W(D_4)\ltimes S_3.
	\end{equation*}
The generators of the ring $J(F_4)$ can then be obtained from the $D_4$ forms by imposing the extra $S_3$ symmetry. 

In terms of Weyl invariant polynomials, see appendix \ref{app:Weyl_invariant_polynomials}, this symmetry is straightforward to impose. As $S_3$ permutes the three external legs of the $D_4$ Dynkin diagram, this symmetry simply permutes the Weyl invariant polynomials $p_1,p_3,p_4$. Hence, $J(F_4)\subset J(D_4)$ is the sub-ring of Jacobi forms invariant under permutations of $p_1,p_3,p_4$. 

By looking directly at the expansion in terms of Weyl invariant polynomials, the following are the generators of $J(F_4)$:
\begin{align}
\label{eq:F4forms}
\phi_{0,1}^{F_4}&=\left(\phi_{0,1}^{D_4}-\frac{2}{3}E_4\phi_{-4,1}^{D_4}\right)\circ i\\
\phi_{-2,1}^{F_4}&=\phi_{-2,1}^{D_4}\circ i\\
\phi_{-6,2}^{F_4}&=\left(\phi_{-6,2}^{D_4}-\frac{1}{18}\phi_{-2,1}^{D_4}\phi_{-4,1}^{D_4}\right)\circ i\\
\phi_{-8,2}^{F_4}&=\left(\left(\phi_{-4,1}^{D_4}\right)^2+3\left(\omega_{-4,1}^{D_4}\right)^2\right)\circ i\\
\phi_{-12,3}^{F_4}&=\left(\phi_{-4,1}^{D_4}\left(\omega_{-4,1}^{D_4}\right)^2-\frac{1}{9}\left(\phi_{-4,1}^{D_4}\right)^3\right)\circ i
\end{align}

\subsection{$G_2$ Jacobi forms}
The $G_2$ and $A_2$ root lattices are the same and the Weyl groups differ only by the involution $x\mapsto -x$ which is in $W(G_2)$ but not in $W(A_2)$. The only generator of $J(A_2)$ not invariant under this transformation is $\phi_{-3,1}$ which gets maps to minus itself. Therefore we can get a set of generators of $J(G_2)$ by squaring this form.

\begin{align}
    \phi_{0,1}^{G_2}&=\phi_{0,1}^{A_2}\\
    \phi_{-2,1}^{G_2}&=\phi_{-2,1}^{A_2}\\
    \phi_{-6,2}^{G_2}&=(\phi_{-3,2}^{A_2})^2.
\end{align}

\section{Specialization formulas}
\label{app:specializationformulas}
\subsection{$A$ series}
The derivative of a modular form is only quasi-modular form. The term proportional to $E_2$ in \ref{eq:Zoperator} is precisely what is needed to cancel the anomalous modular transformation of the other term. This second term is important because if it weren't there, the action of $\cZ$ in a Jacobi modular form wouldn't give another Jacobi modular form. However, at this point, we only want to find the proportionality constant between the different forms in figure \ref{fig:jacobiA}, therefore it is enough to focus on the first term. 

Consider then, the form $\phi_{-k,1}^{A_n}$ with $k=0,2,\dots,n$ restricted to the subspace $x_{n+1}=0$
\begin{align*}
    \phi_{-k,1}^{A_n}|_{x_{n+1}=0}&=\left.\frac{1}{(2\pi i)^{n+1-k}}\left(\sum_{i=1}^{n+1}\frac{\partial}{\partial x_i}\right)^{n+1-k}\prod_{i=1}^{n+1}\alpha(x_i)\right|_{x_{n+1}=0}+\dots\\
    &=(n+1-k)\frac{1}{(2\pi i)^{n+1-k}}\alpha'(0)\left(\sum_{i=1}^{n}\frac{\partial}{\partial x_i}\right)^{(n-1)+1-k}\prod_{i=1}^{n+1}\alpha(x_i)+\dots\\
    &=-(n+1-k)\phi_{-k,1}^{A_{n-1}},
\end{align*}
where $\dots$ means a term proportional to $E_2$ and we used $\alpha(0)=0$ and $\alpha'(0)=-2\pi i$.

We repeat the formula for reference
\begin{equation}
    \phi_{-k,1}^{A_n}|_{x_{n+1}=0}=\left\{\begin{array}{cc}
     0    &k=n+1  \\
    -(n+1-k)\phi_{-k,1}^{A_{n-1}}     & k=0,2,3,\dots,n
    \end{array}\right.
\end{equation}

\subsection{$B$ series}
The Weierstrass function is divergent at the origin. Then one has to be careful when going to $x_n=0$. The series expansion of the Weierstrass function is given by
\begin{equation}
    \wp(x_n)=-x_n^{-2}+o(x_n^2),
\end{equation}
then
\begin{equation*}
    x_n^{2k}\wp^{(2k-2)}(x_n)|_{x_n=0}=-(2k-1)!.
\end{equation*}
Using this, and $\left.\frac{\alpha^2(x_n)}{x_n^2}\right|_{x_n=0}=1$ in equation \ref{eq:Bforms} we find that
\begin{equation*}
    \phi^{B_n}_{-2i,1}|_{x_n=0}=
    \left\{\begin{array}{cc}
     0    &i=n  \\
    (2n-1)\phi^{B_{n-1}}_{-2i,1}   & i=0,\dots,n-1
    \end{array}\right.
\end{equation*}

Because the denominator for theories with $C_n$ gauge algebra goes to 0 after Higgsing, we give the reduction of the lowest weight form of $B_n$ to the smallest order in $x_n$

\begin{equation*}
    \phi_{-2n,1}^{B_n}=\frac{1}{2(n-1)}\alpha^2(x_n)\phi_{-2(n-1)}^{B_{n-1}}=\frac{x_n^2}{2(n-1)}\phi_{-2(n-1)}^{B_{n-1}}+o(x_n^3)
\end{equation*}

\subsection{$C/D$ series}
For $n\geq 4$, by definition of the forms there is no extra factor and the results in figure \ref{fig:jacobiBD} are exact. For $n=3$, the last line in the figure, our bases for $D_4$ and $C_3$ forms are different. A calculation shows:
\begin{align*}
    \phi^{D_4}_{0,1}|_{x_4=0}&=\frac{4}{3} E_4 \phi_ {-4,1}^{C_3}+2 \phi _{0,1}^{C_3}\\
    \phi^{D_4}_{-2,1}|_{x_4=0}&=-6\phi^{C_3}_{0,1}|_{x_4=0}\\
    \phi^{D_4}_{-4,1}|_{x_4=0}&=2\phi^{C_3}_{-4,1}|_{x_4=0}\\
    \phi^{D_4}_{-6,2}|_{x_4=0}&=-\phi^{C_3}_{-6,2}|_{x_4=0}\\
    \omega^{D_4}_{-4,1}|_{x_4=0}&=0
\end{align*}

\subsection{$F_4, C_4$ and $D_4$} 
The root lattices for $F_4,C_4$ and $D_4$ are isomorphic; the Weyl groups on the other hand decrease in order from $F_4$ via $C_4$ to $D_4$,
\begin{equation*}
    O(D_4)=\text{Weyl}(F_4)\supset\text{Weyl}(B_4)\supset\text{Weyl}(D_4)\,.
\end{equation*}
The formulas going from $F_4$ to $D_4$ and $B_4$ were already given, but we repeat them explicitly:

\begin{align*}
\phi_{0,1}^{F_4}&=\left(\phi_{0,1}^{C_4}-\frac{2}{3}E_4\phi_{-4,1}^{C_4}\right)\circ i=\left(\phi_{0,1}^{D_4}-\frac{2}{3}E_4\phi_{-4,1}^{D_4}\right)\circ i\\
\phi_{-2,1}^{F_4}&=\phi_{-2,1}^{B_4}\circ i=\phi_{-2,1}^{D_4}\circ i\\
\phi_{-6,2}^{F_4}&=\left(\phi_{-6,2}^{C_4}-\frac{1}{18}\phi_{-2,1}^{C_4}\phi_{-4,1}^{C_4}\right)\circ i=\left(\phi_{-6,2}^{D_4}-\frac{1}{18}\phi_{-2,1}^{D_4}\phi_{-4,1}^{D_4}\right)\circ i\\
\phi_{-8,2}^{F_4}&=\left(\left(\phi_{-4,1}^{C_4}\right)^2+3
\phi_{-8,2}^{C_4}\right)\circ i=\left(\left(\phi_{-4,1}^{D_4}\right)^2+3\left(\omega_{-4,1}^{D_4}\right)^2\right)\circ i\\
\phi_{-12,3}^{F_4}&=\left(\phi_{-4,1}^{D_4}\phi_{-8,2}^{C_4}-\frac{1}{9}\left(\phi_{-4,1}^{C_4}\right)^3\right)\circ i=\left(\phi_{-4,1}^{D_4}\left(\omega_{-4,1}^{D_4}\right)^2-\frac{1}{9}\left(\phi_{-4,1}^{D_4}\right)^3\right)\circ i.
\end{align*}

\subsection{$C_3$ to $G_2$}
We just have some coefficients in the relations in figure \ref{fig:jacobiB3}. An explicit calculation gives 

\begin{align*}
    \phi_{0,1}^{C_3}|_{\sum x=0}&=-4\phi_{0,1}^{G_2}\\
    \phi_{-2,1}^{C_3}|_{\sum x=0}&=-2\phi_{-2,1}^{G_2}\\
    \phi_{-4,1}^{C_3}|_{\sum x=0}&=0\\
    \phi_{-6,2}^{C_3}|_{\sum x=0}&=\phi_{-6,2}^{G_2}\\
\end{align*}

    \section{Weyl invariant polynomials} \label{app:Weyl_invariant_polynomials}
    
    The most straightforward path towards determining the coefficients in the expansion of the numerator $\cN$ of $Z_k$ in the ansatz \eqref{eq:ZkAnsatz} is to expand $Z_k$ in the exponentiated K\"ahler parameters and compare coefficients to the corresponding expansion of the topological string partition function. We have found it computationally advantageous to first express all quantities at a given order in $q$ and $\gs$ in terms of Weyl invariant polynomials.
    
    A Weyl invariant polynomial for the algebra $\mg$ is a function $p:\mathfrak h^*\to \mathbb C$ such that
    \begin{equation*}
        p(x+\alpha^\vee)=p(x), \quad \forall \alpha^\vee \in \LambdaCR(\mg)
    \end{equation*}
    and
    \begin{equation*}
        p(w x)=p(x), \quad \forall w \in \weyl( \mg).
    \end{equation*}
    It is shown in \cite{lorenz2006multiplicative,Bourbaki} that a set of generators for the ring of $\weyl(\mg)$ invariant polynomials is given by
    \begin{equation*}
        p_i(x)=\sum_{\omega \in \weyl(\mg)\{\omega_i\}}\exp[2\pi i (\omega,x)], \quad \omega_i \text{ a fundamental weight}.
    \end{equation*}
   
    The Weyl invariant polynomials are also useful in imposing Dynkin diagram symmetry. This symmetry permutes simple roots of the gauge algebra or, equivalently, fundamental weights. It hence acts by permutation on the $p_i$, rendering the construction of invariant polynomials straightforward. For instance, Weyl and Dynkin diagram symmetric polynomials for the Lie algebra $D_4$, consist of all polynomials in the corresponding $p_i$ which are invariant under permutations of $p_1,\,p_3$ and $p_4$ (the index 2 being assigned to the central node).

    \section{Rank 1 Higgsing trees} \label{app:Higgsing_trees}
     
    The manifolds underlying rank 1 6d theories are elliptic fibrations
     \begin{equation*}
        \xymatrix{ E \ar[r]&X\ar[d]^\pi\\ 
        &B\\}
    \end{equation*}
    We assume that the fibration contains a global section, so that $X$ can be described by a Weierstrass model. Defining $X$ as the zero set of an anti-canonical section of the weighted projective space
    \begin{equation} \label{eq:ambient_space}
        \mathbb{P}^{2,3,1} (2 K_B \oplus 3 K_B \oplus \cO)\,,
    \end{equation}
    where $\mathcal{O}$ and $K_B$ denote the trivial and the canonical line bundle respectively of the base $B$ respectively, guarantees that it is both Calabi-Yau and an elliptic fibration. With $x$, $y$, $z$ denoting sections of the vector bundle being projectivized, a generic such section can be written as 
    \begin{equation} \label{eq:weierstrass}
        y^2 = x^3 + f x z^4 + g z^6\,,
    \end{equation}
    with $f$ and $g$ sections of $\mathcal{O}(-4K_B)$ and $\mathcal{O}(-6K_B)$ respectively. The zero set of the section \eqref{eq:weierstrass} is generically singular. The singular locus lies along the zero set of the discriminant of the fibration, given by $\Delta = 4 f^3 + 27 g^3 \in \mathcal{O}(-12K_B)$. Along the locus $\Delta = 0$, the elliptic fiber degenerates. In F-theory parlance, this signals the presence of D7-branes. The singularities can be resolved by successive blow-ups. For this process to preserve the Calabi-Yau condition, the vanishing order of $(f, g, \Delta)$ along any divisor in the base must be strictly smaller than $(4,6,12)$. As a result, the possible singularities along a divisor in $B$ must be of Kodaira type. The resolution of the singularity can be worked out explicitly using Tate's algorithm \cite{Bershadsky:1996nh,Katz:2011qp}.
  
    For rank 1 models, the base can be chosen as the total space of the line bundle $\cO(-n) \rightarrow \IP^1$ (which coincides with the normal bundle of the base curve of the Hirzebruch surface $\mathbb{F}_n$). The bases that lead to good F-theory models have $0 \leq n \leq 8$ or $n=12$ \cite{Morrison:2012js}. The zero set of the generic section of the anti-canonical bundle of the projective bundle \eqref{eq:ambient_space} is singular for $n>2$. The generic singularity leads to the maximally Higgsed or non-Higgsable model, in the terminology of \cite{Morrison:2012js}. Upon specializing to a subset of sections (this is the process of specialization of complex structure referred to in the body of this paper), the singularity can be enhanced, leading to gauge theories with higher rank gauge symmetry and charged matter.
    
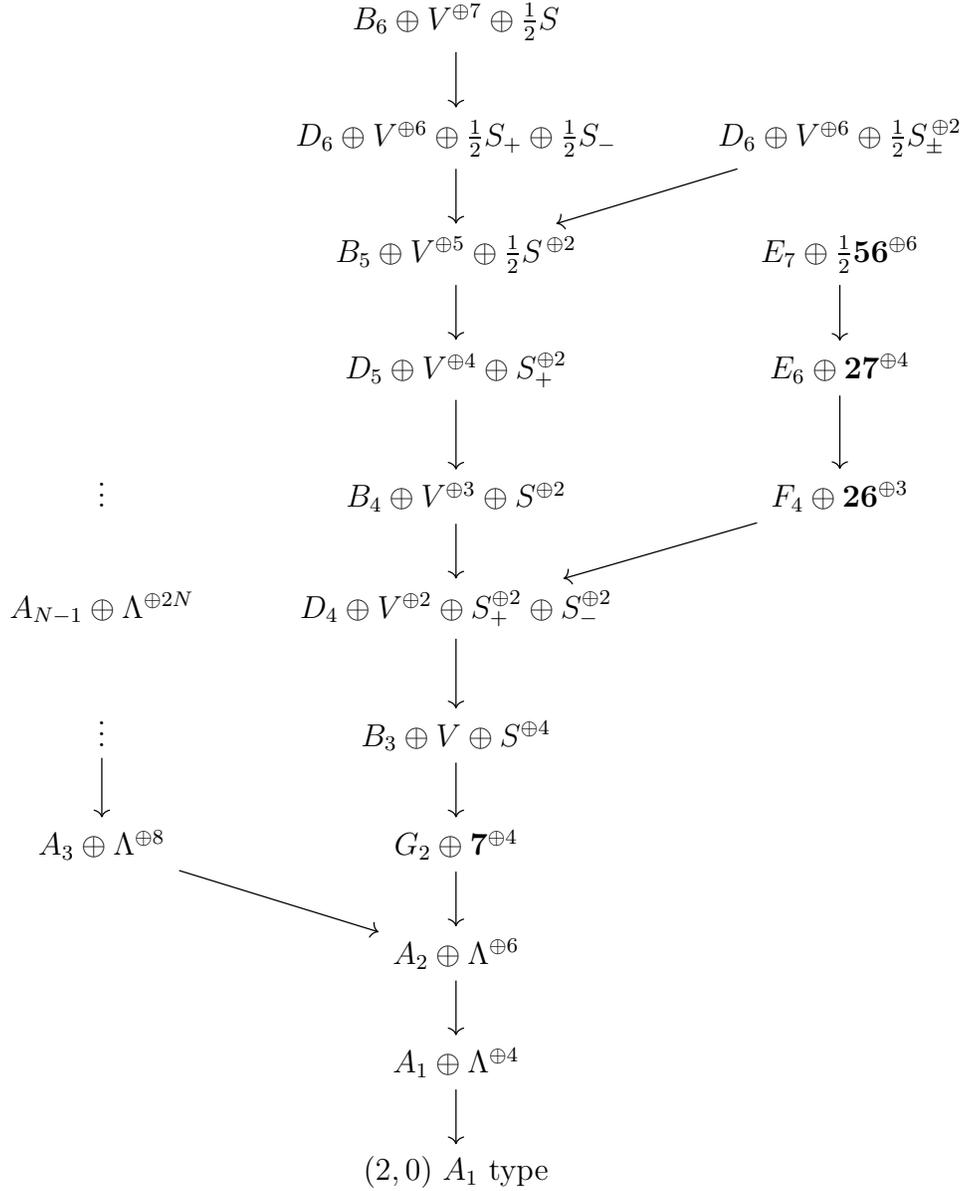
\begin{figure}[ht]
        \centering
        \begin{tikzcd}
        &B_6 \oplus V^{\oplus 7} \oplus \tfrac12 S \ar[d]&\\
        &D_6 \oplus V^{\oplus 6} \oplus \tfrac{1}{2} S_+ \oplus \tfrac12 S_-\ar[d]&D_6 \oplus V^{\oplus 6} \oplus \tfrac{1}{2} S_\pm^{\, \oplus 2}\ar[dl]\\
        &B_5 \oplus V^{\oplus 5} \oplus \tfrac{1}{2} S^{\,\oplus 2}\ar[d]&E_7 \oplus \tfrac{1}{2}{\bf 56}^{\oplus 6}\ar[d]\\
        &D_5 \oplus V^{\oplus 4} \oplus S_+^{\oplus 2} \ar[d]&E_6 \oplus {\bf 27}^{\oplus 4}\ar[d]\\
        \vdots&B_4 \oplus V^{\oplus 3} \oplus S^{\oplus 2}\ar[d] &F_4\oplus {\bf 26}^{\oplus 3}\ar[dl]\\
        A_{N-1} \oplus \Lambda^{\oplus 2N}&D_4 \oplus V^{\oplus 2} \oplus S_+^{\oplus 2}\oplus S_-^{\oplus 2}\ar[d]&\\
        \vdots\ar[d]&B_3 \oplus V \oplus S^{\oplus 4}\ar[d]&\\
        A_3 \oplus \Lambda^{\oplus 8}\ar[dr]&G_2 \oplus {\bf 7}^{\oplus 4}\ar[d]&\\
       &A_2 \oplus \Lambda^{\oplus 6}\ar[d]&\\
       &A_1 \oplus \Lambda^{\oplus 4}\ar[d]&\\
         &(2,0)\ A_1\text{\ type}
        \end{tikzcd}
        \caption{Infinite length Higgsing tree over the base $\cO(-2) \rightarrow \IP^1$, with the M-string theory at its root.}\label{fig:higgsM}
    \end{figure}

    We give a few examples of the resulting Higgsing trees in figure \ref{fig:higgsA3}, \ref{fig:higgsD4} and \ref{fig:higgsM}, following \cite{Bershadsky:1996nh, DelZotto:2018tcj}. The base of the trees corresponds to the maximally Higgsed models, and each successive node corresponds to a further specialization of the complex structure and resolution of the ensuing singularity. The nodes are labelled by the gauge group and matter content of the corresponding F-theory compactification. We denote, following \cite{DelZotto:2018tcj}, the fundamental representation of $A_{N-1}$ by $\Lambda$ stands for; $V$ and $S$ ($S_{\pm}$) denote the vector and spinor (Weyl spinors) representation of $B_N$ ($D_N$) respectively. For exceptional Lie algebras, we label the irreducible representation by its dimension. 

    \begin{figure}[ht]
        \centering
        \begin{tikzcd}
        D_6\oplus V^{\oplus 5} \oplus \frac{1}{2} S_\pm\ar[d]&\\
        B_5\oplus V^{\oplus 4} \oplus \frac{1}{2} S\ar[d]& E_7  \oplus (\tfrac{1}{2} {\bf 56})^{\oplus 5}\ar[d]\\
        D_5 \oplus V^{\oplus 3} \oplus S_+\ar[d]& E_6  \oplus {\bf 27}^{\oplus 3} \ar[d]\\
        B_4 \oplus V^{\oplus 2} \oplus S\ar[d] & F_4 \oplus {\bf 26}^{\oplus 2} \ar[dl]\\
        D_4 \oplus V \oplus S_+ \oplus S_-\arrow[d]&\\
        B_3 \oplus S^{\oplus 2} \arrow[d]&\\
        G_2 \oplus {\bf 7} \arrow[d]&\\
         A_2&
        \end{tikzcd}
        \caption{Finite length Higgsing tree over the base $\cO(-3) \rightarrow \IP^1$, with the maximally Higgsed gauge group $A_2$ at its root.}\label{fig:higgsA3}
    \end{figure}
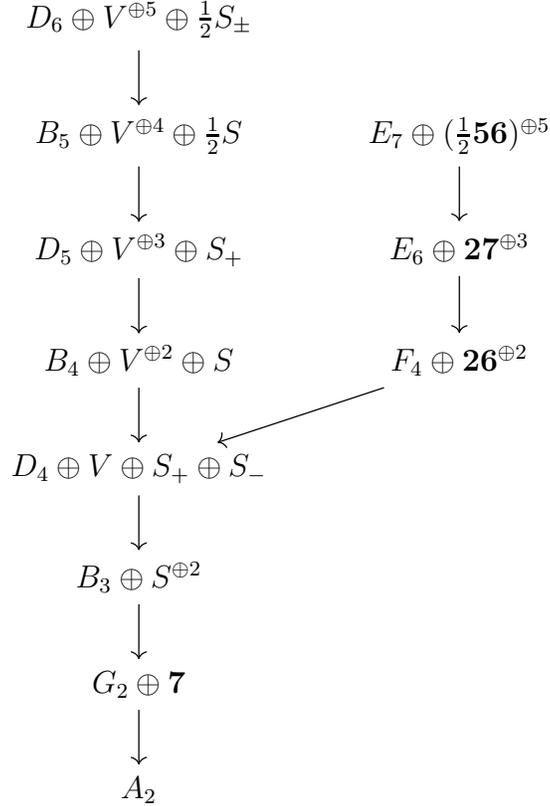

    \begin{figure}[ht]
        \centering
        \begin{tikzcd}
        \vdots\\
         D_{N+1}\oplus V^{\oplus (2N-6)}\ar[d]&\\
        B_N  \oplus V^{\oplus (2N-7)}&\\
        \vdots\ar[d]& E_7 \oplus \tfrac{1}{2}{\bf 56}^{\oplus 4}\ar[d]\\
        D_5\oplus V^{\oplus 2}\ar[d]&E_6 \oplus {\bf 27}^{\oplus 2}\ar[d]\\
        B_4\oplus V \ar[d]&F_4\oplus {\bf 26}\ar[dl]\\
         D_4&
        \end{tikzcd}
        \caption{Infinite length Higgsing tree over the base $\cO(-4) \rightarrow \IP^1$, with the maximally Higgsed gauge group $D_4$ at its root.}\label{fig:higgsD4}
    \end{figure}

   \clearpage
    \bibliography{biblio}

\end{document}